\documentclass[a4paper,12pt,twoside,onecolumn,openright,oldfontcommands]{memoir}

\usepackage{times}
\usepackage{graphicx}
\usepackage{cite}
\usepackage{bm}
\usepackage{amssymb,latexsym,pifont}
\usepackage{chngcntr}
\usepackage[a4paper,bookmarks,breaklinks,colorlinks=true,pdfstartview=Fit,citecolor=green,urlcolor=blue]{hyperref}
\hypersetup{%
  pdftitle  = {Anomalous transport in magnetized shear flow},
  pdfauthor = {Zlatan Dimitrov},
  pdfsubject= {PhD Thesis}
}
\usepackage[]{amsmath}
\usepackage{pdfpages}

\usepackage{tikz}
\usetikzlibrary{%
    decorations.pathreplacing,%
    decorations.pathmorphing%
}

\newcommand{\Eqref}[1]{Eq.~(\ref{#1})}
\newcommand{\Fref}[1]{Fig.~\ref{#1}}
\newcommand{\alf}{{Alfv\'en~}}
\newcommand{\apj}{{ApJ~}}
\newcommand{\onlinecite}[1]{\cite{#1}}

\newcommand{\ii}{\mathrm{i}}
\newcommand{\nn}{\nonumber}
\newcommand{\be}{\begin{equation}}
\newcommand{\ba}{\begin{eqnarray}}
\newcommand{\ea}{\end{eqnarray}}
\newcommand{\ee}{\end{equation}}
\newcommand{\citeauthor}[1]{\cite{#1}}

\newcommand{\openone}{\leavevmode\hbox{\small1\kern-3.3pt\normalsize1}}

\usepackage{color,calc,soul}
\definecolor{nicered}{rgb}{.647,.129,.149}
\makeatletter
\newlength\dlf@normtxtw
\newsavebox{\feline@chapter}
\newcommand\feline@chapter@marker[1][4cm]{%
  \sbox\feline@chapter{%
    \resizebox{!}{#1}{\fboxsep=1pt%
      \colorbox{nicered}{\color{white}\bfseries\sffamily\thechapter}%
  }}%
  \rotatebox{90}{%
    \resizebox{%
      \heightof{\usebox{\feline@chapter}}+\depthof{\usebox{\feline@chapter}}}%
	      {!}{\scshape\so\@chapapp}}\quad%
  \raisebox{\depthof{\usebox{\feline@chapter}}}{\usebox{\feline@chapter}}%
}
\newcommand\feline@chm[1][4cm]{%
  \sbox\feline@chapter{\feline@chapter@marker[#1]}%
  \makebox[0pt][l]{
    \makebox[1cm][r]{\usebox\feline@chapter}%
}}
\makechapterstyle{daleif1}{

  \renewcommand\printchapternum{\null\hfill\feline@chm[2.5cm]\par}

}
\makeatother
\chapterstyle{daleif1}

\begin{document}

\frontmatter

\title{Anomalous transport in magnetized shear flow}

\author{Zlatan Dimitrov Dimitrov}

\begin{titlingpage}
\thispagestyle{titlingpage}
\begin{center}
%
\vskip 1cm
{\LARGE \textsc{\textbf{\thetitle}}}
\vskip 1cm
{\Large \textsc{\theauthor}}\par
\smallskip
\vskip 0.6cm
\begin{center}
\includegraphics[height=7cm]{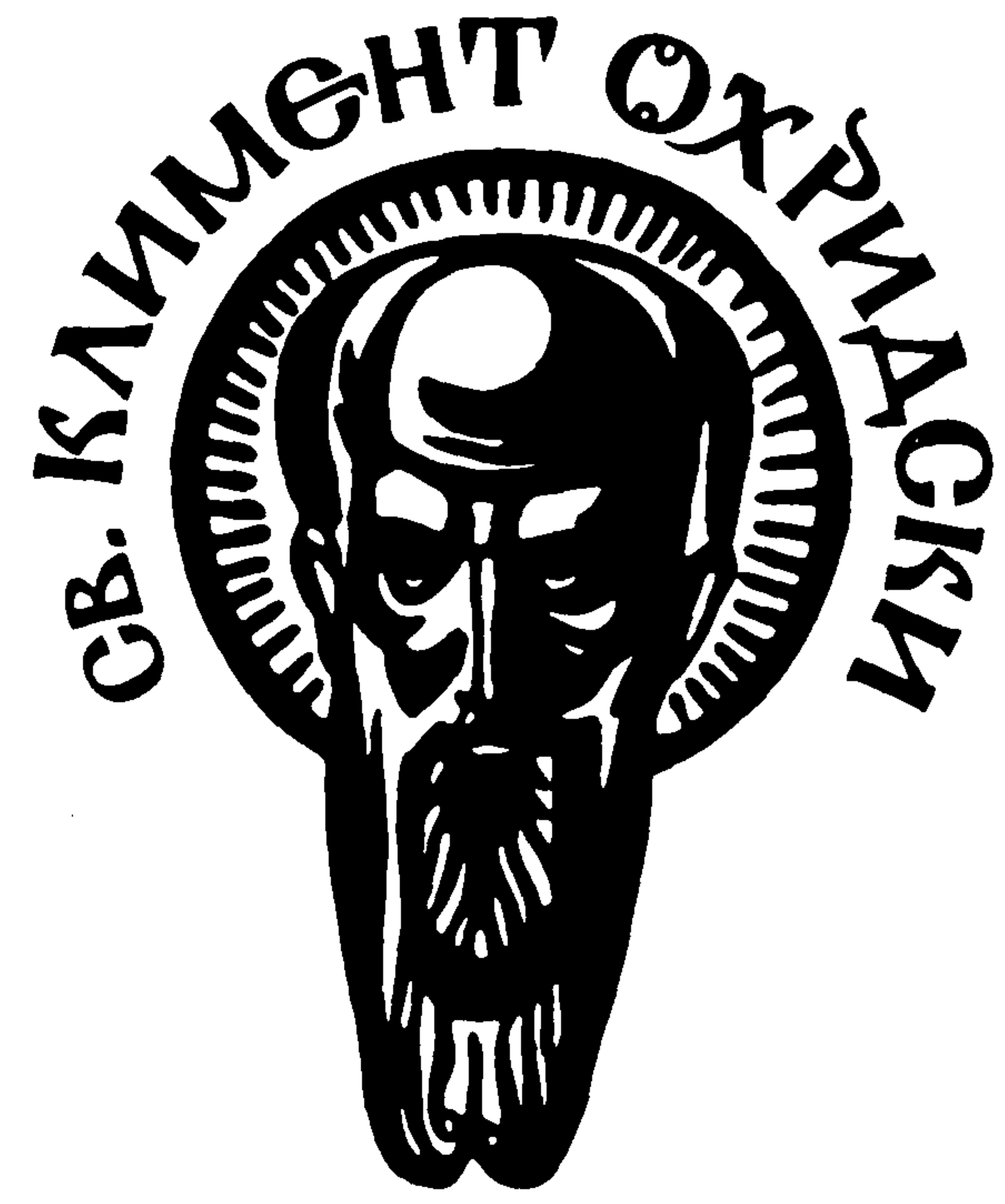}
\end{center}
Thesis submitted for the degree of Doctor of Philosophy\\
of the ``St. Clement of Ohrid'' University of Sofia
\vskip 0.4cm
\textbf{Jury:}
\\

\begin{tabular}{lcl}
\multicolumn{3}{l}{Chairman:} \\
Prof. D. Sc. Stoytcho  Yazadjiev & - & Faculty of Physics, Sofia University \\
\multicolumn{3}{l}{Referees:} \\
Prof. D. Sc. Ivan Zhelyazkov & - & Faculty of Physics, Sofia University \\
Assoc. Prof. Dr. Monio Kartalev & - & Institute of Mechanics, Bulgarian Academy of Sciences \\
\multicolumn{3}{l}{Report:} \\
Assoc. Prof. Dr. Dantchi Koulova-Nenova & - & Institute of Mechanics, Bulgarian Academy of Sciences \\
\multicolumn{3}{l}{Thesis Advisor:} \\
Prof. D. Sc. Todor Mishonov & - & Faculty of Physics, Sofia University
\end{tabular}

\vskip 1.4cm
Sofia \\ 2012\\

\vskip 0.2cm
Version: 28/09/2012

\end{center}
\end{titlingpage}

\thispagestyle{empty}

\maxsecnumdepth{subsubsection}
\setsecnumdepth{subsubsection}
\settocdepth{subsection}

\tableofcontents

\mainmatter
\setsecnumdepth{none}
\maxsecnumdepth{none}
\chapter{Introduction}

After the initial stage of fast expanding of the hot Universe
comes epoch of temperature fall and subsequent formation of
dense clouds of hydrogen \cite{Gamov}.
Due to process of accretion some of this clouds became compact objects.
Initially accretion is spherical, but then take the shape of the disk,
and this 2D disk collect matter more efficient than 3D sphere.
Accretion disks provide the mechanism of redistribution of
angular moment and extraction of potential energy, което leads
to occurrence of compact objects from gas clouds.
To fall moving along a spiral trajectory
on the central gravitating body, a particle rotationg
around must reduce angular moment and energy.
Outward of angular moment and movement of
gas particles in the opposite direction (to the central body)
happens due to of viscose friction. We can think
that the disc is composed of multiple adjacent rings friction
each other, each inner ring has a large
angular velocity to the adjacent outer ring and thus
friction of the drive ring decreased the rate of internal.

If the gravitating body with radius $R$ has mass $M$ then
gravitational potential energy relased
in the process of accretion $m$ will be \cite{AccretionPower}
\begin{equation}
E_{acc}=\frac{GMm}{R}.
\end{equation}
For example, a neutron star this energy would be about $10^{20}$ erg/g.
Energy released for some cases of black holes can reach approximately 40\% $mc^2$, 
while the energy released in nuclear reaction $\mathrm{H}\rightarrow \mathrm{He}$ is of the order
$E_{nuc} \approx 0.007mc^2$ or $6\cdot10^{18}$ erg/g.

Luminosity of the same accretion object is
\begin{equation}
 L_{acc}=\frac{GM\dot{M}}{R},
\end{equation}
where $\dot{M}$ is accretion rate. From observational data we have estiamtion for luminosity
of quasars $L_{acc} \sim 10^{12} L_\odot$

The occurrence of intense dissipation in accretion flows is among
the long-standing unsolved problems in
astrophysics.\cite{Balbus:91,Fridman:08} Revealing how the
magnetized turbulence creates shear stress tensor is of primary
importance to understand the heating mechanism and the transport of
angular momentum in accretion disks. The transport of angular
momentum at greatly enhanced rates is important for the main problem
of cosmogony, that is understanding the dynamics of creation of
compact astrophysical objects.\cite{Balbus:98,Balbus:03} Without a
theory explaining the enhanced energy dissipation in accretion flows
of turbulent magnetized plasma we would have no clear picture of how
our solar system has been created, why the angular momentum of the
Sun is only $2$\% of the angular momentum of solar system, while
carrying 99\% of the solar system's mass, why quasars are the most
luminous sources in the universe. The importance of friction forces
and convection as well the problem of angular momentum
redistribution for the first time was emphasized by von
Weitz\"acker;\cite{Weitzaecker:48} a very detailed bibliography on
the physics of disks is provided in the monograph by Morozov and
Khoperskov.\cite{Morozov:05} Gravitational forces, angular momentum
conservation, and dissipation processes become the main ingredients
of the standard model of accretion disks by Shakura and
Sunyaev\cite{Shakura:73} and Lynden-Bell and
Pringle.\cite{Lynden-Bell:74} Lynden-Bell\cite{Lynden-Bell:69}
suggested that quasars are accretion disks and Shakura and Sunyaev
introduced alpha phenomenology for the stress tensor
\be \sigma_{R\varphi}=\alpha p,\qquad p\sim\rho c_\mathrm{s}^2, \ee
where $\alpha$ is a dimensionless parameter, $p$ is the pressure
$\rho$ is the mass density, $c_\mathrm{s}$ is the sound speed,
and the indices of the tensor come from the cylindrical
coordinate system $(R,\varphi,z)$ related to the disk rotating
around the $z$-axis. As accretion disks can have completely
different scales for protostellar disks, mass transfer disks, and
disks in active galactic nuclei (AGN) it is unlikely that Coriolis
force is the main cause for dissipation, while the bending of the
trajectories is a critical ingredient.  We suppose that the
shear dissipation in magnetized turbulent plasma is a robust and
very general phenomenon which can be analyzed as a local heating for
approximately homogeneous magnetic field and gradient of the
velocity. 

There is almost a consensus that the magnetic field is essential and
should be introduced from the very beginning in the
magnetohydrodynamic (MHD) analysis. The likely importance of MHD
waves on the cosmogony of solar system has been pointed out by
Alfv\'en: ``At last some remarks are made about the transfer of
momentum from the Sun to the planets, which is fundamental to the
theory. The importance of the magnetohydrodynamic waves in this
respect is pointed out.\cite{Alfven:46}''
\maxsecnumdepth{subsubsection}
\setsecnumdepth{subsubsection}
\chapter{Derivation of General Set of Equations}

\section{Model and MHD Equations}
%
Our starting point are the conservation laws for energy and momentum for an incompressible fluid with mass density~$\rho$
\begin{eqnarray}
&&\partial_t(\rho V_i) + \partial_k(\Pi_{ik})=0\,, \\
&&\frac{\partial}{\partial t}\left(\frac{\rho V^2}{2} + \rho\tilde{\varepsilon} + \frac{B^2}{2\mu_0}\right)
+ \mathrm{div}\,\mathbf{q}=0\,,\\
&&\mathrm{div}\mathbf{V}=0, \qquad \rho=\mathrm{const},
\end{eqnarray}
where we have for total stress tensor $\mathbf{\Pi}$ and heat flux $\mathbf{q}$ respectively
\begin{eqnarray}
\label{ch1stress}
\Pi_{ik}&=&\rho V_iV_k + P\delta_{ik} - \eta\left(\frac{\partial V_i}{\partial x_k}+ \frac{\partial V_k}{\partial x_i}\right)
- \frac1{\mu_0}\left(B_iB_k -\frac12\delta_{ik}B^2\right),
\\
\label{ch1heat}
\mathbf{q}&=&\rho\left(\frac{V^2}{2} +\tilde{w} \right)\mathbf{V} - \mathbf{V}\cdot\boldsymbol{\sigma}'
-\varkappa\nabla T
+\frac1{\mu_0}[\mathbf{B}\times(\mathbf{V}\times\mathbf{B})] \nonumber \\
&& - \frac{\varepsilon_0c^2\varrho}{\mu_0}(\mathbf{B}\times\mathrm{curl}\,\mathbf{B}),
\end{eqnarray}
where $\tilde{\varepsilon}$ is the internal energy per unit mass, $\tilde{w}$ is the enthalpy per unit mass,
$\sigma_{ij}^{\prime}\equiv\eta(\partial_iV_k+\partial_kV_i)$ the viscous part of the stress tensor for an incompressible fluid,
$\eta$ is the viscosity, $\varkappa$ is the heat conductivity, $T$ is the temperature, $\varrho$ is the Ohmic resistivity.
The formula are written in SI, for a transition to Gaussian system we substitute $\mu_0=4\pi$ and $\varepsilon_0=1/4\pi,$
i.e. expressions are written in invariant form.

For the  magnetic field's energy density rate of change  we have
\begin{eqnarray}
\frac1{2\mu_0}\frac{\partial}{\partial t} \, B^2 =
\frac1{\mu_0}\mathbf{B}\cdot\frac{\partial B}{\partial t} &=&
\frac1{\mu_0}\mathbf{B}\cdot[\nabla\times(\mathbf{V}\times\mathbf{B}) -\nu_m\nabla\times(\nabla\times\mathbf{B})]\nonumber \\ &=&
\frac1{\mu_0}\nabla\cdot(\mathbf{B}\times(\mathbf{V}\times\mathbf{B}) -\nu_m\mathbf{B}\times(\nabla\times\mathbf{B})).
\end{eqnarray}
We calculate the divergence of the total stress tensor \Eqref{ch1stress}, and using that
\begin{eqnarray}
&&\partial_k\frac1{\mu_0}\left(B_iB_k -\frac12B^2\delta_{ik}\right)=\frac1{\mu_0}\left(\mathbf{B}\cdot\nabla\mathbf{B} - \frac12\nabla B^2\right) \nn\\
&&=\frac1{\mu_0}(\mathbf{B}\times\mathrm{curl}\,\mathbf{B})=\mathbf{j}\times\mathbf{B},
\end{eqnarray}
we obtain the equation of motion for an incompressible plasma,
\begin{equation}
\rho\left(\partial_t\mathbf{V} + \mathbf{V}\cdot\nabla\mathbf{V}\right)= - \nabla P + \mathbf{j}\times\mathbf{B} + \eta\nabla^2\mathbf{V}.
\end{equation}
To close the system of equations we need to use Amp\'ere's law, $\nabla\times\mathbf{B}=\mu_0\mathbf{j},$
and Ohm's law, $\mathbf{E}+\mathbf{V}\times\mathbf{B}=\varrho\mathbf{j},$ and supplement them with Faraday's law
\begin{equation}
\frac{\partial \mathbf{B}}{\partial t}=-\nabla\times\mathbf{E}.
\end{equation}
For the second set of MHD equations we have
\begin{equation}
\frac{\partial \mathbf{B}}{\partial t}=\nabla\times\left(\mathbf{V}\times\mathbf{B}-\frac{\varrho}{\mu_0}\nabla\times\mathbf{B} \right).
\end{equation}
The MHD equations for an incompressible fluid
$\rho=\mathrm{const},$ in homogeneous magnetic field $\mathbf{B}_0$,
shear flow with rate $A$, angular velocity $\mathbf{\Omega}=\Omega\mathbf{e}_z,$ and dimensionless angular velocity $\omega=\Omega/A$ are
\begin{eqnarray}
\label{ch1MHD}
\rho \, \mathrm{D}_t \mathbf{V}
&=& - \nabla P + \frac1{\mu_0}\left(\nabla\cdot\mathbf{B}\right)\mathbf{B} -\nabla\frac{B^2}{2\mu_0}
-2\rho\,\mathbf{\Omega}\times\mathbf{V}
 + \rho\nu_\mathrm{k} \nabla^2\mathbf{V},\\
\label{ch1MHD-B}
\mathrm{D}_t\mathbf{B}&=&\mathbf{B}\cdot\nabla\mathbf{V}+\nu_\mathrm{m}\nabla^2\mathbf{B},
\qquad \mathrm{div} \mathbf{V}=0, \qquad \mathrm{div} \mathbf{B}=0,
\end{eqnarray}
where $\mathrm{D}_t\equiv\partial_t+\mathbf V\cdot\nabla$ is the substantial
(convective) derivative,
$P$ is the pressure,  $\mathbf{j}$ is the current density and
$\nu_\mathrm{k}$ is the kinematic viscosity, The magnetic diffusivity
$\nu_\mathrm{m}=\varepsilon_0c^2\varrho$ is expressed by the Ohmic resistance $\varrho$ and
$\varepsilon_0=1/\mu_0c^2.$
In order to obtain a linear system of dimensionless MHD equations we use the following
anzats for the velocity $\bf V$, the magnetic field $\bf B$, the wave vector $\bf Q$, and the pressure P
\begin{eqnarray}
\label{ch1v_wave_Q}
 \mathbf{V}(t,\bf{r})&=&
\mathbf{V}_\mathrm{shear}(\mathbf{r})+\mathbf{V}_\mathrm{wave}(t,\mathbf{r}),\nn \\
\mathbf{V}_\mathrm{wave} &=& \ii V_\mathrm{A}\sum_{\mathbf{Q}}\mathbf{v}_{\mathbf{Q}}(\tau)\,\mathrm{e}^{\ii\mathbf{Q}\cdot\mathbf{X}}, \quad \mathbf{V}_\mathrm{shear}=Ax\mathbf{e}_y, \\
\label{ch1b_wave_Q}
\mathbf{B}(t,\bf{r})&=&\mathbf{B}_0+\mathbf{B}_\mathrm{wave}(t,\mathbf{r}), \quad \mathbf{B}_\mathrm{wave}(t,\mathbf{r})=
B_0\sum_{\mathbf{Q}}\mathbf{b}_{\mathbf{Q}}(\tau)\,\mathrm{e}^{\ii\mathbf{Q}\cdot\mathbf{X}},\\
\label{ch1p_wave_Q}
P(t,\mathbf{r})&=&P_0+ P_\mathrm{wave}(t,\mathbf{r}),
\quad
P_\mathrm{wave}(t,\mathbf{r})=\rho V_\mathrm{A}^2\sum_{\mathbf{Q}}\mathcal{P}_{\mathbf{Q}}(\tau)\mathrm{e}^{\mathrm{i}\mathbf{Q}\cdot\mathbf{X}}.
\end{eqnarray}
where the sums are actually integrals with respect to 3-dimensional Eulerian wave-vector space with independent coordinates
$Q_x,\,Q_y,\,Q_z$
\begin{equation}
\sum_{\mathbf{Q}}=\iiint_{-\infty,\,-\infty,\,-\infty}^{+\infty,\,+\infty,\, +\infty}
\frac{\mathrm{d} Q_x\mathrm{d} Q_y \mathrm{d} Q_z}{(2\pi)^3}
=\int\mathrm{d}\left(\frac{\mathbf{Q}}{2\pi}\right)=\int\frac{\mathrm{d}^3{Q}}{(2\pi)^3},
\end{equation}
i.e. the sum is a short notation for Fourier integration with omitted differentials, integral limits and $2\pi$ multipliers.
For the static magnetic $\mathbf{B}_0$ field with magnitude
$B_0=\sqrt{B_{0y}^2+B_{0z}^2}$ we suppose a vertical $B_{0z}$ and an
azimuthal $B_{0y}$ components parameterized by an angle $\theta$ and
an unit vector $\boldsymbol{\alpha}$. We also assume that the Alfv\'en velocity
$V_{\mathrm{A}}$ is much smaller than the sound speed $c_{\mathrm{s}}$
\begin{equation}
\mathbf{B}_0=B_0\boldsymbol{\alpha},
\quad \boldsymbol{\alpha}=(0,\,\alpha_y=\sin\theta,\, \alpha_z=\cos\theta),
\quad \mathbf{V}_\mathrm{A}=\frac{\mathbf{B}_0}{\sqrt{\mu_0\rho}},
\quad V_\mathrm{A}=\frac{B_0}{\sqrt{\mu_0\rho}}. \nonumber
\end{equation}
In the equations above we used the dimensionless space-vector
\be
\mathbf{X}=\left(\begin{array}{c} X\\Y\\Z \end{array} \right)
\equiv\frac{\mathbf{r}}{\Lambda}=\frac{A\mathbf{r}}{V_\mathrm{A}},\qquad
\Lambda\equiv\frac{V_\mathrm{A}}{A},\qquad \mathbf{r}=\left(\begin{array}{c} x\\y\\z \end{array} \right)
\ee
and its dimensionless wave-vector counterpart
$\mathbf{Q}$. $\Lambda$ is the characteristic length of the system which we suppose
to be much smaller than space inhomogeneities, e.g.\ the accretion disk thickness.

\section{Wave-vector representation}
%
\subsection{Linear terms}
%
We have a space homogeneous physical system and indispensably its modes bear the character of plane waves.
The purpose of the present section is to find the Fourier transformation of the all the terms in the MHD equations \Eqref{ch1v_wave_Q},
\Eqref{ch1b_wave_Q} and \Eqref{ch1p_wave_Q}.

Let us start, for example, with the pressure. According to \Eqref{ch1p_wave_Q} we have
\begin{eqnarray}
\frac{\nabla P}{\rho}&=&\frac{\ii V_\mathrm{A}^2}{\Lambda} \int
P_\mathbf{Q}(\tau)\mathbf{Q}\,\mathrm{e}^{\ii\mathbf{Q}\cdot\mathbf{X}}\frac{\mathrm{d}^3{Q}}{(2\pi)^3} =
\ii A V_\mathrm{A} \sum_\mathbf{Q} \mathbf{Q}P_\mathbf{Q}(\tau)\,\mathrm{e}^{\ii\mathbf{Q}\cdot\mathbf{X}}
\\
\mathcal{\hat{F}}(\frac{\nabla P}{\rho})&\equiv& \int \mathrm{e}^{-\ii\mathbf{Q}\cdot\mathbf{X}} \frac{\nabla P}{\rho}\,\mathrm{d}^3X
= \ii AV_\mathrm{A}\mathbf{Q}P_\mathbf{Q}(\tau).
\end{eqnarray}
Analogously, according to \Eqref{ch1v_wave_Q} and \Eqref{ch1b_wave_Q}, for the partial time derivatives we obtain
\begin{eqnarray}
&&\partial_t\mathbf{V}=
\ii A V_\mathrm{A}\sum_{\mathbf{Q}}\mathrm{e}^{\ii\mathbf{Q}\cdot\mathbf{X}}\partial_{\tau}\mathbf{v}_{\mathbf{Q}}(\tau), \qquad
\mathcal{\hat{F}} (\partial_t \mathbf{V})= iAV_\mathrm{A}\partial_\tau \mathbf{v}_\mathbf{Q}(\tau), \qquad
\label{ch1time_V}\\
&&\partial_t\mathbf{B}= AB_0\sum_{\mathbf{Q}}\mathrm{e}^{\ii\mathbf{Q}\cdot\mathbf{X}}\partial_{\tau}\mathbf{b}_{\mathbf{Q}}(\tau),
\qquad
\mathcal{\hat{F}}(\partial_t\mathbf{B})=AB_0 \partial_\tau \mathbf{b}_\mathbf{Q}(\tau)\,.
\label{ch1time_B}
\end{eqnarray}

More complicated are the Fourier transformations of the expressions, containing the shear flow $\mathbf{V}_\mathrm{shear}=V_\mathrm{A} X
\mathbf{e}_y$ and the wave variables
\begin{eqnarray}
\mathbf{V}_\mathrm{shear} \cdot \nabla \mathbf{V}_\mathrm{wave} &=&
-A V_\mathrm{A} \int \mathrm{e}^{\ii\mathbf{Q}\cdot\mathbf{X}}XQ_y \mathbf{v}(\tau) \frac{\mathrm{d}^3{Q}}{(2\pi)^3}\,,\\
\mathbf{V}_\mathrm{shear} \cdot \nabla \mathbf{B}_\mathrm{wave} &=&
\ii A B_0 \int \mathrm{e}^{\ii\mathbf{Q}\cdot\mathbf{X}} XQ_y \mathbf{b}(\tau) \frac{\mathrm{d}^3{Q}}{(2\pi)^3}\,.
\end{eqnarray}
Let variables $\mathbf{V}_\mathrm{wave}$ or $\mathbf{B}_\mathrm{wave}$ be presented by their Fourier components
$\psi(\mathbf{X})=\sum_\mathbf{Q}\mathrm{e}^{\mathrm{i} \mathbf{Q}\cdot \mathbf{X}}\psi_\mathbf{Q}$ and $\psi_\mathbf{Q}=
\mathcal{\hat{F}}(\psi(\mathbf{X}))$. Our task is to derive the Fourier transformation $\mathcal{\hat{F}}(\mathbf{X}\psi(\mathbf{X})).$
Using that $\mathbf{X} \mathrm{e}^{\mathrm{i} \mathbf{Q}\cdot \mathbf{X}}=-\mathrm{i}\partial_\mathbf{Q} \mathrm{e}^{\mathrm{i}
\mathbf{Q}\cdot \mathbf{X}}$  and the Gaussian theorem,
$\int_\mathcal{V}\mathrm{d}^3 Q \partial_\mathbf{Q}=\oint_{\partial\mathcal{V}}\mathrm{d}\mathbf{S}$
applied for the whole volume $\mathcal{V}$ in wave-vector space and its boundary $\partial\mathcal{V}$ we can make the partial
integration
\begin{eqnarray}
&&\mathbf{X}\psi(\mathbf{X})=\mathbf{X}\sum_\mathbf{Q}\mathrm{e}^{\mathrm{i} \mathbf{Q}\cdot \mathbf{X}}\psi_\mathbf{Q} \\
&&=-\mathrm{i}\sum_\mathbf{Q}\left\{(\partial_\mathbf{Q} \mathrm{e}^{\mathrm{i} \mathbf{Q}\cdot \mathbf{X}})\psi_\mathbf{Q}
=- \mathrm{e}^{\mathrm{i} \mathbf{Q}\cdot
\mathbf{X}}\partial_\mathbf{Q} \psi_\mathbf{Q}+\partial_\mathbf{Q}\left[\mathrm{e}^{\mathrm{i} \mathbf{Q}\cdot
\mathbf{X}}\partial_\mathbf{Q}\psi_\mathbf{Q}\right]\right\}=\sum_\mathbf{Q} \mathrm{e}^{\mathrm{i} \mathbf{Q}\cdot
\mathbf{X}}\ii\partial_\mathbf{Q}\psi_\mathbf{Q},\nn
\end{eqnarray}
in the limit $$\lim_{Q\rightarrow\infty}{(Q^3\psi_\mathbf{Q})}=0.$$
In such a way we derived the well-known in quantum mechanics operator representation $\hat{\mathbf{X}}=\ii\partial_\mathbf{Q}$
and derived the Fourier transformation
\begin{equation}
\mathcal{\hat{F}}(\mathbf{X}\psi(\mathbf{X}))=\ii \partial_\mathbf{Q}\psi_\mathbf{Q}.
\end{equation}
This expression is analogous to the Fourier transformation of the $\nabla$-operator
\begin{equation}
\mathcal{\hat{F}}(\nabla_{\!\!\mathbf{X}})=\ii\mathbf{Q},
\end{equation}
and gives
\be
\mathcal{\hat{F}}(X\mathbf{e}_y\cdot\nabla_{\!\!\mathbf{X}})=-Q_y\partial_{Q_x}.
\qquad \mathcal{\hat{F}}(\mathbf{V}_\mathrm{shear}\cdot\nabla)=-AQ_y\partial_{Q_x},
\qquad \mathbf{V}_\mathrm{shear} =V_\mathrm{A}X \mathbf{e}_y.
\label{ch1baliga}
\ee
Those relations give that
\begin{equation}
\mathcal{\hat{F}}\left[\mathrm{D}_t^{\mathrm{\,shear}}\equiv\partial_t+\mathbf{V}_{\mathrm{shear}}\cdot\nabla=\partial_t+AX\partial_{_Y}
\right ]
=A\left\{\mathrm{D}_\tau^{\mathrm{\,shear}}\equiv\partial_\tau-Q_y\partial_{Q_x}
=\partial_\tau+\mathbf{U}_{\mathrm{shear}}(\mathbf{Q})\cdot\partial_{\mathbf{Q}}\right\}.
\end{equation}
In other words Fourier transformation of a linearized substantial derivative is again a linearized substantial derivative, but only in the wave-vector space.
For this purpose we introduced the field of shear flow in the wave-vector space
$\mathbf{U}_{\mathrm{shear}}(\mathbf{Q})\equiv-Q_y\mathbf{e}_x$; confer this result with
$\mathbf{V}_\mathrm{shear}/V_\mathrm{A} =X \mathbf{e}_y.$
Returning back to the velocity and magnetic field we arrive at
\ba
&&\mathcal{\hat{F}}[(\partial_t+\mathbf{V}_{\mathrm{shear}}\cdot\nabla)\mathbf{V}_{\mathrm{wave}}]
=\ii AV_\mathrm{A}[\partial_\tau+\mathbf{U}_{\mathrm{shear}}(\mathbf{Q})\cdot\partial_{\mathbf{Q}}]\mathbf{v}_\mathbf{Q}\,,\\
&&\mathcal{\hat{F}}[(\partial_t+\mathbf{V}_{\mathrm{shear}}\cdot\nabla)\mathbf{B}_{\mathrm{wave}}]
= AB_0[\partial_\tau+\mathbf{U}_{\mathrm{shear}}(\mathbf{Q})\cdot\partial_{\mathbf{Q}}]\mathbf{b}_\mathbf{Q}\,.
\ea
For the derivation of these equations we used \Eqref{ch1baliga} and according \Eqref{ch1time_V} and \Eqref{ch1time_B}
$\mathcal{\hat{F}}(\partial_t)=A\partial_\tau.$

Very simple is the Fourier transformation of the dissipative terms which is reduced to the properties of the Laplacian
\begin{eqnarray}
&&\nu_\mathrm{k}\nabla^2\mathbf{V}=-\nu_\mathrm{k}\ii V_\mathrm{A}\int\mathbf{v}_\mathbf{Q}(\tau)Q^2\mathrm{e}^{\ii\mathbf{Q}\cdot\mathbf{X}}\frac{\mathrm{d}^3{Q}}{(2\pi)^3}
=-\frac{\ii V_\mathrm{A}}{\Lambda} \nu_\mathrm{k} \int \mathbf{v}_\mathbf{Q}(\tau)Q^2\mathrm{e}^{\ii\mathbf{Q}\cdot\mathbf{X}}\frac{\mathrm{d}^3{Q}}{(2\pi)^3} \nn \\
&&=-\ii A V_\mathrm{A} \nu'_\mathrm{k} \int Q^2 \mathbf{v}_\mathbf{Q}(\tau)\mathrm{e}^{\ii\mathbf{Q}\cdot\mathbf{X}}\frac{\mathrm{d}^3{Q}}{(2\pi)^3}\,,\nonumber\\
&&\nu_m\nabla^2\mathbf{B}=-\frac{B_0}{\Lambda^2}\nu_m\int Q^2\mathbf{b}_\mathbf{Q}(\tau)\frac{\mathrm{d}^3{Q}}{(2\pi)^3}=
-AB_0\nu_m'\int Q^2\mathbf{b}_\mathbf{Q}(\tau)\frac{\mathrm{d}^3{Q}}{(2\pi)^3}\,,\nonumber\\
&&\frac1{\ii AV_\mathrm{A}}\mathcal{\hat{F}}(\nu\nabla^2\mathbf{V})=-\nu'_\mathrm{kin}Q^2\mathbf{v}(\tau),
\quad\frac1{AB_0}\mathcal{\hat{F}}(\nu_\mathrm{m}\nabla^2\mathbf{B})= -\nu'_\mathrm{m}Q^2\mathbf{b}(\tau), \nn \\
&& \nu_\mathrm{k}^\prime\equiv\frac{A}{V_\mathrm{A}^2}\nu_\mathrm{k}, \quad
\nu_\mathrm{m}^\prime\equiv\frac{A}{V_\mathrm{A}^2}\nu_\mathrm{m}\,.
\end{eqnarray}
Hereafter for all terms coming from the velocity equation \Eqref{ch1MHD} we will separate a factor $\ii AV_\mathrm{A}$ and for all terms from
\Eqref{ch1MHD-B} we will separate a factor $AB_0.$ Those factors will be common for the final equations in the wave-vector space.

For Coriolis force density per unit mass $-2\mathbf{\Omega}\times\mathbf{V}$, we derive
\begin{eqnarray}
&&-2\,\mathbf{\Omega}\times\mathbf{V}
=-2A\omega(-V_y\mathbf{e}_x+V_x\mathbf{e}_y)\nn \\
&& =2A\omega\left(\begin{array}{c}
V_{y} \\
-V_{x} \\
0   \end{array}\right)=
 2 A\omega \left(\begin{array}{c}
Ax + \ii V_\mathrm{A} \int v_{y,\mathbf{Q}}(\tau)\, \mathrm{e}^{\ii\mathbf{Q}\cdot\mathbf{X}}\frac{\mathrm{d}^3{Q}}{(2\pi)^3}\\
-\ii V_\mathrm{A} \int v_{x,\mathbf{Q}}(\tau)\, \mathrm{e}^{\ii\mathbf{Q}\cdot\mathbf{X}}\frac{\mathrm{d}^3{Q}}{(2\pi)^3}\\
0 \end{array}\right)\,,\nonumber\\
&&\mathcal{\hat{F}}(-2\mathbf{\Omega}\times\mathbf{V})=\ii AV_\mathrm{A}2\omega(v_{y,\mathbf{Q}}(\tau)\,\mathbf{e}_x -
v_{x,\mathbf{Q}}(\tau)\,\mathbf{e}_y).
\end{eqnarray}
The centrifugal term $2 \omega A^2 x$ is irrelevant for the wave amplitude equations.

Furthermore, we calculate Lorentz force per unit mass $\mathbf{j}\times\mathbf{B}$, and using $B_0^2/\mu_0=\rho V_\mathrm{A}^2$ we have
\begin{eqnarray}
\left(\left(\frac{\nabla\times\mathbf{B}}{\mu_0}\right)\times\frac{\mathbf{B}_0}{\rho}\right) &=& \frac{\ii B_0}
{\mu_0\rho\Lambda}\int(\mathbf{Q}\times\mathbf{b}_\mathbf{Q})\times(B_{0y}\mathbf{e}_y
+B_{0z}\mathbf{e}_z)\mathrm{e}^{\ii\mathbf{Q}\cdot\mathbf{X}} \frac{\mathrm{d}^3{Q}}{(2\pi)^3} \nn \\
&&=\ii AV_\mathrm{A} \int\left[\left(\mathbf{Q}\times \mathbf{b}_\mathbf{Q}(\tau)\right) \times
\boldsymbol{\alpha}\right]\;\frac{\mathrm{d}^3{Q}}{(2\pi)^3}, \nonumber
\\
\mathcal{\hat{F}}\left(\left(\frac{\nabla\times\mathbf{B}}{\mu_0}\right)\times\frac{\mathbf{B}_0}{\rho}\right) &=&
\ii AV_\mathrm{A}\left[\left(\mathbf{Q}\times \mathbf{b}_\mathbf{Q}(\tau)\right) \times \boldsymbol{\alpha}\right]\,.
\end{eqnarray}

We have also other two zero terms having no influence on the wave dynamics. From momentum equation \Eqref{ch1MHD} and from  equation for
magnetic field \Eqref{ch1MHD-B} we have
\begin{eqnarray}
\mathbf{V}_\mathrm{shear} \cdot\nabla \mathbf{V}_\mathrm{shear} &=& Ax\,\mathbf{e}_y\cdot\nabla Ax\,\mathbf{e}_y
=A^2x\,\mathbf{e}_y\cdot \mathbf{e}_x \,\mathbf{e}_y=0\,,
\\
\mathbf{B}_0\cdot\nabla \mathbf{V}_\mathrm{shear} &=& AB_0(\alpha_y\mathbf{e}_y + \alpha_z\mathbf{e}_z)\cdot
\mathbf{e}_x\,\mathbf{e}_y=0\,.
\end{eqnarray}

For the last linear terms we have
\begin{eqnarray}
&&\mathbf{V}_\mathrm{wave}\cdot\nabla \mathbf{V}_\mathrm{shear}=
\ii AV_\mathrm{A}\int\mathbf{v_\mathbf{Q}(\tau)}\cdot
\mathbf{e}_{x}\,\mathbf{e}_{y}\,\mathrm{e}^{\ii\mathbf{Q}\cdot\mathbf{X}}\frac{\mathrm{d}^3Q}{(2\pi)^3}, \nn \\
&&\mathcal{\hat{F}}(\mathbf{V}_\mathrm{wave}\cdot\nabla \mathbf{V}_\mathrm{shear}) =\ii AV_\mathrm{A} v_{x,\mathbf{Q}}(\tau)\mathbf{e}_y\,,
\\
&&\mathbf{B}_\mathrm{wave}\cdot\nabla \mathbf{V}_\mathrm{shear}= AB_0 \int b_{x,\mathbf{Q}}(\tau)\,\mathbf{e}_y\,\mathrm{e}^{\ii\mathbf{Q}\cdot\mathbf{X}}\frac{\mathrm{d}^3{Q}}{(2\pi)^3},
\nn \\
&&\mathcal{\hat{F}}(\mathbf{B}_\mathrm{wave}\cdot\nabla \mathbf{V}_\mathrm{shear})= AB_0
b_{x,\mathbf{Q}}(\tau)\mathbf{e}_y\,,
\\
&&\mathbf{B}_0\cdot\nabla\mathbf{V}_\mathrm{wave}=\frac{\ii V_\mathrm{A} B_0}{\Lambda}\int (\boldsymbol{\alpha}
\cdot \ii \mathbf{Q})\mathbf{v}_\mathbf{Q}(\tau)\mathrm{e}^{\ii\mathbf{Q}\cdot\mathbf{X}}\frac{\mathrm{d}^3{Q}}{(2\pi)^3}, \nn \\
&&\mathcal{\hat{F}}(\mathbf{B}_0\cdot\nabla\mathbf{V}_\mathrm{wave})=
-AB_0 (\boldsymbol{\alpha}\cdot \mathbf{Q})\mathbf{v}_\mathbf{Q}(\tau)\,.
\end{eqnarray}
All linear terms are well-known from previous investigations of MHD waves in magnetized shear flows.
In the next subsection we will derive the nonlinear terms describing the wave-wave interaction coming from the convective time derivative.

\subsection{Nonlinear wave-wave interaction}
%
In order to derive the nonlinear term in the momentum equation \Eqref{ch1MHD} we calculate
$\mathbf{V}_{\mathrm{wave}}\cdot\nabla\mathbf{V}_{\mathrm{wave}}$ using $\nabla
\mathbf{X}=\frac{A}{V_\mathrm{A}}\openone=\openone/\Lambda$
\begin{eqnarray}
&&\mathbf{V}_{\mathrm{wave}}\cdot\nabla \mathbf{V}_{\mathrm{wave}}=
\sum_{\mathbf{Q}'}\ii V_\mathrm{A}\mathbf{v}(\tau,\mathbf{Q}')\mathrm{e}^{\ii\mathbf{Q}'\cdot\mathbf{X}}\cdot\nabla
\sum_{\mathbf{Q}''}\ii V_\mathrm{A}\mathbf{v}(\tau,\mathbf{Q}'')\mathrm{e}^{\ii\mathbf{Q}''\cdot\mathbf{X}}= \nn \\
&&-\ii A V_\mathrm{A}\sum_{\mathbf{Q}'}\sum_{\mathbf{Q}''}\mathbf{v}_{\mathbf{Q}'}\cdot \mathbf{Q}'' \mathbf{v}_{\mathbf{Q}''}
\mathrm{e}^{\ii(\mathbf{Q}'+\mathbf{Q}'')\cdot\mathbf{X}}\,.
\end{eqnarray}
For the sake of brevity in the last terms we will omit the time argument $\tau$ and write the wave-vector argument $\mathbf{Q}$ as index.
Making the Fourier transformation
\begin{eqnarray}
&&\hat{\mathcal{F}} \left(\mathbf{V}_{\mathrm{wave}}\cdot\nabla\mathbf{V}_{\mathrm{wave}}\right) \equiv
\int\mathrm{d}^3X
\mathrm{e}^{-\ii\mathbf{Q}\cdot\mathbf{X}}\left(\mathbf{V}_{\mathrm{wave}}\cdot\nabla\mathbf{V}_{\mathrm{wave}}\right)\\
\nonumber &&
=-\ii A V_\mathrm{A}\sum_{\mathbf{Q}'}\sum_{\mathbf{Q}''}\mathbf{v}_{\mathbf{Q}'}\cdot \mathbf{Q}''\mathbf{v}_{\mathbf{Q}''}
\delta\left(\frac{\mathbf{Q}'+\mathbf{Q}''-\mathbf{Q}}{2\pi}\right)
=-\ii A V_\mathrm{A}\sum_{\mathbf{Q}'}\mathbf{v}_{\mathbf{Q}-\mathbf{Q}'}\mathbf{v}_{\mathbf{Q}'}\cdot \mathbf{Q}.
\end{eqnarray}
The velocity field
\begin{equation}
\mathbf{V}_\mathrm{wave}(\tau,\mathbf{X})
 =\ii V_\mathrm{A}\int \mathrm{e}^{\mathbf{Q}\cdot\mathbf{X}}{\mathbf{v}}(\tau,\mathbf{Q})\frac{\mathrm{d}^3 X}{(2\pi)^3},
\qquad
\mathbf{B}_\mathrm{wave}(\tau,\mathbf{X})
 = B_0\int\mathrm{e}^{\mathbf{Q}\cdot\mathbf{X}}{\mathbf{b}}(\tau,\mathbf{Q})\frac{\mathrm{d}^3 X}{(2\pi)^3},
\end{equation}
has to be real, hence the Fourier components should be odd for the velocity and even for the magnetic field
\begin{equation}
 \mathbf{v}_{-\mathbf{Q}}=-\mathbf{v}_{\mathbf{Q}},
\qquad
 \mathbf{b}_{-\mathbf{Q}}=\mathbf{b}_{\mathbf{Q}}.
\end{equation}
Analogously, for the Fourier component of the wave-wave interaction part of the Lorentz force $\mathbf{j}_\mathrm{wave} \times
\mathbf{B}_\mathrm{wave}$ we obtain
\begin{eqnarray}
&&\hat{\mathcal{F}}\left(\frac1{\mu_0\rho}(\nabla\times\mathbf{B}
_\mathrm{wave})\times\mathbf{B}_\mathrm{wave} \right)=
\int\mathrm{d}^3X\mathrm{e}^{-\ii\mathbf{Q}\cdot\mathbf{X}}\left(\frac1{\mu_0\rho}(\nabla\times\mathbf{B}
_\mathrm{wave})\times\mathbf{B}_\mathrm{wave} \right) \nn \\
&&= \ii A V_\mathrm{A} \sum_\mathbf{Q'}\left(\mathbf{Q}'\times \mathbf{b}_{\mathbf{Q}'}\right)\times \mathbf{b}_{\mathbf{Q}-\mathbf{Q}'}.
\end{eqnarray}
In such a way we derive the Fourier component of the nonlinear term of the momentum equation
\begin{eqnarray}
\mathbf{N}_{v,\mathbf{Q}}&\equiv&\frac1{\ii A
V_\mathrm{A}}\hat{\mathcal{F}}\left(-\mathbf{V}_{\mathrm{wave}}\cdot\nabla\mathbf{V}_{\mathrm{wave}}
+\frac1{\mu_0\rho}(\nabla\times\mathbf{B}_\mathrm{wave})\times\mathbf{B}_\mathrm{wave} \right)\\
&=&\sum_{\mathbf{Q}'}\left[\mathbf{v}_{\mathbf{Q}-\mathbf{Q}'}\mathbf{v}_{\mathbf{Q}'}\cdot\mathbf{Q}
+(\mathbf{Q}'\times\mathbf{b}_{\mathbf{Q}'})\times\mathbf{b}_{\mathbf{Q}-\mathbf{Q}'}\right].
\end{eqnarray}

Analogously, for the other nonlinear terms $\mathbf{B}_{\mathrm{wave}}\cdot\nabla \mathbf{V}_{\mathrm{wave}}$
and $\mathbf{V}_{\mathrm{wave}}\cdot\nabla\mathbf{B}_{\mathrm{wave}}$ we have
\begin{eqnarray}
\hat{\mathcal{F}}\left(\mathbf{B}_{\mathrm{wave}}\cdot\nabla\mathbf{V}_{\mathrm{wave}}\right)&=&
\int\mathrm{d}^3X
\mathrm{e}^{-\ii\mathbf{Q}\cdot\mathbf{X}}\left(\mathbf{B}_{\mathrm{wave}}\cdot\nabla\mathbf{V}_{\mathrm{wave}}\right)\\\nonumber
&=&-\frac{B_0V_\mathrm{A}}{\Lambda}\sum_{\mathbf{Q}'}\sum_{\mathbf{Q}''}\mathbf{b}_{\mathbf{Q}'}\cdot
\mathbf{Q}''\,\mathbf{v}_{\mathbf{Q}''}
\delta\left(\frac{\mathbf{Q}'+\mathbf{Q}''-\mathbf{Q}}{2\pi}\right)\\\nonumber
&=&-AB_0\sum_{\mathbf{Q}'}\mathbf{b}_{\mathbf{Q}'}\cdot (\mathbf{Q}-\mathbf{Q}')\,\mathbf{v}_{\mathbf{Q}-\mathbf{Q}'},\\
\hat{\mathcal{F}}\left(\mathbf{V}_{\mathrm{wave}}\cdot\nabla\mathbf{B}_{\mathrm{wave}}\right)&=&
\int\mathrm{d}^3X
\mathrm{e}^{-\ii\mathbf{Q}\cdot\mathbf{X}}\left(\mathbf{V}_{\mathrm{wave}}\cdot\nabla\mathbf{B}_{\mathrm{wave}}\right)\\\nonumber
&=&-\frac{B_0V_\mathrm{A}}{\Lambda}\sum_{\mathbf{Q}'}\sum_{\mathbf{Q}''}\mathbf{v}_{\mathbf{Q}'}\cdot
\mathbf{Q}''\,\mathbf{b}_{\mathbf{Q}''}
\delta\left(\frac{\mathbf{Q}'+\mathbf{Q}''-\mathbf{Q}}{2\pi}\right)\\\nonumber
&=&-AB_0\sum_{\mathbf{Q}'}\mathbf{v}_{\mathbf{Q}'}\cdot (\mathbf{Q}-\mathbf{Q}')\,\mathbf{b}_{\mathbf{Q}-\mathbf{Q}'}.
\end{eqnarray}
Those terms participate in the equation for the magnetic field. For their difference we have
\begin{eqnarray}
\mathbf{N}_{b,\mathbf{Q}}&\equiv&\frac1{AB_0}
\hat{\mathcal{F}}\left(\mathbf{B}_{\mathrm{wave}}\cdot\nabla\mathbf{V}_{\mathrm{wave}}-
\mathbf{V}_{\mathrm{wave}}\cdot\nabla\mathbf{B}_{\mathrm{wave}}\right)\\\nonumber
&=&\sum_{\mathbf{Q}'}\left[\mathbf{v}_{\mathbf{Q}'}\cdot (\mathbf{Q}-\mathbf{Q}')\,\mathbf{b}_{\mathbf{Q}-\mathbf{Q}'}
-\mathbf{b}_{\mathbf{Q}'}\cdot (\mathbf{Q}-\mathbf{Q}')\,\mathbf{v}_{\mathbf{Q}-\mathbf{Q}'}\right]\\\nonumber
&=&-\mathbf{Q}\times\sum_{\mathbf{Q}'}(\mathbf{v}_{\mathbf{Q}'} \times \mathbf{b}_{\mathbf{Q}-\mathbf{Q}'})
\end{eqnarray}
As the function in $\mathbf{r}$-space
\begin{equation}
\mathrm{rot}\left(\mathbf{V}_\mathrm{wave}\times\mathbf{B}_{\mathrm{wave}}\right)=
\mathbf{B}_{\mathrm{wave}}\cdot\nabla\mathbf{V}_{\mathrm{wave}}-
\mathbf{V}_{\mathrm{wave}}\cdot\nabla\mathbf{B}_{\mathrm{wave}}
\end{equation}
has zero divergence
\begin{equation}
\mathrm{div}\left[\mathrm{rot}\left(\mathbf{V}_\mathrm{wave}\times\mathbf{B}_{\mathrm{wave}}\right)\right]=0
\end{equation}
its Fourier transform is transversal $\mathbf{Q}\cdot\mathbf{N}_{b,\mathbf{Q}}=0$ and automatically
$\mathbf{N}_{b,\mathbf{Q}}^\perp=\mathbf{N}_{b,\mathbf{Q}}.$

In order to merge the so derived nonlinear terms in the next subsection we will rederive the
linear terms in Lagrangian wave-vector space.

\section{Elimination of pressure in the final MHD equations}
%
It is common in  MHD  to formally seek the limit of a particular expression
for infinite sound speed $c_\mathrm{s}\rightarrow \infty.$
Due to the complexity of the problem this standard approach for consideration weak magnetic fields when $V_\mathrm{A}\ll c_\mathrm{s}$
is inapplicable in our problem and we have to look for direct elimination of the pressure.
After substituting Fourier transformations in \Eqref{ch1MHD} and \Eqref{ch1MHD-B} we obtain
\begin{eqnarray}
\left(\partial_\tau + \mathbf{U}_\mathrm{shear}\cdot\partial_\mathbf{Q} \right)\mathbf{v}_\mathbf{Q} \!\!\! &=& \!\!\!
-v_{x,\mathbf{Q}}\mathbf{e}_y +\mathbf{Q}P_\mathbf{Q} +\left[(\mathbf{Q}\times\mathbf{b}_\mathbf{Q})\times\boldsymbol{\alpha}\right]
+ 2\omega(v_{y,\mathbf{Q}}\mathbf{e}_x - v_{x,\mathbf{Q}}\mathbf{e}_y) \nn \\
&& - \nu'_\mathrm{k}Q^2\mathbf{v}_\mathbf{Q} + \sum_{\mathbf{Q}'}\left[\mathbf{v}_{\mathbf{Q}-\mathbf{Q}'}\mathbf{v}_{\mathbf{Q}'}\cdot\mathbf{Q}
+(\mathbf{Q}'\times\mathbf{b}_{\mathbf{Q}'})\times\mathbf{b}_{\mathbf{Q}-\mathbf{Q}'}\right] \nonumber,\\
\left(\partial_\tau + \mathbf{U}_\mathrm{shear}\cdot\partial_\mathbf{Q} \right)\mathbf{b}_\mathbf{Q} \!\!\!&=&\!\!\!
b_{x,\mathbf{Q}}\mathbf{e}_y -(\mathbf{Q}\cdot\boldsymbol{\alpha})\mathbf{v}_\mathbf{Q} -\nu'_\mathrm{m}Q^2\mathbf{b}_\mathbf{Q}
-\mathbf{Q}\times\sum_{\mathbf{Q}'}(\mathbf{v}_{\mathbf{Q}'} \times \mathbf{b}_{\mathbf{Q}-\mathbf{Q}'})\nn.
\end{eqnarray}
For the sake of brevity we introduce
\begin{eqnarray}
\label{ch1force_short}
\mathcal{F}_\mathbf{Q}&\equiv& Q_y\frac{\partial \mathbf{v}_\mathbf{Q}}{\partial Q_x}-\mathbf{e}_y \mathbf{e}_x \cdot \mathbf{v}_{\mathbf{Q}}
+ [(\mathbf{Q}\times\mathbf{b})\times\boldsymbol{\alpha}] + 2\boldsymbol{\omega}\times\mathbf{v}_\mathbf{Q}  \nn\\
&& +\nu'_\mathrm{k}Q^2\mathbf{v}_\mathbf{Q} +\sum_{\mathbf{Q}'}\left[\mathbf{v}_{\mathbf{Q}-\mathbf{Q}'}\mathbf{v}_{\mathbf{Q}'}\cdot\mathbf{Q}
+(\mathbf{Q}'\times\mathbf{b}_{\mathbf{Q}'})\times\mathbf{b}_{\mathbf{Q}-\mathbf{Q}'}\right].
\end{eqnarray}
Then the equation for the velocity can be rewritten as
\begin{equation}
\label{ch1short_v}
\partial_\tau \mathbf{v}_\mathbf{Q}=P_\mathbf{Q}\mathbf{Q} +\mathcal{F}_\mathbf{Q}.
\end{equation}
In order to express the pressure, we multiply both sides of this equation by $\mathbf{Q}$
\be
\partial_\tau (\mathbf{Q}\cdot\mathbf{v}_\mathbf{Q})= Q^2P_\mathbf{Q}+\mathbf{Q}\cdot\mathcal{F}_\mathbf{Q}.
\ee
The incompressibility condition $\mathbf{Q}\cdot\mathbf{v}_\mathbf{Q}=0$ gives for the pressure the solution of the Poisson equation
\begin{eqnarray}
\mathcal{P}=-\frac{\mathbf{Q} \cdot \mathcal{F}_\mathbf{Q}}{Q^2}
&=&-\frac{1}{Q^2}\left\{ 2\mathbf{Q}\cdot\mathbf{e}_y \mathbf{e}_x \cdot\mathbf{v}_\mathbf{Q}
+2\boldsymbol{\omega}\times\mathbf{v}_\mathbf{Q}
+\mathbf{Q}\cdot\left[\left(\mathbf{Q}\times\mathbf{b}_\mathbf{Q}\right)\times\boldsymbol{\alpha}\right]
\right\} \nonumber\\
&&-\frac{1}{Q^2}
\sum_{\mathbf{Q}'}\left\{\mathbf{Q}\cdot\mathbf{v}_{\mathbf{Q}-\mathbf{Q}'}\mathbf{v}_{\mathbf{Q}'}\cdot\mathbf{Q}
+\left[(\mathbf{Q}'\times\mathbf{b}_{\mathbf{Q}'})\times\mathbf{b}_{\mathbf{Q}-\mathbf{Q}'}\right]\cdot\mathbf{Q}\right\}, \nn
\end{eqnarray}
where we used the obvious vector relations
\begin{eqnarray}
&&\mathbf{Q}\cdot\mathbf{e}_y \mathbf{e}_x \cdot\mathbf{v}_\mathbf{Q}= v_xQ_y,
\qquad \mathbf{Q}\cdot(\boldsymbol{\omega}\times\mathbf{v}_\mathbf{Q})=\omega(Q_yv_x-Q_xv_y), \nn \\
&& \left(\mathbf{Q}\times\mathbf{b}_\mathbf{Q}\right)\times\boldsymbol{\alpha}
=(\mathbf{Q}\cdot\boldsymbol{\alpha})(\mathbf{Q}\cdot\mathbf{b}) - \mathbf{Q}^2(\mathbf{b}\cdot\boldsymbol{\alpha})\,.
\end{eqnarray}

This formula for the pressure we substitute in the \Eqref{ch1short_v} which takes the form
\begin{equation}
\label{ch1uskorenie}
\partial_\tau \mathbf{v}_\mathbf{Q}=\mathcal{F}_\mathbf{Q}^{\perp}
=\mathcal{F}_\mathbf{Q} -\frac{\mathbf{Q}\otimes \mathbf{Q}}{Q^2}\mathcal{F}_\mathbf{Q}=\Pi^{\perp\mathbf{Q}}\mathcal{F}_\mathbf{Q},
\end{equation}
where
\be
\Pi^{\perp\mathbf{Q}}\equiv\openone-\frac{\mathbf{Q}\otimes \mathbf{Q}}{Q^2}=\openone-\mathbf{n}\otimes \mathbf{n},
\qquad \mathbf{n}\equiv \frac{\mathbf{Q}}{Q}
\ee
is the projection operator which applies to the part of a vector, perpendicular to the wave-vector.
In other words the elimination of the pressure conserves the perpendicular part of
the Fourier component of the force $\mathcal{F}_\mathbf{Q}$ in the used dimensionless variables.
The equation \Eqref{ch1uskorenie} means that the velocity field remains orthogonal to the wave vector.
If in the beginning $\mathbf{Q}\cdot\mathbf{v}_\mathbf{Q}(\tau_0)=0$, the evolution gives that
$\mathbf{Q}\cdot\mathbf{v}_\mathbf{Q}(\tau)=0$ for every $\tau>\tau_0.$

Using that for the velocity as applicable for every orthogonal vector
\be
\Pi^{\perp}\partial_\tau\mathbf{v}_\mathbf{Q}=\partial_\tau\mathbf{v}_\mathbf{Q},
\qquad \Pi^{\perp}\mathbf{v}_\mathbf{Q}=\mathbf{v}_\mathbf{Q}
\ee
we can rewrite \Eqref{ch1uskorenie} as
\be
\Pi^{\perp}(\partial_\tau\mathbf{v}_\mathbf{Q} - \mathcal{F}_\mathbf{Q})=0.
\ee

In order to take into account the $Q_y\frac{\partial \mathbf{v}_\mathbf{Q}}{\partial Q_x}$ term in \Eqref{ch1force_short}
we use the obvious relations
\begin{equation}
Q_y\frac{\partial}{\partial Q_x}\left(\mathbf{v}_\mathbf{Q} \cdot \frac{\mathbf{Q}\mathbf{Q}}{Q^2}\right)=
Q_y\frac{\partial \mathbf{v}_\mathbf{Q}}{\partial Q_x} \cdot \frac{\mathbf{Q}\mathbf{Q}}{Q^2}+
\frac{Q_y v_{x,\mathbf{Q}}}{Q^2} \mathbf{Q}=0,
\qquad \mathbf{v}_\mathbf{Q} \cdot \mathbf{Q}=0.
\end{equation}
Now we represent the projection of the advective term $\mathbf{U}_\mathrm{shear}\cdot\partial_\mathbf{Q}\mathbf{v}_\mathbf{Q}$ as
\begin{eqnarray}
\Pi^{\perp\mathbf{Q}}(\mathbf{U}_\mathrm{shear}\cdot\partial_\mathbf{Q}\mathbf{v}_\mathbf{Q})
=-Q_y\frac{\partial \mathbf{v}_\mathbf{Q}}{\partial Q_x}-n_y\mathbf{n}v_{x,\mathbf{Q}}
\end{eqnarray}
and to arrive at the momentum equation in the form where the projection operator exists explicitly only in the nonlinear term
\begin{eqnarray}
&&\left(\partial_\tau + \mathbf{U}_\mathrm{shear}\cdot\partial_\mathbf{Q} \right)\mathbf{v}_\mathbf{Q}(\tau) \nn\\
&&=-v_{x,\mathbf{Q}}\mathbf{e}_y+2n_y\mathbf{n}v_{x,\mathbf{Q}}
+ 2\omega\mathbf{n}(n_yv_{x,\mathbf{Q}}-n_xv_{y,\mathbf{Q}})
+2\boldsymbol{\omega}\times v_{\mathbf{Q}} + (\boldsymbol{\alpha}\cdot\mathbf{Q})\mathbf{b}_\mathbf{Q} \nonumber\\
&&-\nu^\prime_kQ^2\mathbf{v}_\mathbf{Q}+\Pi^{\perp\mathbf{Q}}\sum_{\mathbf{Q}'}\left[\mathbf{v}_{\mathbf{Q}
-\mathbf{Q}'}\mathbf{v}_{\mathbf{Q}'}\cdot\mathbf{Q}
+(\mathbf{Q}'\times\mathbf{b}_{\mathbf{Q}'})\times\mathbf{b}_{\mathbf{Q}-\mathbf{Q}'}\right],
\\
\label{ch1MagneticEvolution}
&&\left(\partial_\tau + \mathbf{U}_\mathrm{shear}\cdot\partial_\mathbf{Q} \right)\mathbf{b}_\mathbf{Q}(\tau) \nn \\
&& = b_{x,\mathbf{Q}}\mathbf{e}_y -(\mathbf{Q}\cdot\boldsymbol{\alpha})\mathbf{v}_\mathbf{Q} -\nu'_\mathrm{m}Q^2\mathbf{b}_\mathbf{Q}
-\mathbf{Q}\times\sum_{\mathbf{Q}'}(\mathbf{v}_{\mathbf{Q}'} \times \mathbf{b}_{\mathbf{Q}-\mathbf{Q}'}),\\
&&\mathbf{v}_\mathbf{Q}(\tau_0)=\Pi^{\perp}\mathbf{v}_\mathbf{Q}(\tau_0),
\quad \mathbf{b}_\mathbf{Q}(\tau_0)=\Pi^{\perp}\mathbf{b}_\mathbf{Q}(\tau_0).
\end{eqnarray}
For numerical calculations the incompressibility conditions
$\mathbf{n}\cdot \mathbf{b}_\mathbf{Q}=0$ and $\mathbf{n}\cdot \mathbf{v}_\mathbf{Q}=0$ can be used as a criterion for the error.

Using the relation
\be
[\mathbf{U}_\mathrm{shear}\cdot\partial_\mathbf{Q}\mathbf{b}_\mathbf{Q}(\tau)]\cdot\mathbf{Q}= Q_yb_{x,\mathbf{Q}},
\ee
one can easily check that the equation for the evolution of the magnetic field \Eqref{ch1MagneticEvolution} can
also be presented as the evolution of its part, perpendicular to the wave-vector
\begin{equation}
\partial_\tau\mathbf{b}_\mathbf{Q}  + \Pi^{\perp}\mathbf{U}_\mathrm{shear}\cdot\partial_\mathbf{Q} \mathbf{b}_\mathbf{Q}
= \Pi^{\perp} \mathbf{e}_y\mathbf{e}_x\cdot\mathbf{b}_{\mathbf{Q}} -(\mathbf{Q}\cdot\boldsymbol{\alpha})\mathbf{v}_\mathbf{Q}
-\nu'_\mathrm{m}Q^2\mathbf{b}_\mathbf{Q}
-\mathbf{Q}\times\sum_{\mathbf{Q}'}(\mathbf{v}_{\mathbf{Q}'} \times \mathbf{b}_{\mathbf{Q}-\mathbf{Q}'}).
\end{equation}
Together with $\mathbf{v}_\mathbf{Q}=\Pi^{\perp}\mathbf{v}_\mathbf{Q}$ this equation automatically
gives $\mathbf{b}_\mathbf{Q}=\Pi^{\perp}\mathbf{b}_\mathbf{Q}$ and $\mathrm{div} \mathbf{B}=0.$

In the matrix form the system of MHD equations reads as
\begin{equation}
\mathrm{D}_\tau\Psi=\mathsf{M}\Psi + \mathsf{N},
\end{equation}
where after some algebra\cite{Biskamp:03}
\begin{eqnarray}
&&\mathsf{N}=\left(\begin{array}{c}
\mathbf{N}_{b,\mathbf{Q}}^{\perp}\\
\mathbf{N}_{v,\mathbf{Q}}^{\perp}
\end{array}\right) \nn \\
&&=\left(\begin{array}{c}
-\mathbf{Q}\times\sum_{\mathbf{Q}'}(\mathbf{v}_{\mathbf{Q}'} \times \mathbf{b}_{\mathbf{Q}-\mathbf{Q}'})\\
\Pi^{\perp\mathbf{Q}}\sum_{\mathbf{Q}'}\left[\mathbf{v}_{\mathbf{Q}-\mathbf{Q}'}\mathbf{v}_{\mathbf{Q}'}\cdot\mathbf{Q}
+(\mathbf{Q}'\times\mathbf{b}_{\mathbf{Q}'})\times\mathbf{b}_{\mathbf{Q}-\mathbf{Q}'}\right]
                 \end{array}\right) \nn \\
&&=\left(\begin{array}{c}
\sum_{\mathbf{Q}'}\left(\mathbf{b}_{\mathbf{Q}'}\mathbf{v}_{\mathbf{Q}-\mathbf{Q}'}-\mathbf{v}_{\mathbf{Q}'}\mathbf{b}_{\mathbf{Q}-\mathbf{Q}'}\right)\cdot\mathbf{Q}\\
\Pi^{\perp\mathbf{Q}}\sum_{\mathbf{Q}'}\left(
\mathbf{v}_{\mathbf{Q}'}\mathbf{v}_{\mathbf{Q}-\mathbf{Q}'}+\mathbf{b}_{\mathbf{Q}'}\mathbf{b}_{\mathbf{Q}-\mathbf{Q}'}
\right)\cdot \mathbf{Q}
                 \end{array}\right)\,\nn
,
\end{eqnarray}
and
\begin{eqnarray}
&&\mathsf{M} =
\left(\begin{array}{ccc|ccc}
-\nu^\prime_\mathrm{m}Q^2 & 0 & 0 & -Q_\alpha& 0 & 0\\
1 & -\nu^\prime_\mathrm{m}Q^2 & 0 & 0 & -Q_\alpha & 0 \\
0 & 0& -\nu^\prime_\mathrm{m}Q^2  & 0 & 0 & -Q_\alpha\\
\hline
Q_\alpha& 0 & 0 & 2n_yn_x(\omega+1)-\nu^\prime_\mathrm{k}Q^2 & -2n_xn_x\omega+2\omega & 0 \\
0 & Q_\alpha & 0 &2n_yn_y(\omega+1) -(2\omega+1) & -2n_xn_y\omega -\nu^\prime_\mathrm{k}Q^2 & 0 \\
0 & 0 & Q_\alpha & 2n_yn_z(\omega+1) & -2n_xn_z \omega  & -\nu^\prime_\mathrm{k}Q^2
\end{array}\right), \nn\\
&&\Psi_\mathbf{Q}=\left(\begin{array}{c}
            b_x\\
	    b_y\\
	    b_z\\
	    v_x\\
	    v_y\\
	    v_z
           \end{array}\right)
\end{eqnarray}
and $Q_\alpha\equiv\mathbf{Q}\cdot\boldsymbol{\alpha}.$

The matrix can also be represented as
\begin{equation}
\mathsf{M} =
\left(\begin{array}{c|c}
\mathsf{M}_{bb} & \mathsf{M}_{bv}\\
\hline
\mathsf{M}_{vb} & \mathsf{M}_{vv} \\
\end{array}\right),\qquad
\Psi_\mathbf{Q}=\left(\begin{array}{c}
            \mathbf{b}\\
	    \mathbf{v}
	    \end{array}\right),
\end{equation}
\begin{equation}
\mathsf{M}_{vv} =
2n_y \left(\begin{array}{ccc}
n_x & 0 & 0\\
n_y & 0 & 0\\
n_z & 0 & 0
\end{array}\right)
-
\left(\begin{array}{ccc}
0 & 0 & 0\\
1 & 0 & 0\\
0 & 0 & 0
\end{array}\right)
+2\omega
\left(\begin{array}{ccc}
n_xn_y & (n_y^2+n_z^2) & 0\\
-(n_x^2+n_z^2) & -n_xn_y & 0\\
n_yn_z & -n_xn_z & 0
\end{array}\right)
-\nu^\prime_\mathrm{k}Q^2\openone,
\end{equation}
\begin{equation}
\mathsf{M}_{vb} = Q_\alpha\openone, \qquad \mathsf{M}_{bv} = -Q_\alpha\openone, \qquad
\mathsf{M}_{bb} =
\left(\begin{array}{ccc}
0 & 0 & 0\\
1 & 0 & 0\\
0 & 0 & 0
\end{array}\right)
-\nu^\prime_\mathrm{m}Q^2\openone.
\end{equation}
With the help of the matrices in this representation in the next section we will make Lyapunov analysis
of the linearized MHD equations.

\section{Lyapunov analysis of the linearized system in Lagrangian variables}
\label{ch1sec:Lyapunov}
%
For small $Q_y$ we may neglect the advective term $\mathbf{U}_\mathrm{shear}\cdot\partial_\mathbf{Q}=-Q_y\partial_{Q_x}.$
Then the linearized MHD equations take the form
\be
\mathrm{D}_\tau\Psi=\mathsf{M}\Psi, \qquad (\mathbf{Q},\mathbf{Q})\cdot\Psi=0.
\ee
To perform an instability analysis we make use of the exponential substitution $\Psi=\exp(\lambda\tau)\psi,$
which leads to an eigenvalue problem with transversality conditions
\be
\mathsf{M}(\mathbf{Q})\Psi_\mathbf{Q}=\lambda\Psi_\mathbf{Q},
\qquad \mathbf{Q}\cdot\mathbf{b}_\mathbf{Q}=0=\mathbf{Q}\cdot\mathbf{v}_\mathbf{Q};
\ee
to make it short we can further omit the index $\mathbf{Q}$
\be
\left(\mathsf{M}-\lambda\openone\right)\Psi=0,
\qquad \mathbf{Q}\cdot\mathbf{b}=0=\mathbf{Q}\cdot\mathbf{v}.
\ee
Should we substitute the incompressibility and transversality conditions
\be
\label{ch1transversal}
v_z=-\frac{Q_x v_x+Q_y v_y}{Q_z}, \qquad b_z=-\frac{Q_x b_x+Q_y b_y}{Q_z},
\ee
in the secular equation, we would end up with an overdetermined system.
To avoid it, we omit the equations which initially have $\lambda b_z$ and $\lambda v_z$ terms, i.e. the 3-rd and the 6-th rows in the secular equation.
In such a way we derive a secular equation for a reduced matrix
\begin{eqnarray}
&&\left(\tilde{\mathsf{M}}-\lambda\openone\right)\psi=0,
\quad
\psi=\left(\begin{array}{c}
            b_x\\
	        b_y\\
	        v_x\\
	        v_y
                      \end{array}
                \right),  \\
&&\mathsf{\tilde{M}} =
\left(\begin{array}{cc|cc}
-\nu^\prime_\mathrm{m}Q^2 & 0 & -Q_\alpha& 0 \\
1 & -\nu^\prime_\mathrm{m}Q^2 & 0 & -Q_\alpha  \\
\hline
Q_\alpha& 0 & 2n_yn_x(\omega+1)-\nu^\prime_\mathrm{k}Q^2 & -2n_xn_x\omega+2\omega  \\
0 & Q_\alpha &2n_yn_y(\omega+1) -(2\omega+1) & -2n_xn_y\omega -\nu^\prime_\mathrm{k}Q^2
\end{array}\right). \nn
\end{eqnarray}
This secular equation
\be
P_4(\lambda; \mathbf{Q},\nu_\mathrm{m}^{\,\prime},\nu_\mathrm{k}^{\,\prime})
\equiv\mathrm{det}\left(\tilde{\mathsf{M}}-\lambda\openone\right)=0
\ee
has 4 eigenvalues and via a calculation of the eigenvectors, we can derive $b_z$ and $v_z$ according to the transversality conditions \Eqref{ch1transversal}.

For an ideal fluid $\nu_\mathrm{m}^{\,\prime}=0=\nu_\mathrm{k}^{\,\prime}$. Omitting the viscosity terms
we have a relatively simple form for the secular equation
\ba
\nn
&&P_4(\lambda; \mathbf{Q},\nu_\mathrm{m}^{\,\prime}=0,\nu_\mathrm{k}^{\,\prime}=0) \nn \\
&=&
\lambda^4
-2n_yn_x\lambda^3
+\left\{\left[(4-8n_y^2)n_x^2+4-4n_y^2\right]\omega^2 +
\left[(2-8n_y^2)n_x^2-4n_y^2+2\right]\omega +2Q_\alpha^2\right\}\lambda^2 \nn \\
&&-2Q_\alpha^2n_yn_x\lambda
+ 2Q_\alpha^2(n_x^2+1)\omega+Q_\alpha^4=0.
\ea
As we pointed out these eigenvalues give only a WKB approximation for the dynamics of MHD variables $\psi(\tau).$
For the special case of $Q_y=0$, which corresponds to an axial-symmetric motion,
with a rotation along the z-axis, the secular equation gives directly the increments of the linearized MHD equations.
\ba
&&P_4(\lambda; Q_y=0, \nu_\mathrm{m}^{\,\prime}=0,\nu_\mathrm{k}^{\,\prime}=0)  \\
&&=\lambda^4+
2\left[Q_\alpha^2 + (1 +2\omega)(n_x^2+1)\omega\right]\lambda^2
+2Q_\alpha^2(n_x^2+1)\omega+Q_\alpha^4=0. \nn
\ea
The most restricted case is for wave-vectors parallel to the rotation axis
$\mathbf{Q}=Q\mathbf{e}_z$ when $Q_\alpha=Q_z \cos{\theta}$
\ba
\label{ch1MRI}
&&P_4(\lambda; Q_x=0, Q_y=0, \nu_\mathrm{m}^{\,\prime}=0,\nu_\mathrm{k}^{\,\prime}=0) \\
&&=\lambda^4 +2\left[Q_\alpha^2 + (1 +2\omega)\omega\right]\!\lambda^2
+ \left(Q_\alpha^2+2\omega\right)\!Q_\alpha^2=0\nn.
\ea
This is perhaps the most cited bi-quadratic equation in the whole history of science because it describes the magnetorotational instability (MRI) discovered by Velikhov\cite{Velikhov:59} in 1959.
In the astrophysics this equation was recognized and overexposed by many astrophysical grants 30 years later, see equation Eq.~(111) of Ref.\cite{Balbus:98} and historical remarks therein.
If we consider the special case of pure shear $\omega=0$ with $Q_y=0$
this dispersion equation gives the usual Alfv\'en waves
\be
(\lambda^2 + Q_\alpha^2)^2 =0,\qquad \omega=\left|Q_\alpha\right|,
\ee
i.e. the rotation destabilizes the Alfv\'en  waves. The polarization of the magnetic field and the velocity of the Alfv\'en  waves are along the shear flow.

For pure axial magnetic field $\mathbf{B}=B\mathbf{e}_z,$ i.e. $\boldsymbol{\alpha}=(0,\,0,\,1),$ and $Q_\alpha=Q_z.$ The matrix reduction is then given by simply erasing the z-components and taking into account only the x- and the y-projections of the equations of motions
\begin{equation}
\tilde{\mathsf{M}}_\mathrm{MRI} =
\left(\begin{array}{cc|cc}
 0 & 0 & -Q_z& 0 \\
 1 & 0 & 0 & -Q_z \\
\hline
Q_z & 0 & 0 & 2\omega  \\
0 & Q_z & -(2\omega+1) & 0 \\
\end{array}\right),\qquad
\psi=\left(\begin{array}{c}
            b_x\\
	    b_y\\
	    v_x\\
	    v_y
           \end{array}\right).
\end{equation}
The secular equation is the equation for MRI \Eqref{ch1MRI} with $Q_\alpha=Q_z.$

The projection method can be generalized in the general case if
we introduce 2 unit vectors perpendicular to the wave-vector $\mathbf{e}_\mathbf{Q}=\mathbf{Q}/Q$
\ba
&\left|2\right>&=\mathbf{e}_2=\frac{\mathbf{e}_z\times\mathbf{e}_\mathbf{Q}}{|\mathbf{e}_z\times\mathbf{e}_\mathbf{Q}|}=
\frac1{\sqrt{Q_x^2+Q_y^2}}\left(\begin{array}{c} -Q_y\\Q_x \\0  \end{array}\right),\nn \\
&\left|1\right>&=\mathbf{e}_1=\frac{\mathbf{e}_2\times\mathbf{e}_\mathbf{Q}}{|\mathbf{e}_2\times\mathbf{e}_\mathbf{Q}|}=
\frac1{\sqrt{Q_x^2+Q_y^2}\sqrt{Q_x^2+Q_y^2+Q_z^2}}\left(\begin{array}{c} -Q_xQ_z\\Q_yQ_z \\-Q_y^2-Q_x^2\end{array}\right),\nn
\ea
and also the corresponding bra-vectors
\ba
&\left<1\right|&=
\frac{\left(-Q_xQ_z,Q_yQ_z,-Q_y^2-Q_x^2\right)}{\sqrt{Q_x^2+Q_y^2}\sqrt{Q_x^2+Q_y^2+Q_z^2}},\\
&\left<2\right|&=
\frac{\left(-Q_y, Q_x, 0\right)}{\sqrt{Q_x^2+Q_y^2}}.
\ea
For the degenerated case of $Q_x=0=Q_y$ we can regularize by chosing $Q_x=\iota$ and $Q_y=0.$ Then the limit $\iota\rightarrow 0$ gives the regularization $\left|1\right>=\mathbf{e}_x$ and $\left|2\right>=\mathbf{e}_y.$
For all matrices $\mathsf{M}_{\alpha,\beta}$ where $\alpha,\,\beta= b,\,v$ we calculate the matrix elements in the two-dimensional space
\be
\left(\overline{\mathsf{M}}_{\alpha,\beta}\right)_{\mathrm{j}\,\mathrm{j}^\prime}=\left<\mathrm{j}\right|\mathsf{M}_{\alpha,\beta}\left|\mathrm{j}^\prime\right>,
\qquad \mbox{where}\;\;\mathrm{j},\,\mathrm{j}^\prime=1,\,2.
\ee
In such a way we obtain a reduced 4$\times$4 matrix
\be
\overline{\mathsf{M}} =
\left(\begin{array}{c|c}
 \overline{\mathsf{M}}_{bb} & \overline{\mathsf{M}}_{bv}\\
\hline
\overline{\mathsf{M}}_{vb} & \overline{\mathsf{M}}_{vv} \\
\end{array}\right)
\ee
whose eigenvectors are automatically perpendicular to $\mathbf{Q},$ simply because we have used the orthogonal to $\mathbf{Q}$ space.

As a rule the linearized analysis is made in Lagrangian, moving, wave-vector space
\be
\label{ch1Psi_Lagrange}
\mathrm{d}_\tau\mathbf{K}(\tau)=\mathbf{U}_\mathrm{shear}(\mathbf{K}(\tau)),
\ee
with a time-dependent wave-vector
\be
K_x=K_{x,0}-K_y(\tau-\tau_0),\qquad K_y=\mathrm{const},\qquad K_z=\mathrm{const}
\ee
for each MHD wave.

In these coordinates for linearized waves the substantial time derivative $\mathrm{D}_\tau^{\mathrm{shear}}=\mathrm{d}_\tau$
is reduced to a usual time derivative and the separation of variables gives a system of ordinary independent equations for every MHD wave
\be
\mathrm{d}_\tau\Psi_\mathbf{K}(\tau)=\mathsf{M}(\mathbf{K}(\tau))\Psi_\mathbf{K}(\tau),
\qquad \mathbf{K}(\tau)\cdot\mathbf{v}_\mathbf{K}(\tau)=0,
\qquad \mathbf{K}(\tau)\cdot\mathbf{b}_\mathbf{K}(\tau)=0.
\ee
In this linearized case it is possible to exclude $b_z$ and $v_z$. In such a way we arrive at a simple-for-programming system of 4 equations
\ba
&& \mathrm{d}_\tau\psi_\mathbf{K}(\tau)=\mathsf{\tilde{M}}(\mathbf{K}(\tau))\psi_\mathbf{K}(\tau), \nn \\
&& b_z=-(K_x(\tau)b_x+K_yb_y)/K_z, \quad v_z=-(K_x(\tau)v_x+K_yv_y)/K_z.
\ea

For small $K_y$ one can apply WKB approximation supposing exponential time dependence of the MHD variables $\Psi(\tau)\propto \exp(\lambda\tau)$ and the wave amplitudes.
In the WKB approximation the energy amplification between $\tau=-\infty$ and $\tau=+\infty$ is given by the eigenvalue $\lambda$ with the maximal real part
\be
G\approx\exp\left(2\int_{-\infty}^{\infty}\mathrm{d}\tau\, \mathrm{Re}\,\lambda_{\mathrm{max}}(\mathbf{K}(\tau))\right).
\ee
For the case of MRI with nonzero $B_z$ the amplification factors are so giant that the linear analysis
makes no sense because the nonlinear terms become rather important and we have a nonlinear saturation of the MRI.
This saturation simulates strong turbulence for small wave-vectors, but definitely
for large wave-vectors $|K_y|\gg 1$ at $\tau \rightarrow\infty$ we have a wave type turbulence with a given frequency.

We have to mention that the linearized case of pure shear is exactly integrable in terms of Heun functions\cite{Mishonov:09,Mishonov:07}.
Investigating numerically this case with $\omega=0$ and $B_z=0$ in his PhD work\cite{Chagelishvili:93}
T.~Hristov discovered in 1990 the amplification of slow magnetosonic waves (SMW) by shear flows.
Applied to the physics of accretion disks this amplification works even for purely azimuthal magnetic fields and gives a scenario for weak magnetic turbulence related to amplification of SMW. We had to wait 30 years of incubation period, cf.\cite{Dessler:70}, for the SMW amplification to be recognized as an important for the astrophysics phenomenon.
In the next section we will consider how to proceed with the solution of MHD equations.

\section{Energy density and power density} %
%
Our first step is to calculate the energy of plane MHD waves with time-dependent amplitudes.
Using that
\begin{equation}
 \int \mathrm{e}^{\ii\mathbf{Q}\cdot\mathbf{X}}\mathrm{d}^3X=(2\pi)^3\delta(\mathbf{Q})
\end{equation}
for the energy we obtain
\begin{equation}
\frac12\int\left(\rho\mathbf{V}^2_\mathrm{wave} + \frac1{\mu_0}\mathbf{B}^2_\mathrm{wave}\right)\mathrm{d}^3X=\rho V_\mathrm{A}^2
\sum_\mathbf{Q}\epsilon_{_\mathbf{Q}}, \qquad \epsilon_{_\mathbf{Q}}\equiv\frac12(\mathbf{v}_\mathbf{Q}^2+\mathbf{b}_\mathbf{Q}^2),
\end{equation}
i.e. the energy density is
\be
\rho V_\mathrm{A}^2 \iiint
\frac12\left[\mathbf{v}_\mathbf{Q}^2(\tau)+\mathbf{b}_\mathbf{Q}^2(\tau)\right]\,
\frac{\mathrm{d}Q_x\mathrm{d}Q_y\mathrm{d}Q_z}{(2\pi)^3}.
\ee
Analogously, with the help of the viscous stress tensor $\sigma^\prime_{ik}$ we express
the volume density of the wave heating
\begin{eqnarray}
Q^{\mathrm{wave}}_{\mathrm{kin}}&=&\int\sigma^\prime_{ik}\partial_k V^{\mathrm{wave}}_i \mathrm{d^3}x=
\frac12\int\sigma^\prime_{ik}(\partial_k V^{\mathrm{wave}}_i+\partial_iV^{\mathrm{wave}}_k)\mathrm{d^3}x \nn \\
&=&\frac{\eta}{2}\int(\partial_k V^{\mathrm{wave}}_i+\partial_i V^{\mathrm{wave}}_k)^2\mathrm{d^3}x \nn \\
&=&\frac{\eta V_\mathrm{A}^2}{2\Lambda^2}\int\left(\sum_\mathbf{Q} Q_iv_k\mathrm{e}^{\ii\mathbf{Q}\cdot \mathbf{X}} + \sum_\mathbf{Q'} Q'_kv_i\mathrm{e}^{\ii\mathbf{Q'}\cdot \mathbf{X}} \right)^{\!\!2} \mathrm{d^3}x
= \rho AV^2_\mathrm{A}\nu'_\mathrm{k}\sum_Q Q^2 v_\mathbf{Q}^2. \nonumber
\end{eqnarray}

Similarly for the Ohmic part of the energy dissipation rate we have
\begin{equation}
Q^{\mathrm{wave}}_\mathrm{Ohm}=\mathbf{j}\cdot\mathbf{E}
=\frac{1}{\mu_0^2\sigma_{_\mathrm{Ohm}}}(\mathrm{rot}\mathbf{B}_\mathrm{wave})^2
=\frac{B_0^2}{\mu_0^2\sigma_{_\mathrm{Ohm}}}\left(\sum_\mathbf{Q} \nabla\times \mathbf{b}_\mathbf{Q}\mathrm{e}^{\ii\mathbf{Q}\cdot\mathbf{X}}\right)^{\!\!2}
=\rho AV^2_\mathrm{A}\nu'_\mathrm{m}\sum_\mathbf{Q} Q^2b^2_\mathbf{Q}.
\end{equation}
The dissipation rate of a laminar shear flow is given according to Newton's formula
\begin{equation}
Q^{\mathrm{shear}}_{\mathrm{kin}}=\frac{\eta}{2}\int(\partial_k V^{\mathrm{shear}}_i
+\partial_i V^{\mathrm{shear}}_k)^2\mathrm{d^3}x=\frac{\eta}{2}A^2(\delta_{k,x}\delta_{i,y}+\delta_{i,x}\delta_{k,y})^2=\eta A^2.
\end{equation}
Now we can calculate the total energy dissipation
$Q_\mathrm{tot}=Q^{\mathrm{shear}}_{\mathrm{kin}}+Q^{\mathrm{wave}}_{\mathrm{kin}}+Q^{\mathrm{wave}}_\mathrm{Ohm}$,
the viscosity and the effective viscosity
$\eta_\mathrm{eff}$
\begin{eqnarray}
\eta=\frac{Q^{\mathrm{shear}}_{\mathrm{kin}}}{A^2}, \qquad \eta_\mathrm{eff}=\rho\nu_\mathrm{eff}=\frac{Q_\mathrm{tot}}{A^2}.
\end{eqnarray}
In this way we can express the effective kinematic viscosity by the dimensionless Fourier components of the velocity and the magnetic field
\begin{eqnarray}
\nu_\mathrm{eff}(\tau)=\nu_\mathrm{k} + \nu_\mathrm{k}\sum_{\mathbf{Q}}Q^2\mathbf{v}_\mathbf{Q}^2(\tau)
+ \nu_\mathrm{m}\sum_{\mathbf{Q}} Q^2\mathbf{b}_\mathbf{Q}^2(\tau).
\end{eqnarray}
For example, if we have static probability distribution functions for the velocity and the magnetic field,
the enhancement factor of the effective viscosity is given by the time-averaged squares of the Fourier components for $\tau\gg1$
\begin{eqnarray}
\label{ch1viscosity}
\frac{\eta_\mathrm{eff}}{\eta}=
1 + \sum_{\mathbf{Q}} Q^2\left<\mathbf{v}_\mathbf{ Q}^2\right>
+ \frac{\nu_\mathrm{m}}{\nu_\mathrm{k}}\sum_{\mathbf{ Q}} Q^2 \left<\mathbf{b}_\mathbf{Q}^2\right>;
\end{eqnarray}
this important parameter determines the work of the accretion discs as a machine for making of compact astrophysical objects.
The most simple scenario is to have the solution of the static equations for the i-th iteration of $\Psi_\mathbf{Q}$ and to calculate the next (i+1)-th iteration
\begin{eqnarray}
\partial_{\overline{\tau}}  \mathbf{v}^{(\mathrm{i}+1)}_\mathbf{Q}\!\!\! &=&\!\!\!-Q_y\frac{\partial \mathbf{v}^{(\mathrm{i}+1)}_\mathbf{Q}}{\partial Q_x} =
-v^{(\mathrm{i}+1)}_{x,\mathbf{Q}}\mathbf{e}_y+2\frac{Q_y v^{(\mathrm{i}+1)}_{x,\mathbf{Q}}}{Q^2}\mathbf{Q}  \\
&&+ 2\omega\left[\mathbf{n}(n_y v^{(\mathrm{i}+1)}_{x,\mathbf{Q}}-n_x
v^{(\mathrm{i}+1)}_{y,\mathbf{Q}})+(v^{(\mathrm{i}+1)}_{y,\mathbf{Q}}\mathbf{e}_x-v^{(\mathrm{i}+1)}_{x,\mathbf{Q}}\mathbf{e}_y)\right]  \nonumber\\
&&+(\boldsymbol{\alpha}\cdot\mathbf{Q})\mathbf{b}^{(\mathrm{i}+1)}_\mathbf{Q} -\nu^\prime_kQ^2\mathbf{v}^{(\mathrm{i}+1)}_\mathbf{Q}
+\Pi^{\perp\mathbf{Q}}\sum_{\mathbf{Q}'}\left[\mathbf{v}^{(\mathrm{i})}_{\mathbf{Q}'}\otimes\mathbf{v}^{(\mathrm{i})}_{\mathbf{Q}
-\mathbf{Q}'} + \mathbf{b}^{(\mathrm{i})}_{\mathbf{Q}'}\otimes\mathbf{b}^{(\mathrm{i})}_{\mathbf{Q}-\mathbf{Q}'}\right]\cdot\mathbf{Q},\nn \\
\partial_{\overline{\tau}}  \mathbf{b}^{(\mathrm{i}+1)}_\mathbf{Q}\!\!\! &=&\!\!\!-Q_y\frac{\partial \mathbf{b}^{(\mathrm{i}+1)}_\mathbf{Q}}{\partial Q_x} = b^{(\mathrm{i}+1)}_{x,\mathbf{Q}}\mathbf{e}_y
-(\mathbf{Q}\cdot\boldsymbol{\alpha})\mathbf{v}^{(\mathrm{i}+1)}_\mathbf{Q} -\nu'_\mathrm{m}Q^2\mathbf{b}^{(\mathrm{i}+1)}_\mathbf{Q} \nn \\
&&+\sum_{\mathbf{Q}'}\left[\mathbf{b}^{(\mathrm{i})}_{\mathbf{Q}'}\otimes\mathbf{v}^{(\mathrm{i})}_{\mathbf{Q}
-\mathbf{Q}'} - \mathbf{v}^{(\mathrm{i})}_{\mathbf{Q}'}\otimes\mathbf{b}^{(\mathrm{i})}_{\mathbf{Q}-\mathbf{Q}'}\right]\cdot\mathbf{Q},
\\
\partial_{\overline{\tau}}&\equiv& -Q_y\frac{\partial }{\partial Q_x}, \; \overline{\tau}\equiv-\frac{Q_x}{Q_y},
\quad \mathrm{D}_\tau=\partial_{\tau}+\partial_{\overline{\tau}},\nn \\
&& \mbox{for independent variables}\quad(\overline{\tau},Q_y,Q_z), \; Q_x=-Q_y\overline{\tau} .
\end{eqnarray}
For cold protoplanetary disks the Ohmic resistivity of weakly ionized gas is very high and the effective viscosity
is dominated in \Eqref{ch1viscosity} by $\nu_\mathrm{m}/\nu_\mathrm{k}$ term, in other words the viscosity of the protoplanetary disks is created by Ohmic dissipation. Completely opposite is the situation for the hot almost completely-ionized Hydrogen plasma in quasars.
The Ohmic resistivity is negligible and the effective viscosity is created by the Fourier components of the MHD waves $\left<\mathbf{v}_\mathbf{ Q}^2\right>.$
Only for small wave-vectors the MHD turbulence remains strong turbulence. At large wave-vectors
we have weak wave turbulence with wave-vectors going to infinity.
In the next section we will consider the stability conditions which have to be checked.

\section{Stability} %
%
The linear Lyapunov analysis which we outlined in Sec.\ref{ch1sec:Lyapunov} gives the idea what we have to do when we obtain
the static solution $\Psi_\mathbf{Q}^{(0)}=(\mathbf{b}_\mathbf{Q}^{(0)},\mathbf{v}_\mathbf{Q}^{(0)})$.
In order to investigate the stability of this static solution we have to consider a small time-dependent deviation from this solution
$\Psi_\mathbf{Q}^{(1)}(\tau)=(\mathbf{b}_\mathbf{Q}^{(1)}(\tau),\mathbf{v}_\mathbf{Q}^{(1)}(\tau))$.
In this case, neglecting the quadratic terms with respect to $\Psi_\mathbf{Q}^{(1)}$,
we find that the nonlinear terms in the MHD equations are linear integral operators in $\mathbf{Q}$-space
\begin{eqnarray}
&& \mathsf{\hat{N}'}\Psi_\mathbf{Q}^{(1)} \nn \\
&&=\left(\begin{array}{c}
\sum_{\mathbf{Q}'}\left[
\mathbf{Q}\cdot\mathbf{v}^{(1)}_{\mathbf{Q}'}\otimes\mathbf{v}^{(0)}_{\mathbf{Q}-\mathbf{Q}'}
+
\mathbf{Q}\cdot\mathbf{v}^{(0)}_{\mathbf{Q}'}\otimes\mathbf{v}^{(1)}_{\mathbf{Q}-\mathbf{Q}'}
+
(\mathbf{Q}'\times\mathbf{b}^{(0)}_{\mathbf{Q}'})\times\mathbf{b}^{(1)}_{\mathbf{Q}-\mathbf{Q}'}
+
(\mathbf{Q}'\times\mathbf{b}^{(1)}_{\mathbf{Q}'})\times\mathbf{b}^{(0)}_{\mathbf{Q}-\mathbf{Q}'}
\right]
\\
-\mathbf{Q}\times\sum_{\mathbf{Q}'}(
\mathbf{v}^{(0)}_{\mathbf{Q}'} \times \mathbf{b}^{(1)}_{\mathbf{Q}-\mathbf{Q}'}
+
\mathbf{v}^{(1)}_{\mathbf{Q}'} \times \mathbf{b}^{(0)}_{\mathbf{Q}-\mathbf{Q}'}
)
                 \end{array}\right).\nn
\end{eqnarray}
We obtain new terms in the eigenvalue problem which finally is reduced to the problem of obtaining the
maximal eigenvalue of an integral equation in which the coefficients are solutions of the static MHD equations.
Now let us analyze the perspectives.

\begin{figure}
\hspace{-2cm}\includegraphics[height=15cm]{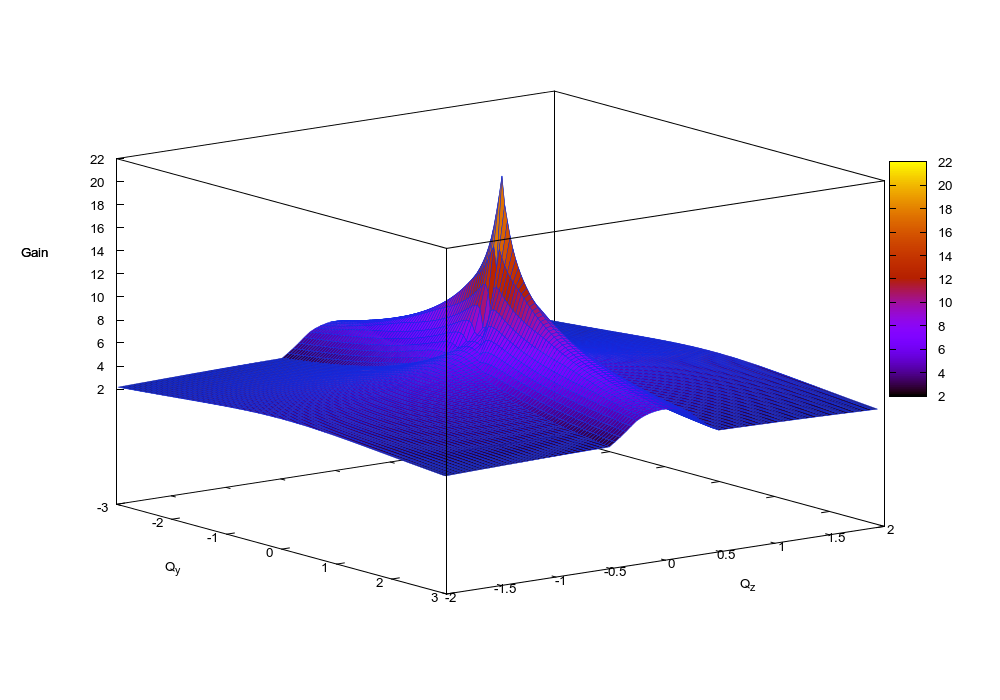}
\caption{Gain for the case of $\omega=0,\quad \theta=0$}
\end{figure}

\begin{figure}
\hspace{-2cm}\includegraphics[height=14cm]{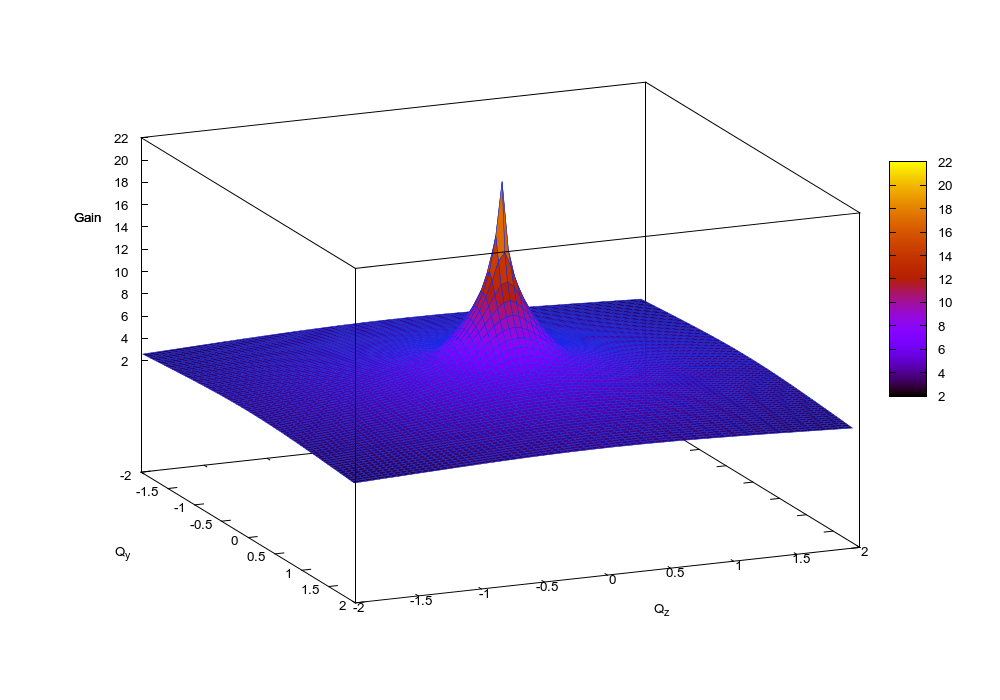}
\caption{Gain for the case of $\omega=0,\quad \theta=\frac{\pi}{2}$}
\end{figure}

\begin{figure}
\hspace{-2cm}\includegraphics[height=15cm]{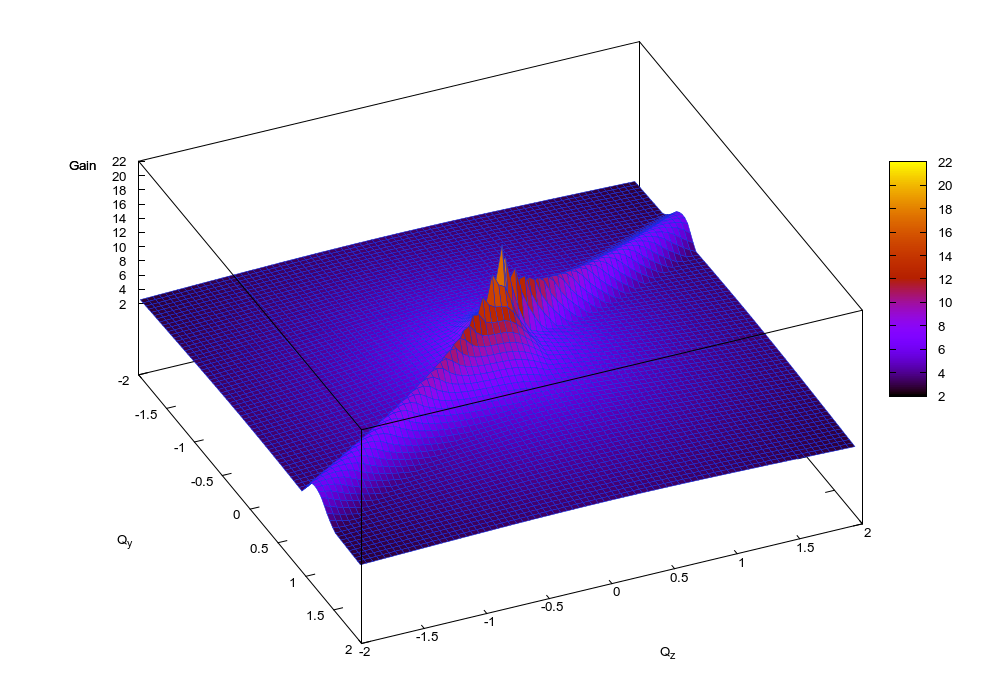}
\caption{Gain for the case of $\omega=0,\quad \theta=\frac{\pi}{3}$}
\end{figure}

\begin{figure}
\hspace{-3cm}\includegraphics[height=15cm]{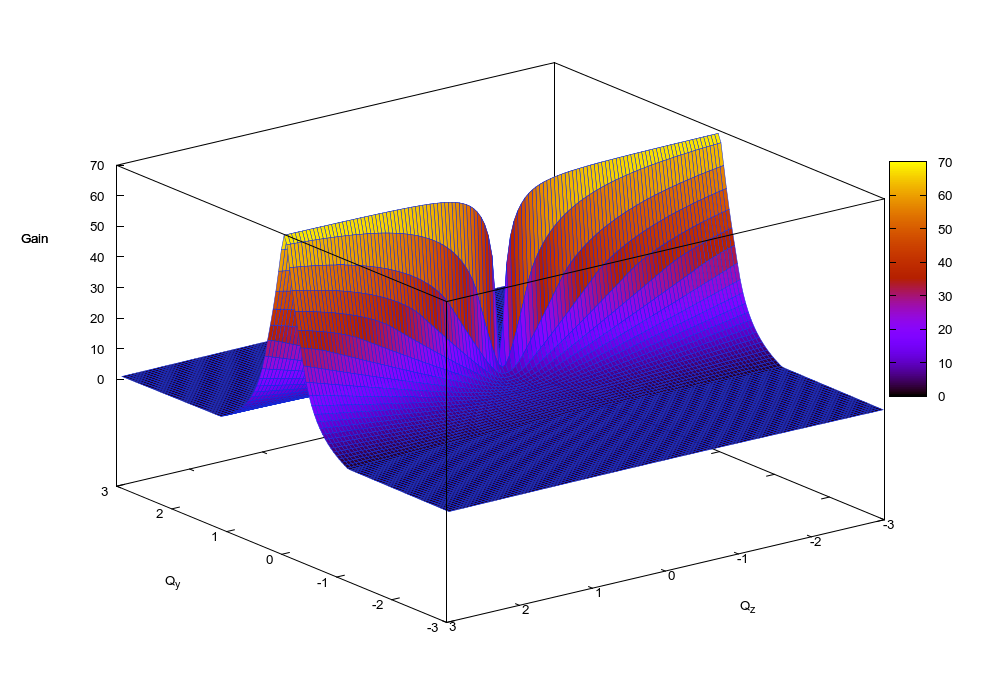}
\caption{Gain for the case of $\omega=-\frac23,\quad \theta=\frac{\pi}{2}$}
\end{figure}

\begin{figure}
\hspace{-4cm}\includegraphics[height=14cm]{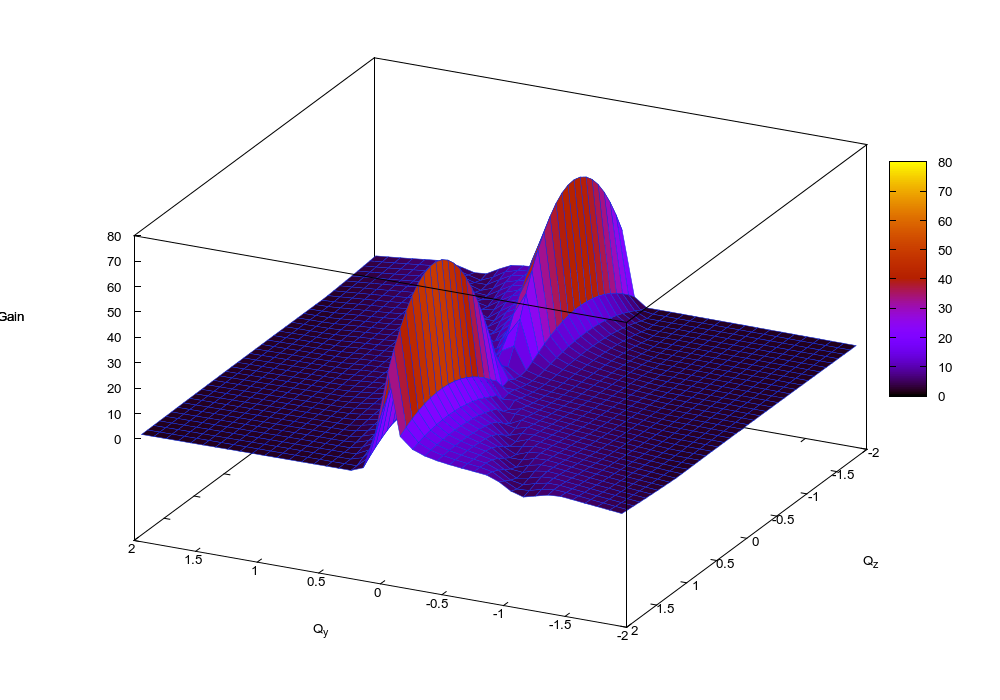}
\caption{Gain for the case of $\omega=-\frac23,\quad \theta=\frac{\pi}{3}$}
\end{figure}

\chapter{Linear Analysis}

\section{Introduction}
%
The aim of the present chapter is to derive the equations which have to be
numerically solved in order to calculate the 
many orders of magnitude enhancement of the effective viscosity. 
We generalize the magnetohydrodynamic equations used in the original PhD work by
Hristov \cite{Chagelishvili:93}
and arrive to the conclusion that divergence SMW amplification coefficient is
the mechanism which trigger the self-sustained turbulence in accretion disks. 
As the problem is a statistical one for short wavelength SMW our starting point
in the next section \ref{ch2sec:model} is the MHD equations in rectangular
Cartesian coordinates. 
In this section are given the main MHD equations \Eqref{ch2VelocityEvolution} and
\Eqref{ch2MagneticEvolution} 
in the wave-vector space. 
Linearized equations considered in sec.~\ref{ch2sec:linear} is the first step of
the analysis of the derived equations. 
The phase portraits considered in sec.~\ref{ch2sec:portraits} is the important tool
for our understanding of SMW amplification -- the heart of the disk turbulence,
cf.~page 28~\cite{Balbus:98}.
Test for the accuracy of the numerical calculation is the
analytical solution of SMW amplification for the case of pure shear described in
sec.~\ref{ch2sec:Solution}.
The well-known magnetorotational instability (MRI) and the corresponding
deviation equations are rederived in sec.~\ref{ch2sec:MRI} as special case of
linearized MHD equations for SMW.
It is considered in the final section~\ref{ch2sec:conclusions} that the derived
equations open the perspective 
for creation of self-sustained wave turbulence for the interesting for
astrophysics case of large Reynolds numbers.

\section{Model and description of the basic equations}
\label{ch2sec:model}
%
First of all let us clarify that for astrophysical applications we suppose
hydrogen plasma. 
For completely ionized plasma  we have well known expression for the 
magnetic diffusivity $\nu_\mathrm{m}$ and kinematic viscosity $\nu_\mathrm{k}$
\begin{eqnarray}
\nu_\mathrm m&=&\varepsilon_0 c^2 \varrho
=\frac{e^2c^2m_e^{1/2}\mathcal{L}_e}{0.6\times 4\pi T_e^{3/2}}
\ll \nu_\mathrm{k}=\frac{\eta}{\rho}
=\frac{0.4\,T_p^{5/2}}{e^4n_pM_p^{1/2}\mathcal{L}_p},\quad \rho
=n_p m_p,\qquad \varepsilon_0=\frac1{4\pi},\\
\mathcal{L}_p&=&\ln\left(\frac{\lambda_{\mathrm D}T_p}{e^2}\right),\quad
\mathcal{L}_e=\ln\left(\frac{\lambda_{\mathrm D}T_e}{e^2}\right),\quad
\frac{1}{\lambda_\mathrm D^2}=4\pi
e^2\left(\frac{n_e}{T_e}+\frac{n_p}{T_p}\right),\quad
e^2=\frac{q_e^2}{4\pi\varepsilon_0}.
\end{eqnarray}
At hight temperatures the magnetic Prandtl number is 
$\mathrm{P}_\mathrm{m}\equiv\nu_\mathrm{k}/\nu_\mathrm{m}\propto T^4\gg1.$
Here we use the self-explaining notations for the hydrogen plasma:
the masses of electron $m_e$ and proton $M_p$, the number of electrons $n_e$
and $n_p$ protons per unit volume, the density $\rho$ and the
viscosity $\eta$, electron $T_e$ and proton temperature $T_p$, Coulomb
logarithms $\mathcal{L}_e$ and $\mathcal{L} _p$, Debye screening
length $\lambda_\mathrm D,$ and the coefficient in Coulomb interaction
$e^2.$ We will suppose equal temperatures $T_e=T_p$ and of charge neutrality
$n_p=n_e.$
The bare viscosity is determined by momentum transport of the elementary
particles of the fluid: 
molecules, atoms, ions, protons and electrons is known as molecular viscosity.

For weak magnetic fields the \alf velocity
$V_{\mathrm{A}}$ is much smaller than the sound $c_{\mathrm{s}}$ speed
\begin{equation}
V_{\mathrm{A}}=\frac{B_0}{\sqrt{\mu_0\rho}}
\ll c_{\mathrm{s}}=\sqrt{\gamma\frac{P}{\rho}},
\qquad \gamma\equiv \frac{c_p}{c_v}=\frac{5}{3}.
\end{equation}
This strong inequality for weak magnetic fields lead that
for slow magnetosonic waves (SMW) we can use the incompressible fluid
$\mathrm{div} \mathbf{V}=0 $ approximation.
In this approximation the slow magnetosonic waves have the same dispersion
as \alf waves (AW) and are often called pseudo-\alf waves.

We suppose the geometry of an accretion disk and choose the $z$- axis along the
axis of rotation with angular velocity $\mathbf{\Omega}.$
The $y$-axis we choose parallel to the shear flow which is local
$\mathbf{e}_\varphi$ direction.
The local $x$-axis is chosen to be along the local $r$ direction. The modulus of
the shear velocity depends on distance to the central compact object $r.$
In the so chosen coordinates for the shear flow we have
\be \mathbf{V}_\mathrm{shear}=Ax\mathbf{e}_y,\quad \mathbf{e}_x=\mathbf{e}_r,
\quad \mathbf{e}_y=\mathbf{e}_\varphi ,\ee
where $A$ is the shear rate with dimension of frequency.

Introducing  the field of shear flow in wave-vector space (``momentum space'' in
the quantum mechanics)
and substantial Lagrangian derivative  
\begin{equation}
\label{ch2ushear}
\mathbf{U}_\mathrm{shear}(\mathbf{Q})\equiv-Q_y\mathbf{e}_x,\quad 
\mathrm{D}_\tau^{\mathrm{\,shear}}\equiv\partial_\tau+\mathbf{U}_{\mathrm{shear}
}(\mathbf{Q})\cdot\partial_{\mathbf{Q}}=\partial_\tau-Q_y\partial_{Q_x}
\end{equation}
the MHD equations in wave-vector space read \cite{Chagelishvili:93,Dimitrov:11}
 as
\begin{eqnarray}
\mathrm{D}_\tau^{\mathrm{\,shear}}\mathbf{v}_\mathbf{Q}(\tau) &=&
-v_{x,\mathbf{Q}}\mathbf{e}_y+2n_y\mathbf{n}v_{x,\mathbf{Q}}
+ 2\omega_\mathrm{c}\mathbf{n}\,(n_yv_{x,\mathbf{Q}}-n_xv_{y,\mathbf{Q}})
+2\boldsymbol{\omega_\mathrm{c}}\times \textbf{v}_{\mathbf{Q}} +
(\boldsymbol{\alpha}\cdot\mathbf{Q})\,\mathbf{b}_\mathbf{Q} \nonumber\\
&&-\nu^\prime_\mathrm{k}Q^2\mathbf{v}_\mathbf{Q}
+\Pi^{\perp\mathbf{Q}}\cdot\sum_{\mathbf{Q}'}\left[\mathbf{v}_{\mathbf{Q}'}
\otimes\mathbf{v}_{\mathbf{Q}
-\mathbf{Q}'} +
\mathbf{b}_{\mathbf{Q}'}\otimes\mathbf{b}_{\mathbf{Q}-\mathbf{Q}'}\right]
\cdot\mathbf{Q},
\label{ch2VelocityEvolution}
\\
\label{ch2MagneticEvolution}
\mathrm{D}_\tau^{\mathrm{\,shear}}\mathbf{b}_\mathbf{Q}(\tau) &=&
b_{x,\mathbf{Q}}\mathbf{e}_y
-(\mathbf{Q}\cdot\boldsymbol{\alpha})\,\mathbf{v}_\mathbf{Q}
-\nu'_\mathrm{m}Q^2\mathbf{b}_\mathbf{Q}\nonumber \\
&&+\Pi^{\perp\mathbf{Q}}\cdot\sum_{\mathbf{Q}'}\left[\mathbf{b}_{\mathbf{Q}'}
\otimes\mathbf{v}_{\mathbf{Q}
-\mathbf{Q}'} -
\mathbf{v}_{\mathbf{Q}'}\otimes\mathbf{b}_{\mathbf{Q}-\mathbf{Q}'}\right]
\cdot\mathbf{Q},
\end{eqnarray}
where velocity and magnetic field are perpendicular to the wave-vector 
\begin{equation}
\mathbf{v}_\mathbf{Q}(\tau_0)=\Pi^{\perp}\mathbf{v}_\mathbf{Q}(\tau_0),
\quad
\mathbf{b}_\mathbf{Q}(\tau_0)=\Pi^{\perp}\mathbf{b}_\mathbf{Q}(\tau_0),\quad
\Pi^{\perp\mathbf{Q}}\equiv\openone-\frac{\mathbf{Q}\otimes
\mathbf{Q}}{Q^2}=\openone-\mathbf{n}\otimes \mathbf{n},
\;\,\, \mathbf{n}\equiv \frac{\mathbf{Q}}{Q},
\end{equation}
and the $\Pi^{\perp\mathbf{Q}}$ is the projection operator.

In the next section we will analyze the linear terms known from other works on
space plasmas. 

\section{Linear terms}
\label{ch2sec:linear}
%
The linearized MHD equations in many special cases of magnetized shear flows
are investigated in great details. One of the purposes of the present section
is to give different tests for the validity of the general equations which was
written in Eulerian time independent wave-vector space $\textbf{Q}$.

If the nonlinear terms are negligible the linearized equations is better to be
analyzed in Lagrangian description in wave-vector space.
Let us introduce time-dependent wave-vector 
\be
\label{ch2ktau}
\mathrm{d}_\mathrm{\tau}\mathbf{K}(\tau)
=\mathbf{U}_\mathrm{shear}(\mathbf{K}(\tau)).
\ee
This equation according \Eqref{ch2ushear} has the solution 
\ba 
&&K_z=\mathrm{const},\quad K_y=\mathrm{const},\quad K_x=K_{x,0} -K_y \tau,
\quad K_{x,0}=\mathrm{const}, \\
&&K_\alpha\equiv\boldsymbol{\alpha}\cdot\mathbf{K}=K_y\sin{\theta}+K_z\cos{\theta}. \nn
\ea
For $K_y\neq 0$ we can parameterize $\tau_0=K_{x,0}/K_y$ then we have $K_x(\tau)
= -(\tau-\tau_0)K_y$.
Considering only one wave without restrictions we can choose $\tau_0=0.$

Now Lagrangian derivative is converted to ordinary time derivative
\be \mathrm{D}_\mathrm{\tau}^\mathrm{shear} \rightarrow
\mathrm{d}_\mathrm{\tau}, \ee
and in the MHD equations we have formally to replace the index $\mathbf{Q}$ by 
$\mathbf{K},$ i.e.
\be\mathbf{v}_\mathbf{Q} \rightarrow \mathbf{v}_\mathbf{K}\rightarrow\mathbf{v},
\qquad 
\mathbf{b}_\mathbf{Q} \rightarrow \mathbf{b}_\mathbf{K}\rightarrow\mathbf{b},
\qquad 
\mathbf{Q} \rightarrow \mathbf{K}.\ee
After this substitution the linearized system \Eqref{ch2VelocityEvolution} and
\Eqref{ch2MagneticEvolution} takes the form
\begin{eqnarray}
\mathrm{d}_\tau\mathbf{v}&=&-v_x\mathbf{e}_y+2n_y\mathbf{n}v_x+K_\alpha\mathbf{b
}
-\nu^\prime_\mathrm{k}K^2\mathbf{v} +
2\omega_\mathrm{c}\left[\mathbf{n}(n_yv_x-n_xv_y)+(v_y\mathbf{e}_x-v_x\mathbf{e}
_y)\right],
\label{ch2dv_final}\\
\mathrm{d}_\tau\mathbf{b}&=&b_x\mathbf{e}_y - K_\alpha\mathbf{v} -
\nu^\prime_\mathrm{m}K^2\mathbf{b},
\qquad \mathbf{K}\cdot\mathbf{b}=0, \qquad \mathbf{K}\cdot\mathbf{v}=0,
\label{ch2db_final}
\end{eqnarray}
and the expression for the pressure taking into account
\Eqref{ch2db_final} reads as
\begin{equation}
\label{ch2Pressure_K}
\mathcal{P}=-\frac{1}{K^2}\left\{2K_yv_x +
(\mathbf{b}\cdot\boldsymbol{\alpha})\mathbf{K}^2 
+ 2\omega_\mathrm{c}(K_yv_x-K_xv_y) \right\}.
\end{equation}

The simplest derivation of the linearized system \Eqref{ch2dv_final},
\Eqref{ch2db_final} and \Eqref{ch2Pressure_K}
is to substitute the single wave anzatz
\begin{eqnarray}
\label{ch2v_wave}
 \mathbf{V}(t,\textbf{r})&=&Ax\mathbf{e}_y +
 \ii V_\mathrm{A}\mathbf{v}(\tau)\mathrm{e}^{\ii\mathbf{k}\cdot\mathbf{r}},
 \qquad \;\;\; v \ll 1,\\
 \label{ch2b_wave}
\mathbf{B}(t,\textbf{r})&=&B_0\boldsymbol{\alpha}
 + B_0\mathbf{b}(\tau)\mathrm{e}^{\ii\mathbf{k}\cdot\mathbf{r}}, \qquad \quad
\;\; b
\ll 1,\\
\label{ch2k_wave}
\mathbf{k}(t)&=&-k_y\tau\mathbf{e}_x+k_y\mathbf{e}_y+k_z\mathbf{e}_z,\qquad \tau
\equiv At.
\end{eqnarray}
into MHD equations \Eqref{ch1MHD} and \Eqref{ch1MHD-B} in coordinate space and to
eliminate the pressure. 

In the next section we will analyze the phase the amplification of SMW in
Lagrangian variables.

\section{Phase portraits of amplification of SMW}
\label{ch2sec:portraits}
%

\begin{figure}
\includegraphics[height=15cm]{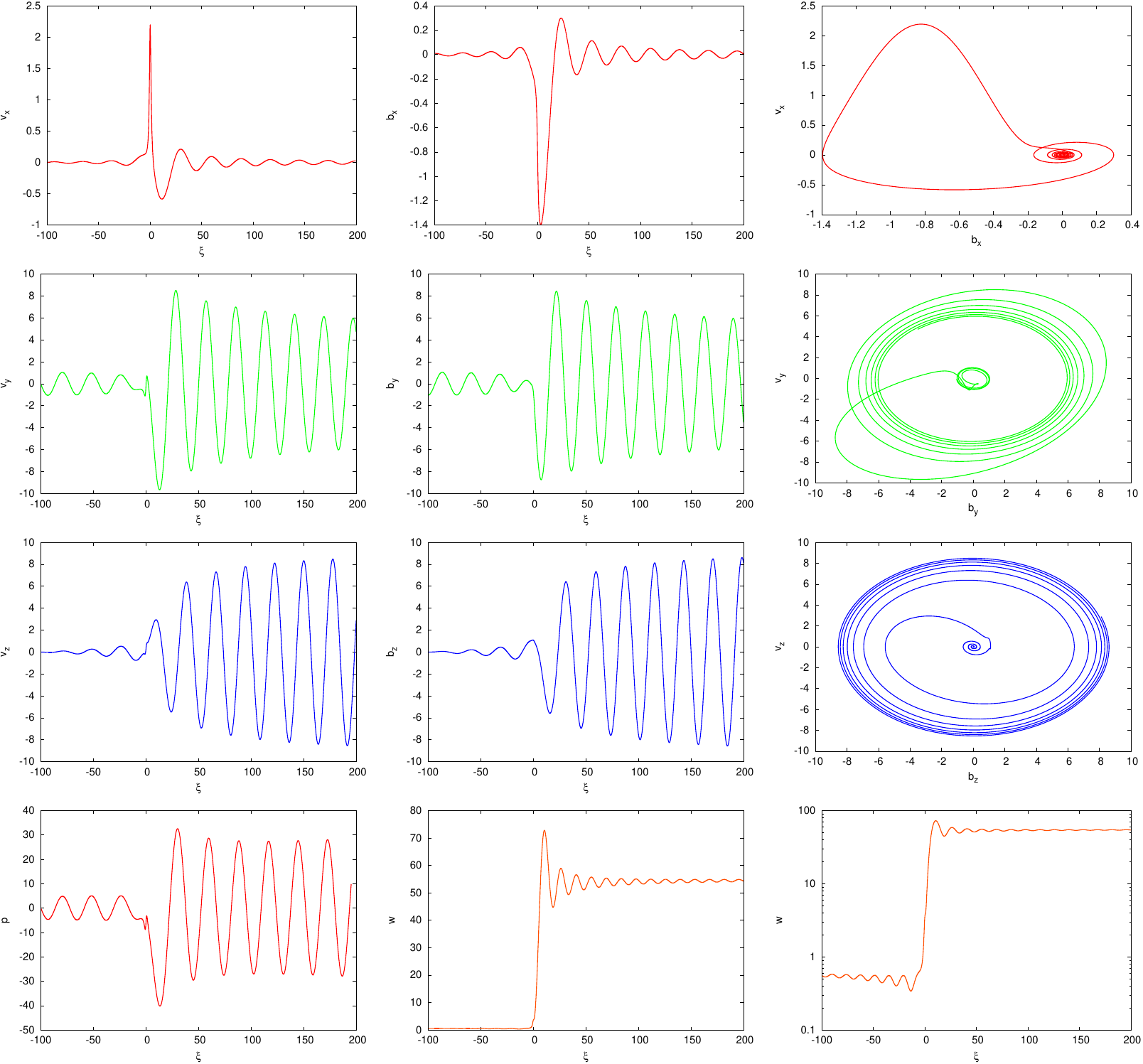}
\caption{
Velocity $\mathbf v(\xi)$, magnetic field $\mathbf b(\xi)$, pressure $p$, total
wave energy $w$ and
phase portraits of radial $e_r=e_x$ (upper row), azimuthal $e_\varphi=e_y$
(middle row) and axial $e_z=e_z$ (bottom row) components of the fields.
The area of the asymptotic cycles is proportional to the energy of each
component. The increasing of the area of azimuthal component $(v_y,\;b_y)$
describes the amplification of SMW -- the main effect investigated in the
present work.
The phase portraits of the right hand side trivialize the analytically derived
amplification of SMW.
For different projections we have transmissions from zero or finite values of
the asymptotic area describing energy distribution of the wave.}
\label{ch2phase_portrait}
\end{figure}

On \Fref{ch2phase_portrait} are shown numerical solutions of the system of
equations \Eqref{ch2dv_final} and \Eqref{ch2db_final} corresponding to
initial conditions:
\begin{eqnarray}
\label{ch2initial_conditions}
&&v_x=0.0003162,\;v_y=-0.033329,\; v_z=0.0000008,\; K_y=0.17,\nn \\
&&b_x=0.0099995,\;b_y=-1.054045,\; b_z=0.0000158,\; K_z=0.1.
\end{eqnarray}
For the radial motion the velocity and the magnetic field start at $\xi=-\infty$
and finish at $\xi=\infty$ with zero values. Only at the moment of wave
amplification $\xi=0$ and $K_x=0$ we have
significant radial components $(v_x,\, b_x)$. For azimuthal motion $(v_y,\,
b_y)$,
which describes the SMW, we assume an initial amplitude, which is
amplified by shear flow. This is the main feature of the phenomenon which we
explore.
Axial components $(v_z,\,b_z)$, which describe AW,
has no amplification; at $\xi=-\infty$ they have zero value. Nonzero area
of asymptotic cycle in the plane $(v_z,\,b_z)$ corresponds to the
conversion of SMW in the AW. The ratio of dimensionless wave energy density
$w=\frac{1}{2}(\mathbf{v}^2 + \mathbf{b}^2)$ at $\xi=\infty$ and $\xi=-\infty$
describes the wave gain $G=w(\infty)/w(-\infty) $.

In order to check the validity of the linearized equations in the next section
we present the solvable case without rotation.

\section{Analytical solution for pure shear of an inviscid magnetized fluid}
\label{ch2sec:Solution}
For non-rotating $\omega_\mathrm{c}=0$ ideal
$\nu^\prime_\mathrm{m}=0=\nu^\prime_\mathrm{k}$ fluid is the MHD system 
\Eqref{ch2dv_final} and \Eqref{ch2db_final} reduced to
two second order equations. 
This result was initially \cite{Mishonov:09} derived only for toroidal static
magnetic fields $B_{0,z}=0$.
Introducing new variables 
\begin{eqnarray}
K_\perp\equiv\sqrt{K_y^2+K_z^2},\quad 
\quad\tilde{K}\equiv\frac{K_\perp}{K_y}K_\alpha\quad\xi\equiv\frac{K_y}{K_\perp}
\tau,\quad \psi(\xi)\equiv b_x(\xi)\sqrt{1+\xi^2}, 
\end{eqnarray}
we obtain 
Schr\"odinger type equation with solution expressed by
confluent Heun functions
\begin{eqnarray}
\label{ch2Schrodinger}
&&\mathrm{d}_\xi^2\psi+\left[\tilde{K}^2-\frac{1}{(1+\xi^2)^2}\right]\psi=0,
\quad
\psi(\xi)=C_\mathrm g \psi_\mathrm g(\xi) + C_\mathrm u \psi_\mathrm u(\xi),
\qquad \\
&&\psi_\mathrm g\!\!=\!\sqrt{1+\xi^2}
\,\mathrm{HeunC}(0,-\frac{1}{2},0,-\frac{\tilde{K}^2}{4},\!\frac{1+\tilde{K}^2}{
4},-\xi^2),\\
&&\psi_\mathrm u\!\!=\!\xi\sqrt{1+\xi^2}
\,\mathrm{HeunC}(0,+\frac{1}{2},0,-\frac{\tilde{K}^2}{4},\!\frac{1+\tilde{K}^2}{
4},-\xi^2),
\end{eqnarray}
where $C_\mathrm g$ and $C_\mathrm u$ are arbitrary constants.
Formal correspondence to quantum mechanics is by replacements 
\begin{equation}
\frac{2m}{\hbar^2}U(\xi)\rightarrow\frac{1}{(1+\xi^2)^2},\quad
\frac{2m}{\hbar^2}E\rightarrow \tilde{K}^2.
\end{equation}

For the $z$-component of the magnetic field we have equation similar to forced
harmonic oscillator
\begin{equation}
\label{ch2bz}
\left(\mathrm{d}_{\xi}^2 + \tilde{K}^2\right)b_z = 2K_z \frac{v_x}{1+\xi^2}.
\end{equation}
In such a way using also the condition $\mathbf{K}\cdot\mathbf{b}=0$
we obtain the analytical solution for the magnetic field.
The dimensionless magnetic field $\mathbf{b}$ expressed by the effective
$\psi$-function
\begin{eqnarray}
\label{ch2b_solution}
&&\!b_x=\frac{\psi(\xi)}{\sqrt{1+\xi^2}},\qquad \chi(\xi) \equiv
\tilde{C}_\mathrm{g}\cos(L\xi) +
\tilde{C}_\mathrm{u}\frac{\sin(\tilde{K}\xi)}{\tilde{K}}, \\
&&b_y=\!-\frac{2K_z^2}{K_y\tilde{K}}\!\int_{-\infty}^{\xi}
\sin[\tilde{K}(\xi-\xi^{\prime})]
\frac{v_x(\xi^{\prime})}
{1+\xi^{\prime 2}}\mathrm{d}\xi^{\prime}
+\frac{\tilde{K}}{K_y}\frac{\xi\psi(\xi)}{\sqrt{1+\xi^2}}
-\frac{K_z}{K_y}\chi(\xi), \nn \\
&&b_z
= \frac{2K_z}{\tilde{K}} \int_{-\infty}^{\xi}
\sin[\tilde{K}(\xi-\xi^{\prime})]
\frac{v_x(\xi^{\prime})}
{1+\xi^{\prime 2}}\mathrm{d}\xi^{\prime}+\chi(\xi).\nn
\end{eqnarray}
\begin{figure}
\includegraphics[height=5.5cm]{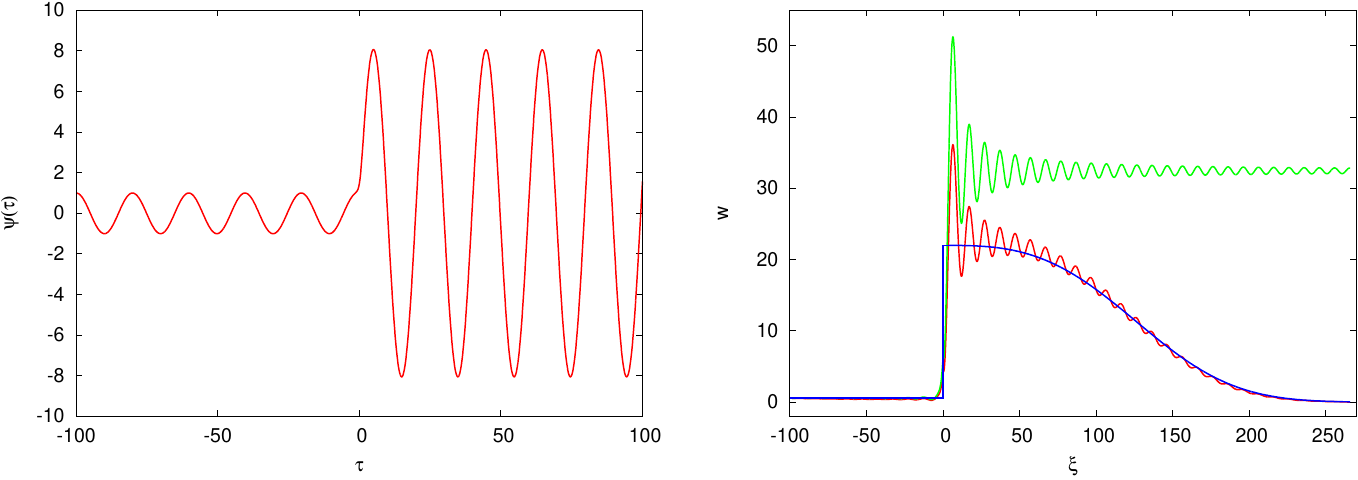}
\caption{\emph{Left}: Solution of effective Schroedinger equation for
$\psi(-100)=1,
\mathrm{d}_\tau\psi(-100)=0,$ and $K_\perp=\sqrt{0.1}\,$. When $\tau=0$ is
observed
almost saltatory increase of the amplitude of oscillation which is related to
the amplification of SMW.
The analytical solution in framework of Heun functions reproduces the
amplification
discovered by numerical analysis \cite{Chagelishvili:93}.
\emph{Right}: Wave energy as a function of time with damping pulsations around
the average value.
One can see that viscosity leads to attenuation of the waves and further heating
of plasma.
Enveloping curve corresponds to $\exp\left[-\int{\nu
k^2(t)\,\mathrm{d}t}\right]$.
The continuous line demonstrates that attenuation of the waves for the case of
small dissipation is
well-described by WKB approximation.
}
\label{ch2Enegy_decay}
\label{ch2psi_time}
\end{figure}
A typical solution is shown in \Fref{ch2psi_time}.
The velocity amplitudes is represented by the derivatives of the magnetic fields
\be
v_z=-\frac1{\tilde{K}}\mathrm{d}_{\xi}b_z,\qquad
v_x=-\frac1{\tilde{K}}\mathrm{d}_{\xi}b_x,\qquad v_y=\frac{K_\perp}{K_y}\xi v_x
-\frac{K_x}{K_y}v_z,
\ee
this is actually a consequence of the Alvf\'en theorem that magnetic field is
frozen in a fluid with negligible Ohmic resistivity. 

\begin{figure}
\includegraphics[height=5.5cm]{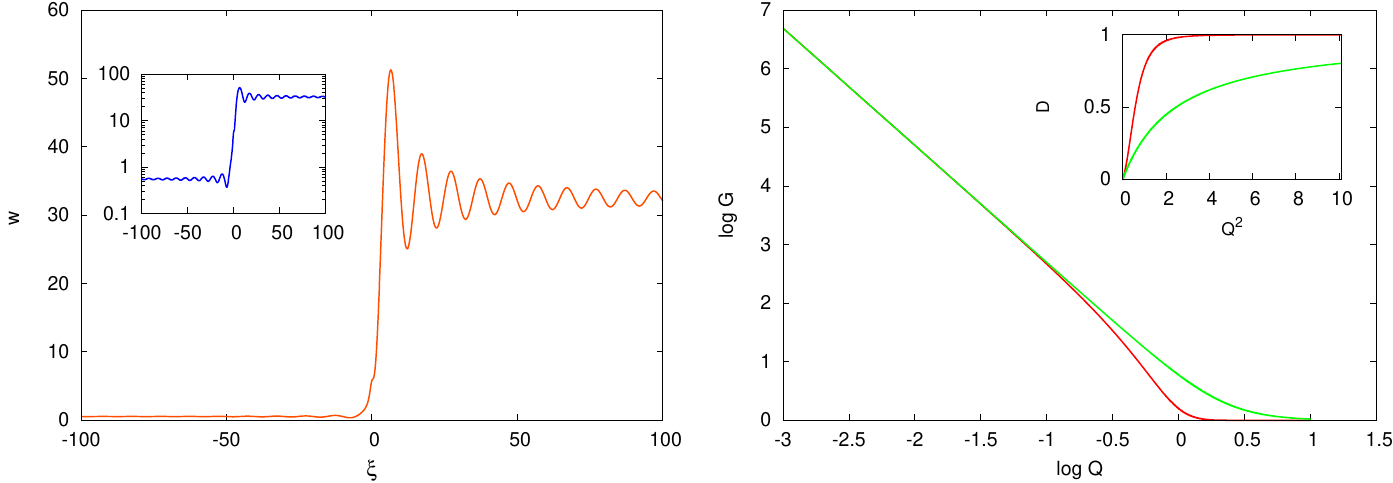}
\caption {\emph{Left}: wave energy as a function of time. On inset figure energy
of wave is shown in logarithmic scale. \emph{Right}: logarithm of the ratio of
amplification of waves as a function of the logarithm of the wave vector. Simple
asymptotic formula \Eqref{ch2gain} (right curve) accurately describes
long-wavelengths
asymptote of the analytical solution.
The left hand side analytically reproduces the energy amplification
observed for first time by the numerical calculations.
The right hand side is new and reproduces the long-wavelength
asymptotic for the amplification coefficient.
}
\label{ch2Energy_and_gain}
\end{figure}
For incompressible fluid the difference between AW and SMW is only in
polarization,
but effect of amplification of SMW exists even if compressibility is taken into
account.  

We consider propagation of plane waves in a homogeneous medium. 
The variables $\mathbf{v}(\xi)$ and $\mathbf{b}(\xi)$ are time dependent 
but for zero dissipation at infinite time we have well defined asymptotics of
the amplitudes.
At those infinite times we have fixed frequency which corresponds to the
dispersion of SMW.
In such a way the analytical solution describes the amplification of SMW from
$-\infty$ to $+\infty.$

A typical time dependence of the
wave energy density $w(\xi)$ is given in \Fref{ch2Energy_and_gain} (left).
For the case of in-plane static magnetic field $B_{0,z}=0$ the amplification of
the SMW is 
directly determined by the amplification of the amplitude of the $\psi$-function
for which we have the 
asymptotic derived by the $\delta$-function method
\begin{equation}
\mathcal{G}(|K_\perp| \ll 1) \approx \frac{\pi^2}{2K_\perp^2} + 1
=\frac12\frac{\pi^2 A^2}{V_\mathrm{A}^2(k_\varphi^2+k_z^2)} +1\gg 1,
\label{ch2gain}
\end{equation}
and the details of derivation are given in \cite{Mishonov:09}.

On the right of the same figure asymptotic formula
for amplification at large wavelengths
and the exact solution are presented. Formally the use of this asymptotic is
reduced to
the application of the method of $\delta$-like potential approximation in
quantum mechanics,
in our case to substitution
\begin{equation}
\label{ch2delta_approximation}
\frac{1}{(1+\xi^2)^2} \rightarrow \pi\delta(\xi)
\end{equation}
in the effective Schr\"odinger equation \Eqref{ch2Schrodinger}. The asymptotic
accuracy of
the $\delta$-potential method at long-wave limit reveals that the phenomenon of
amplification of SMW cannot be understood in the framework of WKB method.

The presence of small viscosity leads to damping of waves determined by the
total viscosity 
$\nu_\mathrm{tot}=\nu_\mathrm{k}+\nu_\mathrm{m},$
and we have damping factor 
$\exp\left[-\int{\nu_\mathrm{tot} k^2(t)\,\mathrm{d} t}\right]$ 
which leads to the time dependence of the amplitude  
\begin{equation}
\label{ch2asymptotics}
\psi \approx D_\mathrm{f}\,\theta(\tau)\cos(K_y\tau+\phi_\mathrm{f})
\exp \left(-\nu^\prime_\mathrm{tot} K_y^2 \tau^3/3\right), \qquad
\nu^{\prime}_\mathrm{tot} \ll 1, \qquad |\tau| \gg 1.
\end{equation}
The influence of viscosity on the waves is illustrated in
\Fref{ch2psi_time}-\emph{Right},
where two solutions with zero, nonzero viscosity, and
enveloping simple exponential function are depicted.
In short we revealed that wave amplification can be observed even without any
rotation and very large amplification factors can be not related to MRI. 
Once amplified the \alf waves decay as independent free waves; 
one can say that they are asymptotically free.  

The derived asymptotic formula for the gain \Eqref{ch2gain} gives infinite wave
amplification for $K_y=0.$
As we will see in the next section this axisymmetric case for the accretion
disks 
give the well known MRI for nonzero rotation and axial magnetic field.

\section{Amplification of the MHD Waves} %
%
The odd and even solutions have asymptotics at $\xi\rightarrow
\infty$
\begin{equation}
\label{ch2asymptotics}
\psi_\mathrm g \approx D_\mathrm g\cos(K_\perp \xi+\delta_\mathrm g),\qquad
\psi_\mathrm u \approx D_\mathrm u\cos(K_\perp \xi+\delta_\mathrm u),
\end{equation}
where the asymptotic phase shifts $\delta_\mathrm g(K_\perp^2),\;\delta_\mathrm u(K_\perp^2)$
depend on the effective energy.
For a sufficiently large wave-vectors we have the asymptotic
\begin{equation}
\delta_\mathrm g(K_\perp^2 \gg 1)\approx0, \qquad
\delta_\mathrm u(K_\perp^2 \gg 1)\approx-\frac{\pi}{2}.
\end{equation}

As we will see later, the averaged amplification coefficient of the
energy of MHD waves $\mathcal{A}(\delta_\mathrm g,\delta_\mathrm u)$
depends only on the asymptotic phases of the solutions but not on the
amplitudes $D_\mathrm g$ and $D_\mathrm u$.

\subsection{Auxiliary Quantum Mechanical Problem}%
%
Temporarily introducing imaginary exponents
$\exp(\ii\mathbf{k}\cdot\mathbf{r})$ instead of
$\sin(\mathbf{k}\cdot\mathbf{r})$ and
$\cos(\mathbf{k}\cdot\mathbf{r})$ gives significant simplification
of the analytical calculations related to the physics of the waves. In
this subsection we will consider $\psi$ in as a complex
function in order to make easier any further analysis of the MHD
amplification coefficient. In order to analyze the effective
MHD equation we will solve the quantum-mechanical
counterpart of our MHD problem, that is a tunneling through a barrier
$2m U/\hbar^2=1/(1+\xi^2)^2,$ supposing that $\psi$ is a complex
function.
Consider an incident wave with a unit amplitude, a reflected wave with
amplitude $R$ and a transmitted wave with amplitude $T$
\begin{eqnarray}
&&\psi(\xi\rightarrow -\infty)\approx \exp(\ii K_\perp \xi)
  + {R}\exp(-\ii K_\perp\xi),\\
&&\psi(\xi\rightarrow +\infty)\approx  T\exp(+\ii K_\perp\xi).
\end{eqnarray}
Using the asymptotics of the eigen-functions
\be
\label{ch2asg}
\psi_\mathrm g\approx \left\{ \begin{array}{l}
D_\mathrm g\cos(K_\perp\xi-\delta_\mathrm{g})
 \qquad\mbox{for  } \xi\rightarrow -\infty,\\
D_\mathrm g\cos(K_\perp\xi+\delta_\mathrm g)
 \qquad\mbox{for  } \xi\rightarrow +\infty,
\end{array}\right.
\ee
and
\be
\label{ch2asu}
\psi_\mathrm u\approx \left\{ \begin{array}{l}
 -D_\mathrm u\cos(K_\perp\xi-\delta_u)
 \quad\;\;\;\mbox{for  } \xi\rightarrow -\infty,\\
\;\;\, D_\mathrm u\cos(K_\perp\xi+\delta_\mathrm u)
  \qquad\mbox{for  } \xi\rightarrow +\infty
\end{array}\right.
\ee
as well as the general condition
\begin{equation}
\psi(\xi)=C_\mathrm g^\mathrm{(q)} \psi_\mathrm g(\xi)
+ C_\mathrm u^\mathrm{(q)} \psi_\mathrm u(\xi),
\end{equation}
we compare the coefficients in front of $\exp(iK_\perp\xi)$ and $\exp(-iK_\perp\xi)$
for $\xi\rightarrow -\infty$ and $\xi\rightarrow \infty$.  The
solution to a simple matrix problem yields
\be
C_\mathrm g^\mathrm{(q)}D_\mathrm g=\exp(\ii\delta_\mathrm g),\qquad
C_\mathrm u^\mathrm{(q)}D_\mathrm u=-\exp(\ii\delta_\mathrm u)
\ee
Then for the tunneling amplitude we get
\be
T=|T|\mathrm{e}^{\ii(\delta_\mathrm u+\delta_\mathrm g-\pi/2)}
\ee
and finally for the tunneling coefficient we obtain
\be
\mathcal{D}=|T|^2=\mathrm{s_{ug}^2},\qquad
\mathrm{s_{ug}}\equiv\sin(\delta_\mathrm u-\delta_\mathrm g).
\ee
The convenience of the tunneling coefficient is that it varies in
the range $0\le\mathcal{D}\le1$. In the next two subsections we will
present the SMW amplification coefficient as a function of the
tunneling coefficient $\mathcal{K}=2/\mathcal{D}-1$ on the analogy
of Heisenberg's ideas in quantum mechanics where the statistical
properties of the scattering problem depend only on the phases of
the $S$-matrix.

The phases $\delta_\mathrm{g},\;\delta_\mathrm{u}\in (-\pi,\;\pi)$
and amplitudes $D_\mathrm{g}$ and $D_\mathrm{u}$
can be determined continuing the exact wave functions with WKB asymptotics
\Eqref{ch2asymptotics}. Thus we obtain
\begin{eqnarray}
&&\psi=D\cos(K_\perp\xi+\delta)\,,\qquad
\tilde\delta=-K_\perp\xi-\arctan\frac{\mathrm{d}_\xi\psi(\xi)}{K_\perp\psi(\xi)}\,,
\nn\\
&&\delta=\tilde\delta-2\pi\times
\mathrm{int}\left(\frac{\tilde\delta+\pi}{2\pi}\right)\in (-\pi,\;\pi)\,,
\qquad D=\frac{\psi(\xi)}{\cos(K_\perp\xi+\delta)}\,
,\end{eqnarray}
at some sufficiently large $\xi\gg 1+2\pi/K_\perp$.  Here $\mathrm{int}(\dots)$
stands for the integer part of a real number. When programming we have to
use the two-argument $\arctan$ function
\be
\arctan(y,x)=\arctan(y/x) + \frac{\pi}{2}\,\theta(-x)\,\mathrm{sgn}(y)\in (-\pi,\;\pi).
\ee
The accuracy of this continuation is controlled by the Wronskian
from the asymptotic wave functions
\be
\label{ch2Wronskian}
W(\psi_\mathrm{g},\psi_\mathrm{u})(\xi)
=QD_\mathrm{g}D_\mathrm{u}\sin(\delta_\mathrm{g}-\delta_\mathrm{u})=1.
\ee
%

\subsection{MHD and Real $\psi$}%
%
Imagine that in an ideal plasma we have at $t \to -\infty$ some
plane MHD wave -- our task is to calculate how many times the energy
density increases at $t \to \infty$, and to average this
amplification over all the initial phases of that wave. As
there is no amplification for the $b_z$ component according to
\Eqref{ch2bz} we will concentrate our attention on the $b_x$ component.
The amplification comes from the negative ``friction'' term $\propto
\xi/(1+\xi^2)$. The influence of this friction is
transmitted to the effective potential barrier $\propto
1/(1+\xi^2)$.  For $K_\perp^2<1$ we have an analog of the quantum
mechanical tunneling.

In the current MHD problem $\psi$ is a real variable with
asymptotics
\be
\label{ch2wave-asymp}
\psi \approx \left\{ \begin{array}{l}
\quad \; \cos(K_\perp\xi-\phi_\mathrm{i})
 \qquad\mbox{for  } \xi\rightarrow -\infty,\\
D_\mathrm{f}\cos(K_\perp\xi+\phi_\mathrm{f})
 \qquad\mbox{for  } \xi \rightarrow +\infty,
\end{array}\right.
\ee
i.e., we have an incident wave with a unit amplitude and an initial phase
$\phi_\mathrm{i}$. $D_\mathrm{f}$ is the amplitude and
$\phi_\mathrm{f}$ is the phase of the amplified wave.

Again we present the $\psi$ function as linear combination of even
and odd solutions
\begin{equation}
\psi(\xi)=C_\mathrm g^\mathrm{(c)} \psi_\mathrm g(\xi) + C_\mathrm u^\mathrm{(c)}
\psi_\mathrm u(\xi).
\end{equation}
Here we substitute the asymptotic formulas \Eqref{ch2asg} and
\Eqref{ch2asu}, and the comparison of the coefficient with
\Eqref{ch2wave-asymp} at $\xi\rightarrow -\infty$ gives
\be
C_\mathrm{g}^{\mathrm{(c)}}D_\mathrm{g}=
\mathrm{\frac{s_{iu}}{s_{gu}}}
,\qquad
C_\mathrm{u}^{\mathrm{(c)}}D_\mathrm{u}=
\mathrm{\frac{s_{ig}}{s_{gu}}},
\ee
where
\be
\mathrm{s_{iu}}\equiv\sin(\phi_\mathrm{i}-\delta_\mathrm u),\qquad
\mathrm{s_{ig}}\equiv\sin(\phi_\mathrm{i}-\delta_\mathrm g).
\ee
The comparison of the coefficients at $\xi\rightarrow \infty$
gives for the phase and the amplification of the signal
\begin{eqnarray}
&&\phi_\mathrm{f}=F(\phi_\mathrm{i})\equiv
\arctan
\mathrm{\frac{s_{ig} s_{u} + s_{iu} s_{g}}{s_{ig} c_{u} + s_{iu} c_{g}}},\\
&&\mathcal{A}(\phi_\mathrm{i})\equiv D_\mathrm{f}^2=\frac{\mathcal{N}}{\mathcal{D}},
\end{eqnarray}
where
\begin{eqnarray}
&&\mathcal{N}= \mathrm{ (s_{ig} s_{u} + s_{iu} s_{g})^2
 + (s_{ig} c_{u} + s_{iu} c_{g})^2 },\\
&&\mathrm{s_{g}}=\sin(\delta_g),\qquad
\mathrm{s_{u}}=\sin(\delta_\mathrm u),\\
&&\mathrm{c_{g}}=\cos(\delta_g),\qquad
\mathrm{c_{u}}=\cos(\delta_\mathrm u).
\end{eqnarray}
The reversibility of the dissipation-free motion leads us to
$\phi_\mathrm{i}=F(\phi_\mathrm{f})$, i.e., function $F$
coincides with its inverse function $F(F(\phi))=\phi$.  As time
reverses the wave amplification is converted to attenuation
(damping in some sense)
$\mathcal{A}(\phi_\mathrm{i})\mathcal{A}(F(\phi_\mathrm{i}))=1$.

In unabridged mathematical notations we have the function
\be
F(\varphi)\equiv
\arctan \mathrm{
\frac{\sin{(\varphi - \alpha)} \sin{\beta} + \sin{(\varphi - \beta)}\sin{\alpha}}
{\sin{(\varphi - \alpha)} \cos{\beta} + \sin{(\varphi - \beta)} \cos{\alpha}}},
\ee
defined in the interval $\varphi \in (-\pi/2,\; \pi/2).$  For
arbitrary values of the parameters $\alpha$ and $\beta$
\be
F[F(\phi)]=\phi,
\ee
i.e., this function $F$ coincides with its inverse function $F^{-1}$.
The nonlinear function $F$ has only $2$ immovable points
\be
F(\alpha)=\alpha,\qquad F(\beta)=\beta.
\ee
Defining also
\begin{equation*}
\mathcal{A}(\varphi)\equiv\left\{\left[\sin{(\varphi-\alpha)}\sin{\beta}
              +\sin{(\varphi-\beta)}\sin{\alpha}\right]^2
+\left[\sin{(\varphi-\alpha)}\cos{\beta}
              +\sin{(\varphi-\beta)}\cos{\alpha}\right]^2\right\}
              /\sin^2(\alpha-\beta),
\end{equation*}

we have another curious relation
\be
\mathcal{A}[F(\varphi)]\,\mathcal{A}(\varphi)=1.
\ee

The so derived amplification coefficient
$\mathcal{A}(\phi_\mathrm{i};\delta_\mathrm g,\delta_\mathrm u)$
depends on the initial phase. In the next subsection we will
consider the statistical problem of phase averaging with respect to
the initial phase $\phi_\mathrm{i}$.

\subsection{Phase Averaged Amplification}%
%
For waves generated by turbulence the initial phase is unknown and
one can suppose a uniform phase distribution. That is why for
solving the statistical problem of energy amplification we need to
calculate average values with respect to the initial phase
$\phi_\mathrm{i}$. That idea is coming from the well-known random
phase approximation (RPA) in plasma physics. The phase averaging
already introduces an element of irreversibility because we already
suppose that waves are created with random phases. This is the MHD
analog of the molecular chaos from the theorem for entropy increase
in the framework of the kinetic theory if the probability distributions
are introduced in the initial conditions of the mechanical problem.
In the case of accretion flows we also suppose that turbulence
is a chaotic phenomenon and we have to apply the RPA for investigating
the statistical properties.

The calculation of the integral
\be
\int_{0}^{\pi}\frac{\mathrm{d}\phi_\mathrm{i}}{\pi}\mathcal{N}(\phi_\mathrm{i})
=2-\mathrm{s_\mathrm{ug}^2}=2-\mathcal{D}
\ee
gives for the initial phase averaged gain
\begin{equation}
\label{ch2ampl-trans}
\mathcal{G} \equiv \int_{-\pi/2}^{\pi/2}
\mathcal{A}(\phi_\mathrm{i}) \, \frac{\mathrm{d}
\phi_i}{\pi} = \frac{2}{\mathrm{
\sin^2(\delta_\mathrm{u}-\delta_\mathrm{g})
}}=\frac{2}{\mathcal D}-1.
\end{equation}
In such a way the SMW amplification coefficient $\mathcal G$ is
presented by the tunneling coefficient $\mathcal D$ of the
corresponding quantum problem. Both coefficients are expressed by
the asymptotic phases $\delta_\mathrm g,$ $\delta_\mathrm u$ in
analogy with partial waves phase analysis of the quantum mechanical
scattering problem in atomic and nuclear physics. The axial symmetry
of this result $\mathcal{G}\left(\sqrt{K_y^2+K_z^2}\right)$
significantly simplifies the further statistical analysis.

Let us analyze the physical meaning of the gain coefficient
$\mathcal{G}$.  As
\begin{eqnarray}
&&K_\perp\xi=K_y\tau=V_\mathrm{A}k_y\,t=\omega_\mathrm{A}\mathrm{sgn}(k_y)t,\qquad
\omega_\mathrm{A}(\mathbf{k})=\left|\mathbf{V}_\mathrm{A}\cdot\mathbf{k}\right|
=V_\mathrm{A}|k_y|\ge 0,\\
&&\mathbf{v}_\mathrm{gr}\equiv\frac{\partial\omega_\mathrm{A}}{\partial \mathbf{k}}
=\mathbf{V}_\mathrm{A}\mathrm{sgn}\left(\mathbf{V}_\mathrm{A}\cdot\mathbf{k}\right)
=V_\mathrm{A}\mathrm{sgn}(k_y)\mathbf{e}_y,
\end{eqnarray}
the asymptotics \Eqref{ch2wave-asymp} let us conclude that for
$|t|\rightarrow\infty$ we have only magnetosonic waves with
dispersion coinciding with that of Alfv\'en waves.\cite{Alfven:42}
In the spirit of M.~T.~Weiss quantum interpretation of the classical
Manley--Rowe theorem\cite{LL8,Zhelyazkov:00} one can present the
wave energy $\hbar \omega_\mathrm{A} N$ by a number of quanta, the
number of \textit{alfvenons}: ``The alfvenons introduced in this
Letter\cite{Stasiewicz:06} appear to be effective and spectacular
converters of electromagnetic energy flux into kinetic energy of
particles.'' We use this notion in a slightly different sense, our
former terminology was \textit{alfvons}\cite{Mishonov:07} in our
case. Following this interpretation, the energy gain $\mathcal{G}$
describes the increasing number of quanta
\be
\mathcal{G}\left(\sqrt{k_y^2+k_z^2}\right)
=\frac{\hbar\omega_\mathrm{A}(|k_y|)
\overline{N}_\mathrm{alfvenons}(t\rightarrow+\infty)}
{\hbar\omega_\mathrm{A}(\left|k_y\right|)
\overline{N}_\mathrm{alfvenons}(t\rightarrow-\infty)},
\ee
as in a laser system. In this terminology the mechanism of heating
of quasars can be phrased, namely due to lasing of alfvenons in
shear flows of magnetized plasma.  Laser or rather
maser\cite{Trakhtengerts:08} effects are typical phenomena in space
plasmas. The hydrodynamic overreflection
instability\cite{Fridman:08} and burst-like increase of the wave
amplitude\cite{Rogava:03} are phenomena of similar kind. More
precisely $\mathcal{G}$ is the gain for the $x$--$y$-polarized SMWs,
the energy of mode conversion in z-polarized AWs will be analyzed
elsewhere. The notion amplification is correct for standing waves
but
$2\cos(\mathbf{k}\cdot\mathrm{r})
=\mathrm{e}^{\mathrm{i}\mathbf{k}\cdot\mathrm{r}}
+\mathrm{e}^{-\mathrm{i}\mathbf{k}\cdot\mathrm{r}}$
and amplification is simultaneous for opposite wave-vectors. That is
why some theorists prefer to use overreflection in spite that there
is no rigid object reflecting the waves.

\subsection{Analytical Approximations for Amplification} %
%

For scattering problems by a localized potential at small
wave-vectors $K_\perp\ll1$, when the wavelength is much larger than the
typical size of the nonzero potential, we can apply the
delta-function approximation
\be
\label{ch2delta-Sch}
\frac{2m}{\hbar^2}U(\xi)\equiv \frac{1}{(1+\xi^2)^2}\rightarrow
2Q_0\delta(\xi).
\ee
In this well-known quantum mechanical problem\cite{Greiner:01} the
transmission coefficient is
\begin{eqnarray}
 \label{ch2D-E} 
&&D\approx\frac{K_\perp^2}{K_\perp^2+Q_0^2},\qquad
\delta_\mathrm{u}=-\frac{\pi}{2},\qquad
\delta_\mathrm{g}=-\frac{\pi}{2}-\arctan\frac{K_\perp}{Q_0}, \\
&& \psi_\mathrm{g} =\frac{\cos(|K_\perp\xi|+\delta_\mathrm{g})}{\cos(\delta_\mathrm{g})},
\qquad \psi_\mathrm{u}=\frac{\sin K_\perp\xi}{K_\perp}.
\end{eqnarray}

According to the tradition of the method of potential of zero
radius, the coefficient $Q_0\approx \frac{1}{2}\pi$ is determined by
the behavior of the phases at small wave-vectors. Only qualitatively
this parameter corresponds to the area of the potential
\be
2Q_0=\frac{\pi}{2}Z_\mathrm{ren}=\pi,\qquad
\frac{\pi}{2}=\int_{-\infty}^{\infty} \frac{1}{(1+\xi^2)^2}
\,\mathrm{d} \xi,
\ee
but the renormalizing coefficient $Z_\mathrm{ren} = 2$ which we have
to introduce differs from unity. According to the general relation
\Eqref{ch2ampl-trans} for the amplification we have
\be \label{ch2EnergyGain}
\mathcal{G}-1=2\left(\frac{1}{\mathcal{D}}-1\right)
\approx\frac{2Q_0^2}{K_\perp^2}, \qquad 2Q_0^2= \frac{\pi^2}{2}\approx
4.934\approx 5\,. \ee

The scattering phases $\delta_\mathrm g$ and $\delta_\mathrm u$ can
be obtained by a fit of the asymptotic wave functions \Eqref{ch2asg}
and \Eqref{ch2asu} with the analytical solutions. In such a way we can calculate the wave-vector
dependence of the transmission coefficient as it is depicted in.

Let us derive the analytical formula for $Q_0$: The solutions
\begin{eqnarray}
&\psi_g(\xi)& =\sqrt{1+\xi^2} , \\
&\psi_u(\xi)& = \sqrt{1+\xi^2} \arctan(\xi)
\end{eqnarray}
of effective Schr\"odinger's equation in long wavelength
limit $K_\perp\rightarrow0$
\be \label{ch2Sch0} \mathrm{d}_\xi^2\psi-\frac{1}{(1+\xi^2)^2}\psi=0
\ee
with asymptotics at $\xi\rightarrow\infty$
\begin{eqnarray}
\psi_g(\xi\gg 1) \approx \xi, \\
\psi_u(\xi\gg 1) \approx \frac{\pi}{2}\xi
\end{eqnarray}
have to  be compared with the approximative solutions \Eqref{ch2asg}
and \Eqref{ch2asu} for $\xi\gg1,$ when for $K_\perp\rightarrow0$ we have
$|\delta_g|\approx|\delta_u|\approx\frac{\pi}{2}$ and sinusoids are
almost linear
\begin{eqnarray}
\psi_g(1 \ll \xi \ll \frac{1}{K_\perp}) \approx D_g\sin(K_\perp\xi) \approx D_g K_\perp\xi, \\
\psi_u(1 \ll \xi \ll \frac{1}{K_\perp}) \approx D_u\sin(K_\perp\xi) \approx D_u
K_\perp\xi.
\end{eqnarray}
The comparison of the first derivatives in this region
\begin{eqnarray}
\mathrm{d}_\xi \psi_g(\xi\gg 1) \approx 1 \approx D_g K_\perp,\\
\mathrm{d}_\xi \psi_u(\xi\gg 1) \approx \frac{\pi}{2}\approx D_u K_\perp
\end{eqnarray}
gives
\be D_g\approx\frac{1}{K_\perp}, \qquad D_u\approx \frac{\pi}{2K_\perp}. \ee
Then formula for Wronskian \Eqref{ch2Wronskian} determines the phase
difference
\be \sin(\delta_\mathrm{g}-\delta_\mathrm{u})\approx\frac{2}{\pi}K_\perp
\ee
and transmission coefficient
\be
\mathcal{D}\approx\sin^2(\delta_\mathrm{g}-\delta_\mathrm{u})=\left(\frac{2K_\perp}{\pi}\right)^2.
\ee
The \Eqref{ch2ampl-trans} then gives for the phase averaged
amplification \be
G=\frac{2}{D}-1\approx\frac{\pi^2}{2Q^2}+\mathrm{const} \ee and the
comparison with $\delta$-potential approach \Eqref{ch2EnergyGain} gives
the analytical result for the strength of the $\delta$-potential
$Q_0=\pi/2$ given in \Eqref{ch2EnergyGain}.

Our problem formally coincides with the quantum problem\cite{Landau:32} of
transmission coefficient $\mathcal{D}\propto E$ at low energies
\be
U(x)=\frac{U_0}{[1+(\alpha x)^2]^2},\quad
U_0=\frac{\hbar^2\alpha^2}{2m},\quad
E= U_0 K_\perp^2, \quad
D\approx \left(\frac{2K_\perp}{\pi}\right)^2=\frac{4}{\pi^2}\frac{E}{U_0}\ll 1.
\ee

The zero-radius potential Pad\'e approximant \Eqref{ch2D-E} has an
acceptable accuracy for small wave-vectors. Having the quantum
mechanical transmission coefficient, we can calculate the energy
gain coefficient $\mathcal{G}-1$.
\be
\label{ch2gain}
\mathcal{G}-1\sim\frac{1}{K_\perp^2}
=\frac{A^2}{V^2_\mathrm{A}q^2},\qquad
q^2= k_\perp^2\equiv k_y^2+k_z^2=\mathrm{const}
\ee
Here the subscript $\perp$ means that the amplification depends on the projection of the
wave-vector perpendicular to the shear velocity and magnetic field.

The delta function approximation has a visual interpretation in
classical mechanics as well, supposing that $\psi$ is the
displacement of an oscillator and $\xi$ is the time. The
approximative equation
\be
\mathrm{d}_\xi^2\psi=-K_\perp^2\psi(\xi) + 2Q_0\delta(\xi)\psi(\xi)
\ee
means that at a time moment $\xi=0$ the oscillator is
subjected to a forcing impulse with a magnitude
\begin{equation*}
\label{ch2jump}
\mathrm{d}_\xi\psi(+0)- \mathrm{d}_\xi\psi(-0)=2Q_0\psi(0),
\quad\mbox{or}\quad
v_x(+0)- v_x(-0)=\frac{\pi\,b_x(0)}{\sqrt{K_y^2+K_z^2}},\quad
\mbox{when}\quad K_x(0)=0.
\end{equation*}
If for $K_\perp\ll Q_0$ we have initial oscillations with amplitude $A_\mathrm{i}$
\be \psi=A_\mathrm{i}\cos K_\perp\xi\qquad\mbox{for}\quad \xi<0, \ee
after the push in $\xi=0$ we have oscillations with much increased
amplitude
\be
\psi=A_\mathrm{f}\sin K_\perp\xi,\qquad\mbox{for}\quad \xi>0,\qquad
A_\mathrm{f}\approx \frac{2Q_0}{K_\perp} A_\mathrm{i}\gg A_\mathrm{i}.
\ee
The strong push with an appropriate phase ``amplifies'' the
oscillations; this burst-like increase of the wave amplitude was
observed in numerical investigations of linearized two-dimensional
MHD equations.\cite{Chagelishvili:93,Rogava:03} This phenomenon is
akin to the extremely strong hydrodynamic instabilities due to a
velocity jump; its prediction and discovery both in theory and
experiments are described in Ref.~[\onlinecite{Fridman:08a}]. In
such a way the Alfv\'en's idea of the importance of MHD waves in the
transfer of momentum is reduced to the very simple mathematics for
the jump of the velocity of SMWs \Eqref{ch2jump}. It is instructive to
rewrite this force in the $\mathbf{r}$-space.

\section{MRI and deviation equation as test example of the general equation}
\label{ch2sec:MRI}

\subsection{Secular equations and Lyapunov analysis}

The axisymmetrical case $K_y=0$ has measure zero but it is very instructive to
reveal Velikhov \cite{Velikhov:59} MRI 
which with some delay become one of the most popular notion in astrophysics. 
For $K_y=0$ we have $\mathbf{U}_\mathrm{shear}=0,$ $\mathbf{K}=\mathrm{const},$
and $\mathrm{D}_\tau^\mathrm{shear}=\mathrm{d}_\tau.$
In this case the linearized MHD equations
\Eqref{ch2dv_final} and \Eqref{ch2db_final} 
are a linear system with constant coefficients
having exponential time dependence $\propto\exp(\lambda\tau).$

Supposing that for small $|K_y|\ll1$ we have small shear velocity we have slowly
changing by time wave-vector 
\be
\mathbf{K}(\tau)=\mathbf{K}_0+(\tau-\tau_0)\mathbf{U}_\mathrm{shear}.
\ee
The substitution of this $\mathbf{K}(\tau)$ in the system above gives the WKB
approximation of slowly changing coefficients. 
For ideal fluid $\nu'_\mathrm{k}=0=\nu'_\mathrm{k}$ the corresponding secular
equation simplifies as
\ba
\nn
\label{ch2lambda3D}
&&\lambda^4
-2n_yn_x\lambda^3
+\left\{\left[(4-8n_y^2)n_x^2+4-4n_y^2\right]\omega_{_\mathrm{C}}^2 +
\left[(2-8n_y^2)n_x^2-4n_y^2+2\right]\omega_{_\mathrm{C}}+2K_\alpha^2\right\}
\lambda^2\\
&&-2K_\alpha^2n_yn_x\lambda
+ 2K_\alpha^2(n_x^2+1)\omega_{_\mathrm{C}}+K_\alpha^4=0.
\ea
Let $\lambda_\mathrm{3D}(\mathbf{K})$ is the solution with maximal real part. 
It embraces all cases of MRI analyzed in the famous woks by Balbus and Hawley
 \cite{Balbus:91,Hawley-2,Hawley:92,Balbus-4,Balbus-Oort,Balbus:98}.

For the 2-dimensional axi-symmetric case we have growth rate
\be \lambda_\mathrm{2D} (K_x,K_z)= \lambda_\mathrm{3D}(K_x,K_y=0,K_z), \ee
which obeys already exact the equation
\be
\lambda^4+
2\left[K_\alpha^2 + (1
+2\omega_{_\mathrm{C}})(n_x^2+1)\omega_{_\mathrm{C}}\right]\lambda^2
+2K_\alpha^2(n_x^2+1)\omega_{_\mathrm{C}}+K_\alpha^4=0 .
\ee
The most restricted is the 1-dimensional case 
$\lambda_\mathrm{1D}(K_z)=\lambda_\mathrm{2D} (K_x=0,K_z)$ 
with wave-vector parallel to the rotation axis
$\mathbf{K}=K\mathbf{e}_z$ when $K_\alpha=K_z \cos{\theta}$
when we have the most cited bi-quadratic equation in the astrophysics
\be
\label{ch2MRI}
\lambda^4 +2\left[K_\alpha^2 + (1
+2\omega_{_\mathrm{C}})\omega_{_\mathrm{C}}\right]\!\lambda^2
+ \left(K_\alpha^2+2\omega_{_\mathrm{C}}\right)\!K_\alpha^2=0,
\ee
which describes the Velikhov MRI growth rate.

In order to compare the results we have to use the traditional for the physics
of disks dimensional variables.
Having in orbital velocity of the fluid $V_\varphi(r)=r\Omega(r)$ the shear
describes the component of the viscous stress tensor
\begin{eqnarray}
\label{ch2stress}
\sigma_{r\varphi}=\eta\left(\frac1r\partial_\varphi V_r+\partial_r V_\varphi -
\frac{V_\varphi}r\right)=
\eta A.
\end{eqnarray}
In such a way we can define the shear rate and dimensional angular velocity by
the radial dependence of the angular velocity 
\be
A(r)=r\mathrm{d}_r\Omega\approx \mbox{const.},\quad
\omega_\mathrm{c}=\frac{\Omega}{A}=\frac{\mathrm{d}\ln r}{\mathrm{d}\ln\Omega}.
\ee
For Keplerian disks when $\Omega\propto r^{-3/2}$ we have
$\omega_\mathrm{c}=-\frac23,$ 
while at solar tachocline we have regions with $|\omega_\mathrm{c}|\ll1,$ whose
linearized MHD is well described 
by the analytical solution of $\omega_\mathrm{c}=0$ considered in the former
section \ref{ch2sec:Solution}.

Further we define the usual formulas for:
the Alfv\'en frequency 
\be
\label{ch2AW_dispersion}
\omega_\mathrm{A}=\left|\mathbf{V}_\mathrm{A}\cdot\mathbf{k}\right|=V_\mathrm{A}
k_z\cos\theta,
\ee
where $\mathbf{k}$ is the usual wave-vector, 
the epicyclic frequency \cite{Balbus:91,Balbus:98}
\be
\kappa^2 \equiv
\frac1{r^3}\frac{\mathrm{d}\left(r^4\Omega^2\right)}{\mathrm{d}r}
=\frac{2\Omega}{r}\frac{\mathrm{d}}{\mathrm{d}r}\left(r^2\Omega\right)
= 2\mathbf{\Omega}\cdot\mathrm{rot}\left[r\Omega(r)\mathbf{e}_\varphi\right]
= 2A\Omega + 4\Omega^2,
\ee
and the usual frequency $\tilde{\omega}=iA\lambda.$
In these variables the dispersion relation for MHD modes \Eqref{ch2MRI} takes the
well known form \cite{Balbus:91,Balbus:98}
\be
\label{ch2secular}
\tilde\omega^4 -\tilde\omega^2\left(\kappa^2 + 2\omega_\mathrm{A}^2 \right)
+ \omega_\mathrm{A}^2\left(\omega_\mathrm{A}^2 +
\frac{\mathrm{d}\Omega^2}{\mathrm{d}\ln r} \right)=0,
\quad
\frac{\mathrm{d}\Omega^2}{\mathrm{d}\ln r}=2A\Omega,
\ee
with the solution
\be
\label{ch2slow-fast}
\tilde\omega_{\pm}^2(k_z)=\frac12\left[\kappa^2
+2\omega_\mathrm{A}^2\pm\sqrt{\kappa^4+16\omega_\mathrm{A}^2\Omega^2}\right],
\ee
which describes the growth rate of MRI when $\tilde\omega_{-}^2<0.$ 
The rederivation of the growth rate of MRI is 
an important test for the validity of our new basic 
equations \Eqref{ch2VelocityEvolution} and \Eqref{ch2MagneticEvolution}.
For a short review we mention: the momentum dependence of 
$\lambda_\mathrm{2D}^2(K_x,K_z)$ was depicted in Fig.~1a of \cite{Balbus:91}, 
the dispersion relation was given in Eq.~(2.9) and Eq.~(2.17). 
The momentum dependence of $\lambda_\mathrm{2D}(Kx=\mathrm{const},K_z)$ was
given in the Fig.~8 in 
\cite{Hawley-2} together with numerical analysis performed in coordinate
space. 
In the same work authors came to the conclusion that compressibility 
has no significant effect on the instability. 
Graphical dependence of $\lambda_\mathrm{2D}(K_x=\mathrm{const},K_z)$ was
presented also in Fig.~1 by \cite{Hawley:92}; 
and on Fig.~7 was presented the section in the other direction
$\lambda_\mathrm{2D}(K_x,K_z=\mathrm{const}).$
The nonaxisymmetric dynamic equation Eq.~(2.25) corresponding to our
\Eqref{ch2lambda3D} was given in \cite{Balbus-4}. 
In this article Balbus and Hawley pointed out the important connection, at the
linear level, 
between the MRI and the giant amplification of SMW in the case of Coriolis
force. 
Their WKB formula at the bottom of page 616 of Ref. \cite{Balbus-4} in our
notations for the energy gain reads as
\be
\label{ch2MRI_SMW_Relation}
G\approx\exp\left( 2\int_{-\infty}^{\infty} \mathrm{d} \tau\,
\lambda_\mathrm{3D}(\mathbf{K}(\tau)) \right)
=\exp\left( 2\int_{-\infty}^{\infty} \mathrm{d}
Q_x\,\lambda_\mathrm{3D}(\mathbf{Q})/|Q_y|
\right)\stackrel{\mathrm{def}}{=}\mathrm{e}^{\mu(Q_y,Q_z)/|Q_y|},
\ee
the last equation is the definition for $\mu(Q_y,Q_z).$

Strictly speaking for non axisymmetric case $K_y\ne0$ MRI does not exists 
because we have a linear system with \textit{time dependent coefficients}. The
$\exp(\lambda\tau)$ 
time dependence is a property of differential equations with constant
coefficients. 
However the WKB approximation is good working for Keplerian values of the
angular velocity 
and ``amplification factors can be tens of orders of magnitude''-- page 620 of
\onlinecite{Balbus-4}.
In order to emphasize that SMW can grow by very large factors, Balbus and Hawley
put quotation marks at page 618 and use the notion ``instabilities,'' 
even though the modes ``are technically transient amplification.''
Really for very small $|Q_y|\ll1$ we have 
$\lambda_\mathrm{3D}(\mathbf{Q})\approx\lambda_\mathrm{2D}(Q_x,Q_z).$
In this approximation for the energy gain we obtain
\be
G(Q_y,Q_z)\approx\exp\left( 2\int_{-\infty}^{\infty} \mathrm{d}
Q_x\,\lambda_\mathrm{2D}(Q_x,Q_z)/|Q_y| \right)
=\mathrm{e}^{\mu(0,Q_z)/|Q_y|}.
\ee
The dependence $\exp(2\mu/|Q_y|)$ presents how the amplification coefficient of
SMW can become giant in the linear regime, for $Q_y\rightarrow0.$ One of the
main results of our work is the above relation between the  
amplification of magnetosonic waves and growth rate of the MRI. This relation
is actually WKB approximation and in this sense we can speak about the WKB SMW
amplification. Divergence of the amplification coefficient $G$  at
$Q_y\rightarrow 0$ means that giant amplification for  $|Q_y|\ll1$ is the
precursor of MRI at  $Q_y=0.$
The path of the wave-vector trough the unstable regions is traced in Fig.~1 of
\cite{Balbus-4} and Fig.~17 of \cite{Balbus:98},  where the most of the result
from the original ApJ papers are reviewed. 
If we trace the wave-vector trajectory $\mathbf{K}(\tau)$ substantial growth is
occur for a finite time spent in the unstable region. 
The physics is analogous to the wave amplification by a lasing medium. 
Picturesquely speaking the enhanced transport in accretion disks is created by
the lasing of alfv\'enons. The giant amplification closer to $Q_y=0$ plane is
the mechanism which trigger the self consistent turbulence and anomalous
transport in accretion disks. In order to emphasize the importance of the
exponential growth Balbus and Hawley called it ``instability'' because it is
believed that in many physical situations the nonlinear effects become
important very fast and this smears the difference between the amplification
and instability. That is why one can also say that MRI is the ``heart'' of
turbulence \cite{Balbus:98}. 

However divergence of the amplification is not a property of the rotation. 
Without rotation the infinite amplification of SMW for $K_\perp=0$ was
discovered \cite{Chagelishvili:93} even before MRI to be exhumed for the
astrophysics by \cite{Balbus:91}. This amplification is adequately described
by a $\delta$-function like sudden change of the rigidity
\Eqref{ch2delta_approximation} of an effective oscillator is called ``swing''
amplification by \cite{Fan:97}.

We demonstrated that general equations given in the present work describe both
the ``swing'' SMW amplification and WKB SMW amplification called sometimes
MRI. As further test of the new approach we will consider in the next
sub-section how the \cite{Balbus:98} mechanical model for MRI can be re-derived
as a consequence of our general MHD equations.

\subsection{Deviation equation}

In order to illustrate derivation of the deviation equation we will consider
only the simplest possible case of axial wave-vector in ideal fluid.
The linearized MHD equations 
\Eqref{ch2dv_final} and \Eqref{ch2db_final} 
in this case take the form
\begin{eqnarray}
\label{ch2dot_v}
\mathrm{d}_\tau\mathbf{v}&=&-v_x\mathbf{e}_y + (\boldsymbol{\alpha}
\cdot\mathbf{K})\mathbf{b}+2\omega_\mathrm{c}(v_y\mathbf{e}_x-v_x\mathbf{e}_y),
\\
\mathrm{d}_\tau\mathbf{b}&=&b_x\mathbf{e}_y
-(\boldsymbol{\alpha}\cdot\mathbf{K})\mathbf{v}.
\label{ch2dot_b}
\end{eqnarray}
In order to eliminate the magnetic field we express $\mathbf{b}$ from
\Eqref{ch2dot_v} and substitute in \Eqref{ch2dot_b}.
In the obtained vector equation we present the velocity by the displacement
$\mathbf{v}=\mathrm{d}_t\boldsymbol{\xi}=\dot{\boldsymbol{\xi}}.$
One additional time integration gives
\begin{eqnarray}
\label{ch2deviation}
\ddot{\xi}_r - 2\Omega\dot{\xi}_\varphi &=&-\omega_\mathrm A^2\xi_r,\\
\ddot{\xi}_\varphi + 2\Omega\dot{\xi}_r &=&-\left(\omega_\mathrm
A^2+2A\Omega\right)\xi_\varphi.
\nn
\end{eqnarray}
The mechanical model \cite{Balbus:98,Balbus:03} for this system is an
anisotropic oscillator in rotating coordinate system.

The variables $\xi_r(t)$ and $\xi_\varphi(t)$ are exactly the amplitudes of the
wave induced
deviation $\xi^\alpha$ of the fluid particle from its Keplerian circular orbit.
One can generalize this equation for the general case of time dependent
wave-vector. 
Such type deviation equations are well-known for the motion of particles in
gravitation fields
\be
\mathrm{D}_\tau^2\xi^\alpha =- R^{\alpha}_{\beta\gamma\delta}\mathrm{d}_\tau
x^\beta\, \xi^\gamma\, \mathrm{d}_\tau x^\delta,
\ee
where $\mathrm{d}_\tau x^\beta$ is the velocity and
$R^{\alpha}_{\beta\gamma\delta}$ is the Riemann tensor;
see for example, Eq.~(1.8') from the monograph \cite{Misner:73}. The secular
equation related to
deviation equation are basic tools for investigation of the mechanical stability
introduced by Lyapunov.
In our case the secular equation related to \Eqref{ch2deviation} is reduced to the
MRI equation \Eqref{ch2slow-fast}.
In such a way we checked the important test that our general MHD equations give
as a particular case MRI
predicted by \cite{Velikhov:59} and the mechanical model advocated in
astrophysics by Balbus and Hawley. Lyapunov analysis of general MHD equations
is the simplest way to rederive the MRI. 

Here we wish to analyze the differences in SMW amplification.
In Keplerian case $\omega_\mathrm{c}=-\frac23$.
The MRI arises by the change of the sign of the spring rigidity
$\left(\omega_\mathrm A^2+2A\Omega\right)$
from the second equation \Eqref{ch2deviation}, 
while for the case of pure shear $\omega_\mathrm{c}=0$ we have analogous
interval with negative ``rigidity''
$\left\{\tilde{K}^2-1/[(1+\xi^2)^2]\right\}$ in 
\Eqref{ch2Schrodinger}.

\begin{figure}
\includegraphics[height=12cm]{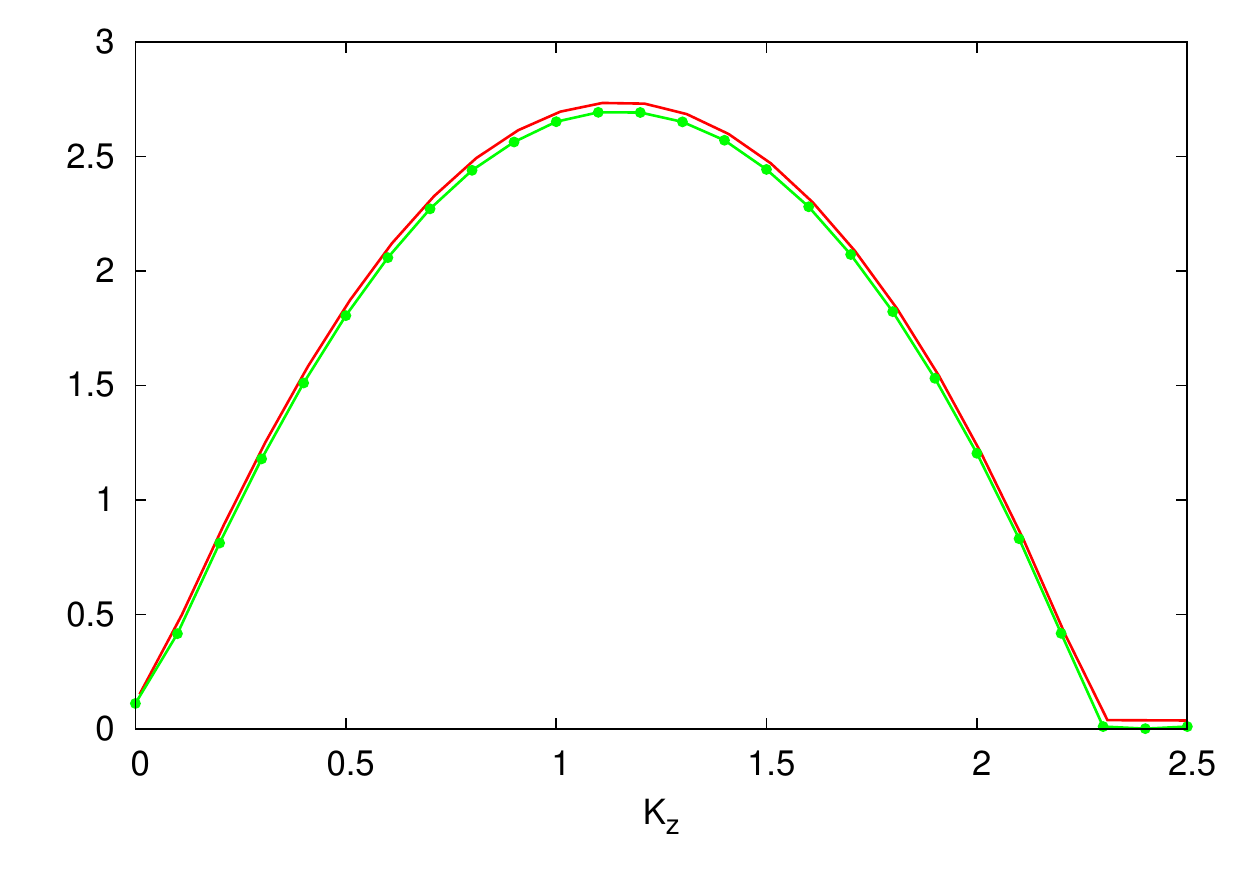}
\caption{The connection between MRI growth rate and giant amplification of SMW
revealed by WKB relation \Eqref{ch2MRI_SMW_Relation} for  $K_y=0.01$.
Doted line is the logarithm of the wave gain coefficient 
$\ln [G=w(\infty)/w(-\infty)]$ multiplied by
$K_y$ component of wave vector.
Thick line represent  \textit{the WKB amplification factor during the growth
phase} \cite{Balbus-4}.
The WKB formula for $\mu(Q_y,Q_z)$~\Eqref{ch2MRI_SMW_Relation} is asymptotically
exact for
$Q_y\rightarrow0.$  
}
\label{ch2fig:amplification}
\end{figure}

Qualitatively we suppose that strong wave amplification can lead to wave
turbulence when we have
self-consistent spectral density of magnetosonic waves. 
We have something like phase transition in an infinite system. 
Wave turbulence is analogous to the fluctuations of the order parameter.
At least on intuitive level this analogy already presents in the theory of the
disks;
in the review by \cite{Balbus:98} the word ``fluctuations'' can be found 49
times.
The excess viscosity is actually the order parameter of the wave turbulent
phase.
In the laminar phase for small Reynolds numbers 
\be
\mathrm{R}\equiv\frac{1}{\nu'_\mathrm{k}}
=\frac{\Lambda V_\mathrm{A}}{\nu_\mathrm{k}}
=\frac{V_\mathrm{A}^2}{A\nu_\mathrm{k}}
<\mathrm{R}_\mathrm{c},
\ee
the excess viscosity is just zero.
The region of Reynolds numbers slightly above the critical one
$\mathrm{R}_\mathrm{c}$ can be investigated by standard numerical methods 
and already self-sustained turbulence is at least qualitatively confirmed. 
Now let us speculate the derived results and perspectives.

\section{Approximate solution}

\begin{eqnarray*}
\ddot{b_x} + \frac{2\tau Q_y^2(1+\omega_c)}{Q_y^2(1+\tau^2)+Q_z^2}\dot{b_x} +\left[Q_\alpha^2 +2\omega_c\left(1-\frac{\tau^2Q_y^2}{Q_y^2(1+\tau^2)+Q_z^2}\right) \right] b_x = 2\omega_c\left(1-\frac{\tau^2Q_y^2}{Q_y^2(1+\tau^2)+Q_z^2}\right) \dot{b_y} \nn \\
\ddot{b_y} - \frac{2\tau Q_y^2\omega_c}{Q_y^2(1+\tau^2)+Q_z^2}\dot{b_x} + Q_\alpha^2 b_y = - \frac{2\tau Q_y^2\omega_c}{Q_y^2(1+\tau^2)+Q_z^2}b_x + 2\left[\frac{(\omega_c+1)Q_y^2}{Q_y^2(1+\tau^2)+Q_z^2} -\omega_c \right] \dot{b_x}
\end{eqnarray*}

\begin{eqnarray}
&& b_x(\tau)= \nn \\
&& C_1 \mathrm{HeunC}(0,-\frac12,\omega_c,-\frac14 \frac{(Q_y^2+Q_z^2)Q_\alpha}{Q_y^2},\frac14 \frac{(Q_\alpha^2+\omega_c+2)Q_y^2+Q_\alpha^2 Q_z }{Q_y^2};\frac{-\tau Q_y^2}{Q_y^2+Q_z^2}) \nn \\
&& +C_2\tau \mathrm{HeunC}(0,-\frac12,\omega_c,-\frac14 \frac{(Q_y^2+Q_z^2)Q_\alpha}{Q_y^2},\frac14 \frac{(Q_\alpha^2+\omega_c+2)Q_y^2+Q_\alpha^2 Q_z }{Q_y^2};\frac{-\tau Q_y^2}{Q_y^2+Q_z^2})\nn \\
&&b_y(\tau)= \nn \\
&& C_3 \mathrm{HeunC}(0,-\frac12,1+\omega_c,-\frac14 \frac{(Q_y^2+Q_z^2)Q_\alpha}{Q_y^2},\frac14 \frac{(Q_\alpha^2+\omega_c+2)Q_y^2+Q_\alpha^2 Q_z }{Q_y^2};\frac{-\tau Q_y^2}{Q_y^2+Q_z^2})\cdot Q^{2(1+\omega_c)} \nn \\
&&+C_4 \tau \mathrm{HeunC}(0,\frac12,1+\omega_c,-\frac14 \frac{(Q_y^2+Q_z^2)Q_\alpha}{Q_y^2},\frac14 \frac{(Q_\alpha^2+\omega_c+2)Q_y^2+Q_\alpha^2 Q_z }{Q_y^2};\frac{-\tau Q_y^2}{Q_y^2+Q_z^2})\cdot Q^{2(1+\omega_c)} \nn \\
&&+ 2\omega_c(Q_y^2+Q_z^2)[Q_y^2(1+\tau^2)+Q_z^2]^{\omega_c}\cdot\mathsf{J} \nn \\
\label{HeunW}
\end{eqnarray}

\begin{eqnarray*}
\label{deviation}
\ddot{\xi}_r - 2\Omega\dot{\xi}_\varphi &=&-\omega_\mathrm A^2\xi_r,\\
\ddot{\xi}_\varphi + 2\Omega\dot{\xi}_r &=&-\left(\omega_\mathrm
A^2+2A\Omega\right)\xi_\varphi.
\nn
\end{eqnarray*}

\begin{eqnarray*}
 b_x(\tau)&=&C_1e^{w^+\tau} + C_2e^{w^-\tau} + C_3e^{-w^+\tau} + C_4e^{-w^-\tau} \\
 b_y(\tau)&=&\frac{-1}{2\omega_cQ_\alpha^2}\left[\frac{-w^-}2(C_2e^{w^-\tau} - C_1e^{-w^-\tau})(4\omega_c^2 + (w^-)^2 + Q_\alpha^2 + 2\omega_c )\right. \\
           &&+ \left.\frac{-w^+}2(C_3e^{w^+\tau} - C_4e^{-w^+\tau})(4\omega_c^2 + (w^+)^2 + Q_\alpha^2 + 2\omega_c )\right]
\end{eqnarray*}

\begin{eqnarray*}
w^{\pm} \equiv \sqrt{-Q_\alpha^2-2\omega_c^2 -\omega_c \pm \tilde{w} },\quad \tilde{w} \equiv |\omega_c|\sqrt{4Q_\alpha^2 + 4\omega_c(\omega_c+1)+1} \\
\widehat{w}^{\pm} \equiv \mathrm{i}\sqrt{Q_\alpha^2+2\omega_c^2 +\omega_c \mp \tilde{w} },\quad \overline{w}^{\pm}\equiv
\widehat{w}^{\pm}\left( Q_\alpha^2 + 4\omega_c^2 +2\omega_c + (w^{\pm})^2 \right)
\end{eqnarray*}

\begin{eqnarray*}
&&\psi(\tau\rightarrow -\infty)\approx \openone (\overline{w}^-\exp(\ii \widehat{w}^-\tau ) + \overline{w}^+\exp(\ii \widehat{w}^+\tau )) \\
  &&+ {R}(\overline{w}^-\exp(-\ii \widehat{w}^-\tau ) + \overline{w}^+\exp(-\ii \widehat{w}^+\tau ),\\
&&\psi(\tau\rightarrow +\infty)\approx  T (\overline{w}^-\exp(\ii \widehat{w}^-\tau ) + \overline{w}^+\exp(\ii \widehat{w}^+\tau )).
\end{eqnarray*}
Using the asymptotics of the eigen-functions
\begin{equation*}
\psi_\mathrm g\approx \left\{ \begin{array}{l}
D_\mathrm g [ \overline{w}^+\cos(\widehat{w}^+\tau-\delta_\mathrm{g}) + \overline{w}^-\cos(\widehat{w}^-\tau-\delta_\mathrm{g})]
 \quad\mbox{for  } \tau\rightarrow -\infty,\\
D_\mathrm g [ \overline{w}^+\cos(\widehat{w}^+\tau+\delta_\mathrm{g}) + \overline{w}^-\cos(\widehat{w}^-\tau+\delta_\mathrm{g})]
 \quad\mbox{for  } \tau\rightarrow +\infty,
\end{array}\right.
\end{equation*}
and
\begin{equation*}
\label{ch2asu}
\psi_\mathrm u\approx \left\{ \begin{array}{l}
 -D_\mathrm u [ \overline{w}^+\cos(\widehat{w}^+\tau-\delta_\mathrm{u}) + \overline{w}^-\cos(\widehat{w}^-\tau-\delta_\mathrm{u})]
 \quad\;\;\;\mbox{for  } \tau \rightarrow -\infty,\\
\;\;\, D_\mathrm u [\overline{w}^+\cos(\widehat{w}^+\tau+\delta_\mathrm{u}) + \overline{w}^-\cos(\widehat{w}^-\tau+\delta_\mathrm{u})]
  \qquad\mbox{for  } \tau \rightarrow +\infty
\end{array}\right.
\end{equation*}
as well as the general condition
\begin{equation*}
\psi(\xi)=C_\mathrm g^\mathrm{(q)} \psi_\mathrm g(\xi)
+ C_\mathrm u^\mathrm{(q)} \psi_\mathrm u(\xi),
\end{equation*}

For the tunneling coefficient we obtain
\begin{equation*}
\mathcal{D}=|T|^2=\mathrm{s_{ug}^2},\qquad
\mathrm{s_{ug}}\equiv\sin(\delta_\mathrm u-\delta_\mathrm g).
\end{equation*}

\begin{equation*}
\label{ch2wave-asymp}
\psi \approx \left\{ \begin{array}{l}
\quad \; \overline{w}^+\cos(\widehat{w}^+\tau-\phi_\mathrm{i})  +\overline{w}^-\cos(\widehat{w}^-\tau-\phi_\mathrm{i})
 \quad\mbox{for  } \tau \rightarrow -\infty,\\
D_\mathrm{f} [\overline{w}^+\cos(\widehat{w}^+\tau+\phi_\mathrm{f})  +\overline{w}^-\cos(\widehat{w}^-\tau+\phi_\mathrm{f})]
 \quad\mbox{for  } \tau \rightarrow +\infty,
\end{array}\right.
\end{equation*}

\begin{eqnarray*}
\phi_\mathrm{f}=F(\phi_\mathrm{i})\equiv
\arctan
\mathrm{\frac{s_{ig} s_{u} + s_{iu} s_{g}}{s_{ig} c_{u} + s_{iu} c_{g}}},\\
\end{eqnarray*}

\begin{eqnarray*}
&&\mathrm{s_{g}}=\sin(\delta_\mathrm{g}),\quad \mathrm{c_{g}}=\cos(\delta_\mathrm{g}),\quad \mathrm{s_{ig}}\equiv\sin(\phi_\mathrm{i}-\delta_\mathrm g),\\
&&\mathrm{s_{u}}=\sin(\delta_\mathrm u),\quad \mathrm{c_{u}}=\cos(\delta_\mathrm u),\quad \mathrm{s_{iu}}\equiv\sin(\phi_\mathrm{i}-\delta_\mathrm u)
\end{eqnarray*}

\section{Discussion and perspectives}
\label{ch2sec:conclusions}
%

We consider that relation between wave amplification and MRI growth rate 
in WKB approximation is one important property of the linearized MHD
equations. This relation demonstrates that exponential big amplification is a
precursor of the MRI and this is a subject of an indisputable mathematics.
Let us now clarify what is common and what is different between swing SMW
amplification and MRI.

Both phenomena are based on the negative square of the frequency of
an effective oscillator which means negative rigidity of the effective spring
if we look from eigenvector coordinates. For SMW negative rigidity arises for
negative  values $\tilde{K}^2-1/(1+\xi^2)^2$ in \Eqref{ch2Schrodinger}. While for
MRI we have negative squares of the frequency in the solution \Eqref{ch2slow-fast}
of the bi-quadratic equation \Eqref{ch2MRI_SMW_Relation}. 

The differences are related to the different physical sense of the effective
time. For the SMW the immanent time 
$\xi=-K_x/\sqrt{K_y^2+K_z^2}=\tau K_y/\sqrt{K_y^2+K_z^2}$ has
kinematic sense in the wave-vector space, while for MRI we have simply $\tau=A
t$. As $K_x(t)= -K_y At$ the SMW time $\xi$ has sense only for $K_y \neq 0$
whilst the effective time of MRI strictly speaking $\tau= At$ is applicable
only for $K_y=0.$ In this sense SMW amplification and MRI are complimentary. 

The different physical nature of the effective times leads also to different
mathematical approximations for long wavelength approximation. For SMW without
rotation we have $\delta$-function approximation \Eqref{ch2delta_approximation} and
some people call this over-reflection \cite{Gogoberidze:04} as \textit{swing
amplification} \cite{Fan:97} caused by a \textit{sudden change} of the rigidity
of an effective oscillator. On the other hand for MRI woks WKB approximation
\cite{Landau:81} and \cite{Migdal:00} for the wave amplification
\Eqref{ch2MRI_SMW_Relation}.

The exponential growth function
\be
\label{ch2growth}
\mu(Q_y,Q_z)=2 \int \lambda_\mathrm{3D}(Q_x,Q_y,Q_z) \mathrm{d} Q_x
\ee
have finite limit for $Q_y=0,$ i.e. $\mu(Q_z)=\mu(0,Q_z)$ describes very well
the exponentially big wave amplifications $G\approx\exp(\mu(Q_z)/|Q_y|)$ for
small $|Q_y|\ll1,$ as it is depicted at Fig.~4. This exponential growth with
many decades of amplitude was the motivation for the change of the terminology:
the exponentially big wave amplification is called simply MRI because the
nonlinear effects can become important before the applicability of the
linearized analysis. From formal point of view however the divergence of
$G(Q_y\rightarrow 0)$ means that big SMW amplification is the precursor of MRI.
We wish to emphasize that for infinite times we have asymptotic wave behavior
and this is the reason to speak about the wave amplification. For big enough
$|Q_x|\gg1$ the waves are in good approximation independent and
in this sense we have a Kraichnan like MHD-wave turbulence. The wave
independence at big wave vectors is analogous to the asymptotic freedom in the
elementary particle physics. 

Once again we wish to emphasize the distinction between the ``SMW swing 
amplification'' and MRI. The SMW swing amplification discovered in
Ref.~\cite{Chagelishvili:93} occurs even without rotation for $\omega=0.$
This amplification is divergent for $Q_\perp=\sqrt{Q_y^2+Q_z^2}=0.$ This is
only a point in $(Q_y,\,Q_z)$ plane, let us call this point $\Gamma$. 

MRI however arises only for \textit{anticyclonic alignment} of the angular speed
$\Omega$ and rotation of shear velocity 
$\omega= \boldsymbol{\Omega}\cdot\mathrm{rot}\mathbf{V}_\mathrm{shear}/A^2<0.$
The dimensionless angular velocity $\omega$ is analogous to the reduced
temperature $\epsilon=(T-T_\mathrm{c})/T_\mathrm{c}$ from the physics of the
phase transitions and critical phenomena \cite{Pokrovskii:79}. For evanescent
negative $\omega$ MRI arises close to the $\Gamma$-point, i.e. infinite
amplification related to MRI arises for $\omega=-0$ at the point where swing SMW
amplification is divergent. In this sense swing SMW amplification is the
precursor of the MRI. Analogously divergent susceptibility at $\epsilon=+0$ is
the precursor of the appearance of the order parameter at $\epsilon=-0.$ 
Below the critical temperature in the physics of second order phase transitions
we have order parameter while for our MHD system we have MRI. For Keplerian
value of the angular velocity $\omega_\mathrm{Kepler}=-\frac23$ the initial
$\Gamma$-point of divergent amplification is extended to a finite interval in
$Q_z$ axis with $Q_y=0,$ see Fig.~4. That is why we can say that
divergent-swing-SMW amplification for $\omega=0$ is the precursor of the
divergent MRI amplification for negative anticyclonic $\omega<0.$ 

As we already cited \cite{Balbus:98} there is already an emerging consensus amidst 
the astrophysicist that MRI is the heart of the disk turbulence. Here we wish to repeat 
the terminology. Strictly speaking instability means exponential amplification which occurs 
only in a manifold with measure zero $Q_y=0.$ As poetic metaphor we can use the notion 
instability (I) if some mode have a time interval with almost exponential growth which is
described in an acceptable accuracy with WKB approximation; in our case \Eqref{ch2growth}
This instability leads to giant amplification of MHD waves which also can be call heart of the turbulence.
However our analytical solution has revealed that divergent SMW amplification 
occurs even in the case of pure shear without rotation $\omega_\mathrm{c}$. Generalizing 
one can say that heart continues even without rotation and the reason of the self-sustained turbulence 
is the divergent wave amplification at small wave-vectors. Due to drift velocity on the momentum space Eq.~(10)
and the small bare viscosity the result of wave amplification and MRI propagates along $Q_x\mathbf{e}_x$ 
direction at very big wave-numbers.

\chapter{Wave Turbulence}

\section{Incorporation of Turbulence as Random Driver of MHD Waves}%
%
Being as efficient as has been demonstrated above, the amplification of SMWs,
\Eqref{ch2gain},
is possibly the dominant physical factor responsible for generating and
maintaining the turbulence in accretion flows.  However the theory of turbulence
is much more
complicated than the theory of linearized waves. That is why here we
provide an illustration how the wave amplification can be
incorporated in the turbulence theory. In order to establish common
set of notions and notations we will recall some basic properties of
the homogeneous isotropic Kolmogorov--Obukhov turbulence.

\subsection{Kolmogorov Turbulence}%
%
Let the velocity be presented by the Fourier integral
\be
\mathbf{V}(\mathbf{r})=\int \mathrm{e}^{\ii\mathbf{k}\cdot\mathbf{r}}
\mathbf{V}_\mathbf{k} \frac{\mathrm{d}^3k}{(2\pi)^3},\qquad
\mathbf{V}_\mathbf{k}
=\int\mathbf{V}(\mathbf{r}) \mathrm{e}^{-\ii\mathbf{k}\cdot\mathbf{r}}
\mathrm{d}^3x.
\ee
The energy per unit mass is
\be
\int \frac{1}{2} V^2(\mathbf{r})\mathrm{d}^3x
=\int \frac{1}{2} V^2_\mathbf{k} \frac{\mathrm{d}^3k}{(2\pi)^3}
=\int \mathcal{E}_\mathbf{k}\frac{\mathrm{d}^3k}{(2\pi)^3},
\ee
where we introduce the spectral density averaged with respect to the
turbulence
\be \mathcal{E}_\mathbf{k}
=\frac{1}{2}\langle V^2_\mathbf{k}\rangle_\mathrm{turbulence}.
\ee
We can introduce also the energy of vortices which contains
Fourier components with wavelength $2\pi/k,$ shorter than some fixed
length $\lambda$
\be
\frac{1}{2}V_\lambda^2
=\int_{k<1/\lambda} \mathcal{E}_k \frac{\mathrm{d}^3k}{(2\pi)^3},
\ee
where for isotropic turbulence $\mathcal{E}_\mathrm{k}
=\mathcal{E}_k$. This energy evaluates turbulent pulsation with size
$\lambda$. Further on we will continue with only order of magnitude
evaluations, hence in the following estimations we will drop
off factors such as $4\pi,$ $\frac{1}{2}$, etc.

According to the Kolmogorov--Obukhov (KO) scenario in the inertial
range the magnitude of the velocity pulsations $V_\lambda$ can
depend only on the turbulent power dissipated per unit mass
$\varepsilon$. There is only one combination with the appropriate
dimension
\be
\label{ch2Kolmogorov}
\varepsilon=
\frac{V_\lambda^2}{\lambda/V_\lambda}
=\mathrm{\frac{energy/mass}{time=length/velocity}=\frac{power}{mass}},
\ee
which yields
\begin{eqnarray}
&&V_\lambda^2\sim\left(\varepsilon\lambda\right)^{2/3}
\sim\int_{k\lambda>1}\mathcal{E}_\mathbf{k}^\mathrm{KO}\mathrm{d}^3k
\sim\int_{k\lambda>1}\frac{\varepsilon^{2/3}}{k^{5/3}}\mathrm{d}k,\qquad
\mathrm{d}^3k\sim k^2\mathrm{d}k,\nn \\
&&E(k)=\int k^2 \mathcal{E}_\mathbf{k}^\mathrm{KO} \mathrm{d} \Omega \!
\approx C_\mathrm{K}\varepsilon^{2/3}k^{-5/3},\qquad C_\mathrm{K}\approx 1.6,
\qquad
\mathcal{E}_\mathbf{k}^\mathrm{KO}\sim\frac{\varepsilon^{2/3}}{k^{11/3}}\,.
\end{eqnarray}
$V_\lambda$ is the amplitude of variation in the velocity pulsation
at distance $\lambda$.  $\mathcal{E}_\mathbf{k}$ is the energy
density in the \textbf{k}-space per unit mass; in the
Kolmogorov--Obukhov picture this is a static variable.

The scaling law $ V_\lambda\sim(\varepsilon \lambda)^{1/3}$
\Eqref{ch2Kolmogorov} is applicable for large enough distances
$\lambda>\lambda_0$, where $\lambda_0$ describes the scale where
dissipation effects become essential.

Let us now consider a magnetosonic wave with a time-dependent
wave-vector
\be
\label{ch2q(t)}
q_x(t)=-A(t-t_0)q_y,\qquad q_y=\mathrm{const}, \qquad q_y=\mathrm{const},
\ee
and time-dependent energy density per unit mass in real space
\be
\label{ch2w}
w(t)=\frac{1}{2}
\langle \mathbf{V}_\mathrm{wave}^2+\frac{\mathbf{B}_\mathrm{wave}^2}{\rho\mu_0} \rangle
=\frac{V_\mathrm{A}^2}{4}
\left[\mathbf{b}^2+\left(\mathrm d_{Q\xi} \mathbf{b}\right)^2\right],
\ee
where $\langle\dots\rangle$ stands for spatial averaging,
$\langle\cos^2(\mathbf{k}\cdot\mathrm{r}) \rangle=\frac{1}{2}$. Then
the energy density in the \textbf{k}-space is
\be
\label{ch2density_of_rain}
\mathcal{E}_\mathbf{k}(t)=w(t)\delta[\mathbf{k}-\mathbf{q}(t)].
\ee
Let us mention that all MHD variables $\mathbf{b}$, $\mathbf{v}$,
$w$ in \Eqref{ch2w}, and $P$ depend on the
effective wave functions $\psi$ and $\chi$ (solutions to the
effective Schr\"{o}dinger equations) through the dimensionless time
$\xi$.
Having in the beginning $t=t_0$ a distribution of the magnetic field
$\mathbf{B}_\mathrm{wave}(\mathrm{r},t_0)$ and velocity with
$\nabla\cdot\mathbf{V}_\mathrm{wave}(\mathrm{r},t_0)=0$, we can
calculate the Fourier components
\begin{eqnarray}
&&\mathbf{v}_\mathbf{k}(t_0)=\ii\int
\frac{\mathbf{V}_\mathrm{wave}(\mathrm{r},t_0)}{V_\mathrm{A}}
\mathrm{e}^{-\ii\mathbf{k}\cdot\mathbf{r}}\mathrm{d}x^3,\\
&&\mathbf{b}_\mathbf{k}(t_0)=\int
\frac{\mathbf{B}_\mathrm{wave}(\mathrm{r},t_0)}{B_0}
\mathrm{e}^{-\ii\mathbf{k}\cdot\mathbf{r}}\mathrm{d}x^3,\\
&&\xi_{0,\,\mathbf{k}}\equiv
-\frac{k_x}{\sqrt{k_y^2+k_z^2}},
\end{eqnarray}
and initial dimensionless time $\xi_{0,\,\mathbf{k}}$. If $k_z=0$
then $\mathrm{sgn}(k_y)\xi_{0,\,\mathbf{k}}=\tau_{0,\,\mathbf{k}}=-k_x/k_y$.  Then we
have to determine the coefficients $C$ in the general solutions for
$\psi$ and $\chi$ using the initial
values at $t_0$
\begin{eqnarray}
&&\mathbf{b}_\mathbf k(\xi)=C_\mathrm{g}\mathbf{b}_\mathrm{g}
          +C_\mathrm{u}\mathbf{b}_\mathrm{u}
          +\tilde{C}_\mathrm{g}\mathbf{b}_\mathrm{\tilde g}
          +\tilde{C}_\mathrm{u}\mathbf{b}_\mathrm{\tilde u},\\
&&\mathbf{v}_\mathbf k(\xi)=C_\mathrm{g}\mathbf{v}_\mathrm{g}
          +C_\mathrm{v}\mathbf{b}_\mathrm{u}
          +\tilde{C}_\mathrm{g}\mathbf{v}_\mathrm{\tilde g}
          +\tilde{C}_\mathrm{u}\mathbf{v}_\mathrm{\tilde u},\\
&&\mathbf{k}\cdot\mathbf{b}_\mathbf{k}
=0=\mathbf{k}\cdot\mathbf{v}_\mathbf{k}.
\end{eqnarray}
In this set we can use only $x$- and $z$-components, and so we
obtain $4$ equation for the constants $C_\mathrm{g}$,
$C_\mathrm{u}$, $\tilde{C}_\mathrm{g}$, and $\tilde{C}_\mathrm{u}$.
The functions $\mathbf{b}_\mathrm{g}(\xi)$ and
$\mathbf{v}_\mathrm{g}(\xi)$ are defined via substituting
$\psi_\mathrm g$, and
analogously $\psi_\mathrm u$, $\chi_\mathrm g$, and
$\chi_\mathrm{u}$.  Then at each moment $t$ we can calculate all
variables in the \textbf{k}-space
\begin{eqnarray}
&&\xi_\mathbf{k}(t)=\xi_{0,\,\mathbf{k}}
+(t-t_0)A\frac{k_y}{\sqrt{k_y^2+k_z^2}},\\
&&\mathbf{b}_\mathbf k(t)=\mathbf{b}_\mathbf k(\xi_\mathbf{k}(t)),\\
&&\mathbf{v}_\mathbf k(t)=\mathbf{v}_\mathbf k(\xi_\mathbf{k}(t)),\\
&&k_{x}^\mathrm{wave}(t)=k_x - (t-t_0)Ak_y.
\end{eqnarray}
Finally, we can return back to the real \textbf{r}-space
\begin{eqnarray}
&&\mathbf{V}_\mathrm{wave}(\mathrm{r},t)
=-\ii V_\mathrm{A} \!\int\!
\mathbf{v}_\mathbf{k}(t)
\,\mathrm{e}^{-\ii[\mathbf{k}\cdot\mathbf{r}-(t-t_0)Ak_yx]}
\,\frac{\mathrm{d}k^3}{(2\pi)^3},\nn\\
&&\mathbf{B}_\mathrm{wave}(\mathrm{r},t)
=B_0 \!\int\!
\mathbf{b}_\mathbf{k}(t)\,
\mathrm{e}^{-\ii[\mathbf{k}\cdot\mathbf{r}-(t-t_0)Ak_yx]}
\,\frac{\mathrm{d}k^3}{(2\pi)^3},\nn\\
&&\mathrm{Re}(\mathrm{e}^{-\ii\,\varphi})=
\cos \varphi,\qquad
\mathrm{Re}(-\ii\mathrm{e}^{-\ii\,\varphi})=-\sin \varphi.
\end{eqnarray}
This evolution of MHD variables is the main detail of the theory of
MHD turbulence in a shear flow.

Consider now an imaginary fluid filling the phase space ${\bf k}$
and $w(t) \equiv  \mathcal{E}_\mathbf{k}(t)$ from
\Eqref{ch2density_of_rain} being the energy density carried by a
droplet of that fluid. As a wave mode initially with wave-vector
${\bf k}$ evolves according to \Eqref{ch2q(t)}, the infinitesimal
phase-fluid droplet associated with that mode moves in the ${\bf
k}$-space. Wave amplification means that the energy density of the
droplets increases by a factor of $G$
\be
G=\frac{w(t\rightarrow+\infty)}{w(t\rightarrow-\infty)}.
\ee
Indeed for $\chi=0$, $k_z=0$, and $\xi\rightarrow\infty$
\be
b_y^2\asymp \psi^2\gg b_x^2+b_z^2,\qquad
v_y^2\asymp \left( \frac{\mathrm{d}_\xi\psi}{Q} \right)^2
\gg v_x^2+v_z^2
\ee
and
\be
w(t\rightarrow\infty)=\frac{1}{4}V_\mathrm{A}^2 D_\mathrm{f}^2,\qquad
w(t\rightarrow -\infty)=\frac{1}{4}V_\mathrm{A}^2.
\ee
For big enough time arguments $|\xi|\gg1$ and purely two-dimensional
waves with $k_z=0$ the motion of the fluid asymptotically
corresponds to a SMW with dispersion coinciding with the AW one
\begin{eqnarray}
&&Q\xi=\omega_{_\mathrm{SMW}}t,\qquad
\omega_{_\mathrm{SMW}}=\omega_{_\mathrm{AW}}= V_\mathrm{A}|k_y|,\nn\\
&&\psi(\xi)\asymp D_\mathrm{f}
\cos(\omega_{_\mathrm{SMW}}t+\phi_\mathrm{f}).
\end{eqnarray}
The Poynting vector, i.e., the energy flux in $\mathbf{r}$-space is
$V_\mathrm{A}w$.

The velocity of the droplet in the \textbf{k}-space according to
\Eqref{ch2q(t)} determines the field of the shear flow in the
\textbf{k}-space
\be
\label{ch2U_shear}
\mathbf{U}=\mathrm{d}_t\mathbf{q}(t)=-Aq_y\mathbf{e}_x,\qquad
\mathbf{U}^\mathrm{shear}_\mathbf{k}=-Ak_y\mathbf{e}_x.
\ee
Looking at a droplet we actually derive the shear flow velocity
field in \textbf{k}-space, $\mathbf{U}^\mathrm{shear}_\mathbf{k}$.

According to the Kolmogorov--Obukhov cascade of energy we have a
constant energy flux through each spherical surface with surface
element $\mathrm{d}\mathbf{f}$ in \textbf{k}-space
\be
\label{ch2surface}
\varepsilon
=\oint
\mathcal{E}_\mathbf{k}^\mathrm{KO}\mathbf{U}^\mathrm{KO}_\mathbf{k}
\mathrm{d}\mathbf{f}
=\mathcal{E}_{k}^\mathrm{KO} {U}^\mathrm{KO}_{k}4\pi k^2,
\ee
which gives
\be
\mathbf{U}^\mathrm{KO}_\mathbf{k}\sim \varepsilon^{1/3}k^{5/3} \mathbf{e}_k,
\qquad
\mathbf{e}_k=\frac{\mathbf{k}}{k},
\ee
i.e., the velocity in \textbf{k}-space has dimension 1/(time$\times$length).
Here we used an important for our further work notion of the
energy flux in the \textbf{k}-space
\be
\mathbf{S}=\mathcal{E}_\mathrm{k}\mathbf{U}_\mathbf{k}
\ee
which is equal to energy density times velocity in the \textbf{k}-space.
This notion is analogous to the Poynting vector being, however,
defined in the \textbf{k}-space. In the Kolmogorov--Obukhov scenario we
have
\be
\frac{\partial}{\partial \mathbf{k}}\cdot \mathbf{S}^\mathrm{KO}
=\varepsilon
\delta(\mathbf{k}).
\ee

In order to approximate the turbulence as an initial source of MHD
waves we have to merge the turbulence with the wave spectral
densities and velocities. The simplest possible scenario is given in
the next subsection.

\subsection{Derivation of Shakura--Sunyaev Phenomenology in\\ %
the Framework of Kolmogorov Turbulence}                       %
%
How vortices create waves is a complicated problem far beyond the
scope of the present study. Here we will give only a model
illustration merging the spectral density of vortices
$\mathcal{E}^\mathrm{turb}_{\mathbf{k}}$ from Kolmogorov turbulence
with spectral density of magnetosonic waves
$\mathcal{E}^\mathrm{wave}_{\mathbf{k}}$
\be
\mathcal{E}^\mathrm{wave}_{\mathbf{k}}
\sim\mathcal{E}^\mathrm{turb}_{\mathbf{k}}
\sim \mathcal{E}_{_\Lambda}=
\varepsilon^{2/3}\Lambda^{11/3},
\ee
on the plane in momentum space
\be
k_x=-\mathrm{sgn}(k_y)\Lambda^{-1},\qquad
k_y^2+k_z^2<\Lambda^{-2},
\ee
where we qualitatively suppose that vortices are converted into
waves. Sign function corresponds to the direction of the shear flow
in the \textbf{k}-space, \Eqref{ch2U_shear}. For $k_y$ we consider that
turbulent vortices have a given spectral density at
$k_x>\Lambda^{-1}$ which is converted to MHD wave energy at the
plane $k_x=\Lambda^{-1}$, and further on this wave energy evolves
according to our solution. In other words, the plane
$k_x>\Lambda^{-1}$ is the boundary between the vortex region and the
beginning of the amplification in the wave region where vortices
have negligible influence. In our qualitative picture we suppose
that vortices create spectral density which further on evolves as
wave spectral density with negligible influence.

The amplification \Eqref{ch2gain} is essential $\mathcal G\gg1$ only within
a cylinder
\begin{equation}
\label{ch2Gain_substitution}
\mathcal{G}(k_y, k_z) -1\sim \frac{1}{\Lambda^2 q^2},\qquad
q^2=k_y^2+k_z^2<\Lambda^{-2}
\end{equation}
with radius $\Lambda^{-1}$. This result with remains unchanged in
amplitude if we include the $J_\mathrm{c,u}$ and $J_\mathrm{s,g}$
terms.

The amplification occurs in the region
$-\Lambda^{-1}<k_x<\Lambda^{-1}$, that is to say from the cylinder
we cut a tube with length $2\Lambda^{-1}$.  In other words, we have a
domain with a shape of a tube in momentum space
\be
\mathcal{V}=\{
k_y^2+k_z^2<\Lambda^{-2},\qquad
|k_x|<\Lambda^{-1}\}.
\ee
In order to calculate the total power of waves $\mathcal{H}$ (per
unit mass) analogously to \Eqref{ch2surface} we will integrate the
energy flux on the surface of the tube
\be
\label{ch2tube}
\mathcal{H}=\oint
\mathcal{E}_\mathbf{k}^\mathrm{wave}\mathbf{U}^\mathrm{shear}_\mathbf{k}
\mathrm{d}\mathbf{f}.
\ee

As the shear in the physical flow results in a drift
of the wave modes along the axis of the tube,  we have to take into
account only the circular surfaces
\be
\epsilon\sim
\int
|U_x|\left[\mathcal{G}(k_y, k_z)-1\right]
\mathcal{E}_{_\Lambda}
\mathrm{d}k_y\mathrm{d}k_z.
\ee
The multiplier $(\mathcal G -1)$ takes into account the difference
between flowing out and flowing in energy fluxes.

We can use polar coordinates
\be
k_z=q \cos \theta,\qquad
k_y=q \sin \theta,\qquad
U_x=Aq \sin \theta.
\ee
Averaging over the angle $\theta$
\be
\langle U_x\rangle = \frac{2}{\pi}Aq
\sim A q,\;
\int_0^\pi\sin \theta \frac{\mathrm{d}\theta}{\pi}=\frac{2}{\pi},\;
\ee
and substituting it in \Eqref{ch2tube}, using
$\mathrm{d}k_y\mathrm{d}k_z=\mathrm{d}(\pi q^2),$ leads to the
simple integral
\be
\mathcal{H}\sim
\int_0^{\Lambda^{-1}}
\frac{A q}{\Lambda^2 q^2}\,
\mathcal{E}_{_\Lambda}
q\mathrm{d}q
\sim \frac{A \mathcal{E}_{_\Lambda}}{\Lambda^{3}}\sim A V_\Lambda^2.
\ee
Then for the volume density of the amplified waves we have
\be
\mathcal{Q}\equiv\rho\mathcal{H}
\sim \rho A V_\Lambda^2
\sim \rho(\varepsilon V_A)^{2/3}A^{1/3}.
\ee
As all waves are finally dissipated, $\mathcal{Q}$
is actually the volume density of plasma heating.

For evanescent Kolmogorov turbulence power
\be
\rho \varepsilon \ll A\rho V_A^2=A B_0^2/\mu_0
\ee
the heating power $\mathcal{H}$ has a critical behavior
\be
\mathrm{d}_{\varepsilon}\mathcal{H}\sim
G_\mathrm{turb}\equiv\frac{\mathcal{H}}{\varepsilon}
\sim\left(\frac{AV_A^2}{\varepsilon}\right)^{1/3}\gg1,
\qquad
\mathcal{H}\gg\varepsilon
\qquad \mbox{for}\quad \varepsilon \rightarrow 0
\ee
which demonstrates that disks can ignite as a star even for very
weak turbulence and magnetic field. The ratio of wave power and
Kolmogorov vortex power $G_\mathrm{turb}$ can be considered as an
amplification coefficient for the turbulence. This energy gain shows
how efficient is the transformation of shear flow energy into waves
or in a broader framework the transformation of gravitational energy
into heat of accretion disks.

For hydrogen plasma $\rho c_\mathrm{s}^2/p=5/3\sim 1.$ Now we can
evaluate the shear stress (as given by the ratio of the volume
density of heating power and the shear frequency)
\be
\sigma=\frac{2\mathcal{Q}}{A}\sim
\rho \left(\frac{\varepsilon V_A}{A}\right)^{2/3}
\sim \rho V_\Lambda^2
\ee
via an effective viscosity
\be
\eta_\mathrm{eff}=\frac{\sigma}{A}\sim
\rho \frac{\left(\varepsilon V_A\right)^{2/3}}{A^{5/3}},\qquad
\nu_\mathrm{eff}=\frac{\eta_\mathrm{eff}}{\rho}
\sim\frac{\mathcal{H}}{A^2}
\sim \frac{\left(\varepsilon V_A\right)^{2/3}}{A^{5/3}}
\ee
and the dimensionless Shakura--Sunyaev friction coefficient
\be
\label{ch2Shakura-Sunyaev}
\alpha\equiv \frac{\sigma}{p}\sim \frac{V_\Lambda^2}{c_\mathrm{s}^2}.
\ee
Including of the energy of $z$-polarized AWs does not modify this
result. Here we wish to emphasize that in our evaluation of the
energy gain, we were concentrated on the wave amplification of the
energy of two dimensional motion in the $x$--$y$ plane. Taking into
account the energy in $z$-direction shows that the heating is even
higher, which is of course in the favor of the concepts.

For an approximately  Keplerian disk rotation the shear rate is half
of the frequency of the orbital Keplerian angular velocity
$A=-\frac{1}{2}\omega_\mathrm{Kepler}$.  In this case for time
$A^{-1}$ the disk rotates per $2$ radians.  For Earth's rotation
along the Sun this time is of the order of one season. In such a way
the length parameter of our problem $\Lambda=V_\mathrm{A}/A$ can be
evaluated as one Alfv\'en season. Then $V_\Lambda$ from the final
result for the Shakura--Sunyaev parameter can be qualitatively
considered as a pulsation of the turbulent velocity for two disk
particles at distance equal to one Alfv\'en season $\Lambda$. Our
theory is formally applicable for $V_\Lambda\ll c_\mathrm{s}$ but
the boundary of its applicability (when compressibility effects stop
the SMWs amplification) allows us to understand that strong disk's
turbulence can lead to Shakura--Sunyaev upper limit $\alpha\sim1$.
Thus the following cascade of events emerges as a likely scenario
for the intense heating in accretion flows: the heating of the bulk
of the disk creates convection. For strong heating the convection is
turbulent. Turbulence generates magnetohydrodynamic waves. Waves are
amplified by the shear flow -- this is the transformation of
gravitational energy of orbiting plasma into waves. Waves finally are
absorbed by the viscosity which creates the heating. The heat is
emitted through the surface of the disk.  This process of formation
of stars and other compact astrophysical objects from nebulas works
continuously -- we have a self-consistent theory for self-sustained
turbulence of the magnetized accretion disks.

The weak point of this scenario is the supposed convective
turbulence which in presence of magnetic fields is unlikely to be of
Kolmogorov type. We consider as much more plausible scenario the
appearance of a self-sustained magneto-hydrodynamical turbulence
considered in the next subsection.

\subsection{Kraichnan Turbulence as a more Plausible Scenario for Accretion Disks}%
%
Magnetic field qualitatively changes the behavior of the fluid.  We
have no waves generated by vortices -- the turbulence in magnetic
field is related to MHD waves. Analogously to the Kolmogorov law
\Eqref{ch2Kolmogorov}, for the Kraichnan turbulence the power of energy
cascade in the dissipation-free regime is given by the wave--wave
interaction
\be
\label{ch2Kraichnan}
\varepsilon=
\frac{(V_\lambda^2)^2}{\lambda V_\mathrm{A}}
=\mathrm{\frac{(velocity)^3}{length}=\frac{power}{mass}}.
\ee
This power is proportional to the intensity of the two interacting waves
and this nonlinear effect for incompressible fluid is due to
the convective term $\mathbf{V}\cdot\nabla\mathbf{V}$ of the substantial
acceleration $\mathrm{D}_t\mathbf{V} = \partial_t\mathbf{V} +
\mathbf{V}\cdot\nabla\mathbf{V}$ of the momentum equation.

The theory of generation of SMWs invokes parallels with the nonlinear
optical phenomena in lasers. The velocity oscillations of two amplified MHD waves
$\mathbf{V}^{(a)}$ and $\mathbf{V}^{(b)}$ create an external driving
force of the new wave with velocity field $\mathbf{V}$. In
the linearized we have to insert a small nonlinear
correction
\be
\rho \partial_t \mathbf{V} =
- \nabla p + \frac{\nabla\times\mathbf{B}}{\mu_0}\times \mathbf{B}
+\rho \mathbf{f},
\qquad \mathbf{f}\equiv \frac{1}{2}\sum_{a,\,b}
\mathbf{V}^{(a)}\cdot\nabla\mathbf{V}^{(b)}.
\ee
Here, in the inhomogeneous term $\mathbf{f}$ we have to perform
summation over all other MHD waves. This external for the wave force
(per unit mass) acts as an external noise and its statistical
properties are determined by the force--force correlator
\be
\hat\Gamma(\mathbf{r}_1,t_1;\mathbf{r}_2,t_2)
=\langle \mathbf{f}(\mathbf{r}_1,t_1)\,\mathbf{f}(\mathbf{r}_2,t_2) \rangle,
\ee
where the averaging is over the waves phases. A scenario of such
type (a Langevin MHD) was described in
Ref.~[\onlinecite{Mishonov:07}]; this approach is similar in the
spirit to the forced burgers turbulence.\cite{Woyczynski98} In the
framework of that scenario the strongly amplified
$|D_\mathrm{f}|\gg1$ MHD waves with asymptotics \Eqref{ch2wave-asymp}
\be
\label{ch2wave-asymp-damping}
\psi \approx
D_\mathrm{f}\,\theta(\xi)\cos(Q\xi+\phi_\mathrm{f})
\exp \left(-\nu^\prime K_y^2 \tau^3/3\right)
\ee
generate new waves and after a statistical averaging we have a
self-consistent theory for magnetic turbulence in a shear flow. So
MHD waves ignite the chain reaction of quasar self-heating. The last
exponential term describes the wave damping when a small viscosity
is taken into account. Damping is significant only for
$t\rightarrow\infty$ when $|k_x|\gg |k_y|$ and the wave-vector is
almost parallel to the magnetic field. In this geometry, the damping
rate of the wave density of AWs and SMWs is proportional to the
square of the frequency\cite{LL8}
\be
w(t)=w(0)\exp \left( -\frac{\omega^2}{V_\mathrm{A}^2}\nu t \right)
=w(0)\exp \left( -\nu k^2 t \right).
\ee
For the time-dependent wave-vector $k^2(t)\approx (k_y At)^2$ in the
argument of the exponent we have to make the replacement
\be
\nu k^2 t \rightarrow \nu \int_0^t k^2(t') \mathrm{d}t'
=\frac{1}{3}\nu^\prime K_y^2 \tau^3,\qquad \nu'
=\frac{\nu A}{V_\mathrm{A}^2}=\frac{1}{\mathcal{R}},
\qquad\mathcal{R}\equiv\frac{\Lambda V_\mathrm{A}}{\nu_\mathrm{k}},
\qquad\mathcal{S}_\mathrm{MRI}\equiv\frac{\Lambda V_\mathrm{A}}{\nu_\mathrm{m}},
\ee
where $\nu'$ is the dimensionless viscosity and
$\mathcal{R}=V_\mathrm{A}^2/A$ is the ``Reynolds number of the
magnetorotational instability (MRI),\cite{Masada:08}'' and
$\mathcal{S}_\mathrm{MRI}$ is the Lundquist number of MRI. After
long enough time $t_\nu$ when
\be
|k_x(t_\nu)|=1/\lambda_\mathrm{a},
\qquad
\lambda_\mathrm{a}=\frac{\nu}{V_\mathrm A},
\ee
MHD waves are completely dissipated.

The details of self-consistent MHD turbulence will be given
elsewhere, but again the wave amplification operates as a turbulence
amplifier. For MHD turbulence one can expect
\be
\label{ch2tusj5}
\sigma_{R\varphi}=\alpha_\mathrm{m}(\nu') p_B,
\qquad p_B=\frac{1}{2}\rho V_\mathrm{A}^2. \ee
The evaluation of the magnetic friction coefficient $\alpha_\mathrm{m}$
as a function of the dimensionless viscosity is a new problem addressed
to the theoretical astrophysics.

\section{Review of used results}

Imagine that \textit{with painstaking numerical work} \cite{Hawley:92} the
\textit{frustratingly complex} system
of MHD equations \Eqref{ch2VelocityEvolution} and \Eqref{ch2MagneticEvolution} is
already solved.
Using this solution
we can express the effective kinematic viscosity
\begin{eqnarray}
\label{ch2wave_nu}
\nu_\mathrm{eff}(\tau)=\nu_\mathrm{k} +
\nu_\mathrm{k}\sum_{\mathbf{Q}}Q^2\mathbf{v}_\mathbf{Q}^2(\tau)
+ \nu_\mathrm{m}\sum_{\mathbf{Q}} Q^2\mathbf{b}_\mathbf{Q}^2(\tau).
\end{eqnarray}
The additional time averaged wave viscosity
\be
\label{ch2wave_viscosity}
\eta_\mathrm{wave}=
\lim_{\Delta \tau\rightarrow\infty}\int_0^{\Delta
\tau}\frac{\mathrm{d}\tau}{\Delta\tau}\left(
\rho\nu_\mathrm{k}\sum_{\mathbf{Q}}Q^2\mathbf{v}_\mathbf{Q}^2(\tau)
+ \rho\nu_\mathrm{m}\sum_{\mathbf{Q}} Q^2\mathbf{b}_\mathbf{Q}^2(\tau)\right)
\ee
is created by the momentum transfer by the SMW.
In such a way the ``alfv\'enons'' are the new particles which can have even
dominant contribution to the viscosity.
We wish to emphasize that this expression is an exact result for used model of
incompressible fluid; for compressible fluid expression is more complicated and
contains the second viscosity.
For the usual hydrodynamics Boussinesq \cite{Boussinesq} and
Prandtl \cite{Prandtl} pointed out the role of vortices for the momentum
transport and creation of the turbulent viscosity. The magnetic field in
conducting plasmas opens the new possibility the role of the vortices to be
substituted by MHD waves. For accretion disks the problem for the effective
viscosity is reduced to calculation of the spectral densities of the
wave turbulence presented by \Eqref{ch2wave_nu} and \Eqref{ch2wave_viscosity}.

The MHD equations are investigated in coordinate space in uncountable set of
numerical works. However development in this direction is almost saturated and
we have to look for other possibilities. Numerical solution in wave-vector
space gives a promising perspective especially if significant part of
calculation of nonlinear terms is made analytically using analytical asymptotics
for large wave-vectors. In this case the MHD modes are almost independent. We
have something like asymptotic freedom known from the elementary particles
physics. MHD modes are generated and amplified at small wave-vectors where
numerical analysis is indispensable. The numerical solution from the finite
wave-vector domain has to be continued by the analytical asymptotic which gives
the possibility for fast calculation of the nonlinear terms. We believe that
further progress can be made by the essential use of the wave-vector space.
That is why one of the goals of the present research is to write the general
nonlinear equation in wave-vector space and to analyze the properties of
the linearized ones. We have to know MHD set of equations which
have to be solved.

The parallel between the propagation of the waves and particles was pointed out
by Hamilton. We recall this analogy just to emphasize the differences.
For the elementary particles of the plasmas the elementary acts of the
scattering are in a very good approximation independent and this facilitates the
kinetic problem. While MHD turbulence is a hidden coherent structure described
by the very difficult for solution system \Eqref{ch2VelocityEvolution} and
\Eqref{ch2MagneticEvolution}. Instead of separate particles we have something like
phase transition in a fluctuating Bose condensate.
For example, in the two-dimensional fluid the equilibrium statistics has a most
interesting structure, despite the simplicity of the energy expression, because
there is an additional important constant of motion: the enstrophy, or
integrated square of the vorticity.
The enstrophy constant leads to equilibria in which a large fraction of the
energy is condensed into the largest spatial scales of motion, a situation
closely analogous to the Bose--Einstein condensation in an ideal boson gas
\cite{Kraichnan}.

The simplest possible scenario is that the system have stable static solution,
$\mathbf{v}_\mathbf{Q}(\tau)=\mathbf{v}_\mathbf{Q}$ and
$\mathbf{b}_\mathbf{Q}(\tau)=\mathbf{b}_\mathbf{Q},$
and analogously to the turbulence above the ocean \cite{Miles:57} we have again
a \textit{cooperative behavior buried in} in the wave turbulence.
In this case it is not necessary to make time averaging and for the enhancement
factor of the viscosity we obtain \cite{Dimitrov:11}
\begin{eqnarray}
\label{ch2viscosity}
Z=\frac{\eta_\mathrm{eff}}{\eta}
=\frac{\sigma_{xy}}{\eta A}
=\frac{\tilde{\mathcal{Q}}}{\eta A^2}
=1 +
\int_\mathbf{Q}
\left(\mathbf{v}_\mathbf{Q}^2
+ \frac{\mathbf{b}_\mathbf{Q}^2}{\mathrm{P_m}}
\right) Q^2\,
\frac{\mathrm{d}Q_x \mathrm{d}Q_y \mathrm{d}Q_z}{(2\pi)^3},
\qquad
\frac{1}{\mathrm{P_m}}
=\frac{\varepsilon_0 c^2 \varrho_\Omega}{\eta/\rho}.
\end{eqnarray}
The viscosity renormalization parameter $Z\gg1$ (for quasars is
possible $\ln Z\simeq 20$) determines the work of the
accretion disks as a machine for making of compact astrophysical objects;
$\sigma_{xy}\equiv\sigma_{r\phi}$ is the time averaged stress tensor and
$\tilde{\mathcal{Q}}$ is the volume density of heating power. The formulas for
the
effective viscosity are the tool to take into account the strong anisotropy
of the wave turbulence, i.e. the anisotropy of the integrand in the expression
above. As we mentioned this result is exact in the framework of MHD for
incompressible fluid. Due to general mechanical theorem the time averaged volume
density of heating power, shear component of stress tensor and homogeneous shear
rate are related with the simple equation
$\tilde{\mathcal{Q}}=\sigma_{xy}A=\eta_\mathrm{eff}A^2$ -- this is the
definition for
effective viscosity $\eta_\mathrm{eff}$ determined by Ohmic resistivity,
bare viscose friction, and Fourier components of the MHD variables.

The last term in \Eqref{ch2viscosity} can be essential only in the initial stages
of the accretion disk where the hydrogen is weakly ionized.
In this case before the re-ionization of the hydrogen after the big bang the
magnetic Prandtl number is small
$\mathrm{P}_\mathrm{m}=\nu_\mathrm{k}/\nu_\mathrm{m}\ll 1.$
In this regime of weakly ionized plasma the viscosity coefficient is determined
Ohmic dissipation.

The nonlinear hydrodynamic terms are known from the monographs on hydrodynamics
\cite{Pope:00} and magnetohydrodynamics \cite{Biskamp:03}.
In our work we use dimensionless imaginary components of the velocity
$\mathbf{v}_\mathbf{Q}(\tau)$
in order the final MHD system to be real.

Our starting point are general magnetohydrodynamic (MHD) equations in wave-vector space

\begin{eqnarray}
\mathrm{D}_{\overline{\tau}}^{\mathrm{\,shear}}\mathbf{v}_\mathbf{Q}(\tau) &=&
-v_{x,\mathbf{Q}}\mathbf{e}_y+2n_y\mathbf{n}v_{x,\mathbf{Q}}
+ 2\omega_\mathrm{c}\mathbf{n}\,(n_yv_{x,\mathbf{Q}}-n_xv_{y,\mathbf{Q}})
-2\boldsymbol{\omega_\mathrm{c}}\times \textbf{v}_{\mathbf{Q}} \\
&& + (\boldsymbol{\alpha}\cdot\mathbf{Q})\,\mathbf{b}_\mathbf{Q} -\nu^\prime_\mathrm{k}Q^2\mathbf{v}_\mathbf{Q} + \mathbf{N}_v,
\label{ch3VelocityEvolution}
\\
\mathbf{N}_v &=& \Pi^{\perp\mathbf{Q}}\cdot\sum_{\mathbf{Q}'}\left[\mathbf{v}_{\mathbf{Q}'}\otimes\mathbf{v}_{\mathbf{Q}
-\mathbf{Q}'} + \mathbf{b}_{\mathbf{Q}'}\otimes\mathbf{b}_{\mathbf{Q}-\mathbf{Q}'}\right]\cdot\mathbf{Q},
\\
&&\mathbf{Q} \cdot \mathbf{N}_v=0, \quad \mathbf{v}_\mathbf{Q}(\overline{\tau}_0)=\Pi^{\perp\mathbf{Q}}\mathbf{v}_\mathbf{Q}(\overline{\tau}_0),\nonumber
\\
\label{ch3MagneticEvolution}
\mathrm{D}_{\overline{\tau}}^{\mathrm{\,shear}}\mathbf{b}_\mathbf{Q}(\tau) &=& b_{x,\mathbf{Q}}\mathbf{e}_y
-(\mathbf{Q}\cdot\boldsymbol{\alpha})\,\mathbf{v}_\mathbf{Q} -\nu'_\mathrm{m}Q^2\mathbf{b}_\mathbf{Q} + \mathbf{N}_b,
\\
\mathbf{N}_b &=& \Pi^{\perp\mathbf{Q}}\cdot\sum_{\mathbf{Q}'}\left[\mathbf{b}_{\mathbf{Q}'}\otimes\mathbf{v}_{\mathbf{Q}
-\mathbf{Q}'} - \mathbf{v}_{\mathbf{Q}'}\otimes\mathbf{b}_{\mathbf{Q}-\mathbf{Q}'}\right]\cdot\mathbf{Q},
\\
&&\mathbf{Q}\cdot \mathbf{N}_b=0, \quad \mathbf{b}_\mathbf{Q}(\overline{\tau}_0)=\Pi^{\perp\mathbf{Q}}\mathbf{b}_\mathbf{Q}(\overline{\tau}_0),\nonumber
\\
\Pi^{\perp\mathbf{Q}} &\equiv &\openone-\mathbf{n}\otimes \mathbf{n},\quad \mathbf{n}\equiv \frac{\mathbf{Q}}{Q},\quad \overline{\tau}\equiv-\frac{Q_x}{Q_y}
\\
\label{ch3ushear}
\mathrm{D}_\tau^{\mathrm{\,shear}} &\equiv& \partial_\tau+\mathbf{U}_{\mathrm{shear}}(\mathbf{Q})\cdot\partial_{\mathbf{Q}}=\partial_\tau-Q_y\partial_{Q_x}= \partial_\tau + \partial_{\overline{\tau}},
\end{eqnarray}

for the velocity and magnetic field

\begin{eqnarray}
\label{ch3v_wave_Q}
\mathbf{V}(t,\bf{r})&=&Ax\mathbf{e}_y + \ii V_\mathrm{A}\sum_{\mathbf{Q}}\mathbf{v}_{\mathbf{Q}}(\tau)\,\mathrm{e}^{\ii\mathbf{Q}\cdot\mathbf{X}},
\\
\label{ch3b_wave_Q}
\mathbf{B}(t,\bf{r})&=&B_0\boldsymbol{\alpha} + B_0\sum_{\mathbf{Q}}\mathbf{b}_{\mathbf{Q}}(\tau)\,\mathrm{e}^{\ii\mathbf{Q}\cdot\mathbf{X}},\\
\end{eqnarray}

\begin{eqnarray}
&& \sum_{\mathbf{Q}}=\int\mathrm{d}\left(\frac{\mathbf{Q}}{2\pi}\right)=\int\frac{\mathrm{d}^3{Q}}{(2\pi)^3},
\quad \boldsymbol{\alpha}=(0,\,\alpha_y=\sin\theta,\, \alpha_z=\cos\theta),\\
&& V_\mathrm{A}=\frac{B_0}{\sqrt{\mu_0\rho}},
\quad \Lambda\equiv\frac{V_\mathrm{A}}{A},
\quad \tau \equiv At.
\end{eqnarray}

\begin{equation}
\nu^\prime_\mathrm{m} = \nu_\mathrm{m}/ \Lambda V_A, \quad \nu^\prime_\mathrm{k} = \nu_\mathrm{k}/ \Lambda V_A,
\quad \nu_\mathrm{m}=\varepsilon_0 c^2 \varrho_{_\Omega},
\quad \nu_\mathrm{k}=\frac{\eta}{\rho}.
\end{equation}
Here we use the self-explaining notations for the \alf velocity $V_A$, shear rate with dimension of frequency $A$,
angular velocity of the fluid $\mathbf{\Omega}=A\omega_\mathrm{c}\mathbf{e}_z,$
kinematic viscosity $\nu_\mathrm{k}$ and magnetic diffusivity $\nu_\mathrm{m}$.
The $\theta$ is the angle between the normal of the shear flow plane $(x,y)$ and magnetic field $\mathbf{B}_0=B_0\boldsymbol{\alpha}$.
This problem is inspired from the physics of accretion disks where $z$ is the rotation axis.
Our goal is to calculate the effective viscosity
\begin{eqnarray}
\nu_\mathrm{eff}(\tau)=\nu_\mathrm{k} + \nu_\mathrm{wave},\quad \nu_\mathrm{wave}\equiv \nu_\mathrm{k}\sum_{\mathbf{Q}}Q^2\mathbf{v}_\mathbf{Q}^2(\tau)
+ \nu_\mathrm{m}\sum_{\mathbf{Q}} Q^2\mathbf{b}_\mathbf{Q}^2(\tau),
\end{eqnarray}
in the spirit of Boussinesq \cite{Boussinesq} and Prandtl \cite{Prandtl},
when for static wave-turbulence we need of the time averaged spectral densities $<\mathbf{v}^2_\mathbf{Q}>$ and $<\mathbf{b}^2_\mathbf{Q}>$.

As the general problem for ``eddy'' viscosity $\eta_{eff}$ is frustratingly complex
we can start with the simplest possible 2D case of ``pure shear'' Fig.~9 [\onlinecite{Balbus:98}],
$\mathbf{\Omega}=0$, toroidal magnetic field $\theta=\pi/2$ and evanescent dissipation
$\nu^{\prime}_\mathrm{tot}=\nu^{\prime}_\mathrm{k}+\nu^{\prime}_\mathrm{m} \ll 1$.
For ideal fluid with $\nu^{\prime}_\mathrm{tot}=0$ the linearized equations have exact solution in the framework
of Heun functions \cite{PoP}, because the MHD problem is reduced to an effective quantum mechanical problem.

We will use this analytical result for the static case
$\mathbf{v}_\mathbf{Q}(\tau)=\mathbf{v}_\mathbf{Q}$ and $\mathbf{b}_\mathbf{Q}(\tau)=\mathbf{b}_\mathbf{Q}$.
For the 2D case of $Q_z=0$, $v_z=0=b_z$ using the substitution $b_x=\psi(\overline{\tau})/\sqrt{1+\overline{\tau}^2}$ we have Schr\"odinger type equation
\begin{equation}
 \mathrm{d}_{\overline{\tau}}^2\psi + \left[Q_y^2 - \frac1{(1+\overline{\tau}^2)^2} \right]\psi=0, \qquad \overline{\tau}\equiv-\frac{Q_x}{Q_y},
\label{ch3Schrodinger}
\end{equation}
whose solution represent the 2D MHD variables
\begin{equation}
b_y(\overline{\tau})  = \overline{\tau} b_x(\overline{\tau}) , \quad v_y(\overline{\tau})  = \overline{\tau} v_x(\overline{\tau}),
\quad v_x= \mathrm{d}_{\overline{\tau}} b_x / Q_y
\end{equation}
for convenience we consider $\overline{\tau}$ and $Q_y$ as independent variables in the 2D wave-vector space.

\section{Interplay between dissipative and nonlinear terms}

Searching a static solution of general MHD equations \Eqref{ch3VelocityEvolution} an \Eqref{ch3MagneticEvolution}
we have express $v_y$ from the $y$-component of \Eqref{ch3MagneticEvolution},
and substitution in the $y$-component of \Eqref{ch3VelocityEvolution} gives

\begin{equation}
\label{ch3bay_ivan}
\mathrm{d}^2_{\overline{\tau}}b_y + \nu_\mathrm{tot}^\prime Q^2\mathrm{d}_{\overline{\tau}}b_y + [Q^2_\alpha + 2\nu_\mathrm{m}^\prime \tau Q_y^2 +\nu_\mathrm{m}^\prime\nu_\mathrm{k}^\prime Q^4]b_y
= \mathrm{d}_{\overline{\tau}}N^y_b - Q_\alpha N_v^y -2Q_\alpha n_y^2v_x + \nu_\mathrm{k}^\prime QN_b^y + (\nu_\mathrm{k}^\prime - \nu_\mathrm{m}^\prime)Q^2b_x
\end{equation}

Now let us analyze all irrelevant for $|\overline{\tau}| \gg 1$ terms:

1) the linear term $v_x/(1+\overline{\tau}^2)$ is responsible for the amplification of SMW at $\overline{\tau} \sim 1$ but is negligible for $|\overline{\tau}| \gg 1$,

2) $\nu_\mathrm{m}^\prime\nu_\mathrm{k}^\prime$ is negligible for evanescent dissipation $\nu_\mathrm{m}^\prime\nu_\mathrm{k}^\prime \ll 1$,

3) $\nu_\mathrm{m}^\prime\tau$ will be never essential because for large enough $|\overline{\tau}| > 1/ \nu_\mathrm{m}^\prime$ the wave amplitude will be exponentially small.

In such a way for $|\overline{\tau}| \gg 1$ we derive
an effective oscillator equation with damping $\propto \nu_\mathrm{tot}^\prime$
and external force created by the nonlinear terms
\begin{equation}
\left[\mathrm{d}^2_{\overline{\tau}} + \gamma(\overline{\tau})\mathrm{d}_{\overline{\tau}} + Q_y^2\right] b_y = F_\mathrm{ext},\quad \gamma(\overline{\tau})=\nu_\mathrm{tot}^\prime Q^2,\quad  F_\mathrm{ext}= \mathrm{d}_{\overline{\tau}}N^y_b - Q_\alpha N_v^y.
\end{equation}

For small friction $\nu_\mathrm{tot}^\prime \ll 1$ the left hand linear operator has the WKB Green function
\begin{equation}
\left[\mathrm{d}^2_{\overline{\tau}} + \gamma(\overline{\tau})\mathrm{d}_{\overline{\tau}} + Q_y^2\right] G(\overline{\tau}-\overline{\tau}_0) = \delta(\overline{\tau}-\overline{\tau}_0)
\end{equation}

\begin{equation}
G(\overline{\tau}-\overline{\tau}_0) \approx \frac1{Q_y}\sin[Q_y(\overline{\tau}-\overline{\tau}_0)]
\exp\left(-\frac12\int^{\overline{\tau}}_{\overline{\tau}_0} \mathrm{d}\tau \gamma(\tau) \right)
\theta(\overline{\tau}-\overline{\tau}_0),
\end{equation}
where
\begin{equation}
\frac12\int_{\overline{\tau}_0}^{\overline{\tau}} \gamma(\tau)\mathrm{d}\tau \approx \frac{\nu_\mathrm{tot}^\prime}6 Q_y^2(\overline{\tau}^3-\overline{\tau}^3_0).
\end{equation}
The solution of the oscillator equation
\begin{equation}
 b_y(\overline{\tau},Q_y)= \int_{-\infty}^{\overline{\tau}} G(\overline{\tau},\overline{\tau}_0)F_\mathrm{ext}(\overline{\tau}_0)\mathrm{d}\overline{\tau}_0
\end{equation}
is the basis of the self-consistent theory of wave turbulence described in the next section.
If we take into account a small dissipation in the MHD equations, we have large $\overline{\tau}\gg1$ asymptotics
for 2D case $Q_z=0$, and evanescent friction $\nu_\mathrm{tot}^\prime\ll1$
\begin{eqnarray}
\psi&\approx&D_\mathrm{f}\,\mathrm{e}^{-\nu_\mathrm{tot}^\prime Q_y^2\overline{\tau}^3/6}\cos(|Q_y|\overline{\tau}+\phi_f),\quad \overline{\tau} > 0 \\
b_y&\approx&\frac{\overline{\tau}\psi}{\sqrt{1+\overline{\tau}^2}} \approx \psi \approx D_\mathrm{f}\,\mathrm{e}^{-\nu_\mathrm{tot}^\prime Q_y^2\overline{\tau}^3/6}\cos(|Q_y|\overline{\tau}+\phi_f), \\
v_y&\approx&\overline{\tau}\frac{(1+\overline{\tau}^2)\mathrm{d}_{\overline{\tau}}\psi - \overline{\tau}\psi}{Q_y(1+\overline{\tau}^2)^{3/2}} \approx -D_\mathrm{f}\,\mathrm{e}^{-\nu_\mathrm{tot}^\prime Q_y^2\overline{\tau}^3/6}\sin(|Q_y|\overline{\tau}+\phi_f).
\end{eqnarray}
In other words, after the sher created amplification the nonlinear terms are negligible and alfv\'enons decay as free particles.
In this WKB asymptotic for the wave energy density in wave-vector space for $\overline{\tau}\gg1$ we have
\be
\mathbf{b}_\mathbf{Q}^2+\mathbf{v}_\mathbf{Q}^2\approx
D_\mathrm{f}^2\,\exp\left(-\int_{0}^{\overline{\tau}} \gamma(\tau)\, \mathrm{d}\tau \right)\approx
D_\mathrm{f}^2\,\mathrm{e}^{-\nu_\mathrm{tot}^\prime Q_y^2\overline{\tau}^3/3},
\ee
and supposing strong amplification $\mathcal{A}\gg1$ we have

\begin{eqnarray}
\label{ch3df}
&& \nu_\mathrm{tot}^\prime \int_{-\infty}^{\infty} Q^2 \left(\mathbf{b}_\mathbf{Q}^2+\mathbf{v}_\mathbf{Q}^2\right)\mathrm{d}Q_x\approx |Q_y|\left.\left(\mathbf{b}^2_\mathbf{Q} + \mathbf{v}^2_\mathbf{Q} \right)\right|_\mathrm{f} =|Q_y|D_\mathrm{f}^2, \\
&& \left.\left(\mathbf{b}^2_\mathbf{Q} + \mathbf{v}^2_\mathbf{Q} \right)\right|_\mathrm{f} = D_\mathrm{f}^2
\label{ch3Df}
\end{eqnarray}

Density of wave energy in $Q$-space immediately after amplification is determined by amplitude of damped magnetosonic wave. Initial energy of magnetosonic wave gradually dissipation in head due to viscous friction.
Thus according \Eqref{ch3df} dissipation power for all $Q_x$ can be presented as energy of amplified wave for $Q_x=0$.
If we use virial theorem whereby in magnetosonic waves we have continuous transformation of magnetic energy into mechanical $<\mathbf{v}_\mathbf{Q}^2> \approx  <\mathbf{b}_\mathbf{Q}^2>$ then for large Prandtl numbers
$\nu_\mathrm{m} \ll \nu_\mathrm{k} \approx \nu_\mathrm{tot}$ equation \Eqref{ch3nu_eff} can be rewrite as

\begin{equation}
\nu_\mathrm{eff} = \frac{\nu_\mathrm{eff}}{\nu_k} = 1 + \nu'_\mathrm{wave} = 1+ \frac12\int_{Q} Q^2(\mathbf{v}_\mathbf{Q}^2 +\mathbf{b}_\mathbf{Q}^2) \frac{\mathrm{d}Q_x\mathrm{d}Q_y\mathrm{d}Q_z}{(2\pi)^3},
\end{equation}

In this approximation of strongly amplified waves with small decay rate we have
\be
\nu'_\mathrm{wave}\approx \frac{1}{2}\frac{1}{(2\pi)^3}\int |\mathbf{U}_\mathrm{shear}(\mathbf{Q})| \left.\left(\mathbf{b}_\mathbf{Q}^2+\mathbf{v}_\mathbf{Q}^2\right)\right|_\mathrm{f} \mathrm{d}Q_y \mathrm{d}Q_z
\ee
or in 2D case taking into account the symmetry
\be
\label{ch3nu_wave2}
\nu'_\mathrm{wave}\approx \frac{1}{(2\pi)^2}\int_0^\infty |\mathbf{U}_\mathrm{shear}(\mathbf{Q})| \left.\left(\mathbf{b}_\mathbf{Q}^2+\mathbf{v}_\mathbf{Q}^2\right)\right|_\mathrm{f} \mathrm{d}Q_y
\approx\frac{1}{4\pi^2}\int_0^\infty Q_y D_\mathrm{f}^2(Q_y)\,\mathrm{d}Q_y.
\ee
In other words the total heating power can be evaluated as the flux of the amplified waves after the amplification plane $Q_x=0$.
Every wave finally gives his energy to the fluid and effective viscosity can be evaluated as surface integral from the energy flux
\be
\label{ch3Pointing}
\mathbf{S}(\mathbf{Q})\equiv
\mathbf{U}_\mathrm{shear}(\mathbf{Q})\left(\mathbf{b}_\mathbf{Q}^2+\mathbf{v}_\mathbf{Q}^2\right)/(2\pi)^3,\quad
\nu'_\mathrm{wave}=\frac12 \oint \mathbf{S}\cdot\mathrm{d}\mathbf{f},
\ee
where $\mathrm{d}\mathbf{f}=\mathrm{d}Q_y \mathrm{d}Q_z$ is the elementary area in momentum space after the wave amplification when waves become independent.
In the next section we will derive a self-consistent chain of the equation of the amplitudes of the amplified waves $ D_\mathrm{f}(Q_y)$ and finally we will express the effective viscosity by the solution using \Eqref{ch3nu_wave}.

Dissipative processes in MHD

\begin{equation}
 \tilde{Q}_\mathrm{tot}= \tilde{Q}_\mathrm{\eta}^\mathrm{wave} + \tilde{Q}_\mathrm{Ohm}^\mathrm{wave} + \tilde{Q}_\mathrm{\eta}^\mathrm{shear} = \eta_\mathrm{eff}A^2
\end{equation}

\begin{equation}
 \tilde{Q}_\mathrm{\eta}^\mathrm{shear} = \frac{\eta}{2}\left<\left(\partial_\mathrm{k}V_\mathrm{i}^\mathrm{shear}  + \partial_\mathrm{i}V_\mathrm{k}^\mathrm{shear} \right)^2\right> = \eta A^2 = \rho V_\mathrm{A}^2 A \nu'_\mathrm{k}
\end{equation}

\begin{equation}
 \tilde{Q}_\mathrm{\eta}^\mathrm{wave} = \frac{\eta}{2}\left<\left(\partial_\mathrm{k}V_\mathrm{i}^\mathrm{wavw}  + \partial_\mathrm{i}V_\mathrm{k}^\mathrm{wave} \right)^2\right> = \rho A V_\mathrm{A}^2 \nu'_\mathrm{k} \sum_{Q} Q^2 \mathbf{v}_\mathbf{Q}^2
\end{equation}

\begin{equation}
 \tilde{Q}_\mathrm{Ohm}^\mathrm{wave} = <\mathbf{j}\cdot\mathbf{E}> = \frac1{\mu_0\sigma_{_\mathrm{Ohm}}}(\mathrm{rot}\, \mathbf{B}^\mathrm{wave})^2 = \rho A V_\mathrm{A}^2 \nu'_\mathrm{m} \sum_{Q} Q^2 \mathbf{b}_\mathbf{Q}^2
\end{equation}

\begin{equation}
 \tilde{Q}=(\rho\nu_\mathrm{eff})A^2=\eta_\mathrm{eff}A^2=\sigma_{xy}A
\end{equation}

\begin{equation}
 \label{ch3nu_eff}
 \nu_\mathrm{eff}= \nu_\mathrm{k} + \nu_\mathrm{k}\sum_{Q}Q^2\mathbf{v}_\mathbf{Q}^2 + \nu_\mathrm{m}\sum_{Q}Q^2\mathbf{b}_\mathbf{Q}^2
\end{equation}

\begin{equation}
\label{ch3renorm}
 Z = \frac{\eta_\mathrm{eff}}{\eta}=\frac{\sigma_{xy}}{\eta A}=\frac{\tilde{Q}_\mathrm{tot}}{\eta A^2}=1+\int_{Q}Q^2\left( \mathbf{v}_\mathbf{Q}^2 + \frac{\mathbf{b}_\mathbf{Q}^2}{P_m}\right) \frac{\mathrm{d}Q_x\mathrm{d}Q_y\mathrm{d}Q_z}{(2\pi)^3},
\end{equation}

\begin{equation}
 \frac{1}{ P_\mathrm{m} } = \frac{ \epsilon_0 c^2 \varrho }{ \eta/\rho }
\end{equation}

$$ P_\mathrm{m} \propto T^4, \quad \frac1{P_\mathrm{m}} \ll 1,\quad \nu'_\mathrm{tot}= \nu'_\mathrm{k} + \nu'_\mathrm{m} \approx \nu'_\mathrm{k} $$

\begin{equation}
\label{ch3nu_eff2}
\nu_\mathrm{eff} \approx \nu_\mathrm{k} + \nu_\mathrm{tot}\sum_{Q}Q^2\mathbf{v}_\mathbf{Q}^2 = \nu_\mathrm{k} + \nu_\mathrm{wave}
\end{equation}

Lets introduce density of wave energy in $Q$-space $w_\mathbf{Q}= \frac12(\mathbf{v}_\mathbf{Q}^2 + \mathbf{b}_\mathbf{Q}^2)$,
drift velocity of magnetosonic waves in $Q$-space $\mathbf{U}_\mathrm{shear}(\mathbf{Q})=-Q_y\mathbf{e}_x$ cf. 146 Sozopol !!!!!   ,
and energy flux in $Q$-space $\mathbf{S}(\mathbf{Q})=\frac12\mathbf{U}_\mathrm{shear}(\mathbf{Q})w_\mathbf{Q}/(2\pi)^3$.
Using \Eqref{ch3df} and virial theorem $<\mathbf{v}_\mathbf{Q}^2> \approx  <\mathbf{b}_\mathbf{Q}^2>$ the equation for the viscosity created by the waves \Eqref{ch3nu_eff2} is

\begin{equation}
\label{ch3nu_wave}
\nu'_\mathrm{wave} =  \frac12\int_{Q} Q^2(\mathbf{v}_\mathbf{Q}^2 +\mathbf{b}_\mathbf{Q}^2) \frac{\mathrm{d}Q_x\mathrm{d}Q_y\mathrm{d}Q_z}{(2\pi)^3} = \oint \mathbf{S}\cdot\mathrm{d}\mathbf{f} =
\frac1{(2\pi)^3} \int_0^\infty |\mathbf{U}_\mathrm{shear}(\mathbf{Q})|\left(\mathbf{b}_\mathbf{Q}^2+\mathbf{v}_\mathbf{Q}^2\right)\mathrm{d}Q_y\mathrm{d}Q_z,
\end{equation}
where $\mathrm{d}\mathbf{f}=-\mathrm{sgn}(Q_y)\mathrm{d}Q_y \mathrm{d}Q_z \mathbf{e}_x$ is the elementary area in $Q$-space.
This formula have simple physical point - dissipated power of waves is equal to the power ``emitted'' waves near plane of amplification.
In 2D case we skip integration on $\mathrm{d}Q_z/2\pi$ and formal substitute $Q_z=0$. Thus for renorm factor of viscosity Z \Eqref{ch3renorm}
formula  \Eqref{ch3nu_wave2} and \Eqref{ch3Df} gives
\begin{eqnarray*}
&&Z= 1+ \frac1{(2\pi)^2\nu'_\mathrm{k}}\int_0^\infty Q_y D_\mathrm{f}^2\mathrm{d}Q_y=1+ \frac{R^2}{(2\pi)^2\nu'_\mathrm{k}}I_D=1+\frac{\Gamma^2(\frac43)}{128\pi^4\nu'_\mathrm{k}}\left(\frac6{\nu'_\mathrm{tot}} \right)^{2/3}I_D,\\
&&I_D=\int_0^\infty Q_y \tilde{D}_\mathrm{f}^2\mathrm{d}Q_y,\quad \tilde{D}=RD.
\end{eqnarray*}
here we use $D_\mathrm{f}^2(-Q_y)=D_\mathrm{f}^2(Q_y)$.

For hot plasmas where $\nu'_\mathrm{tot}\approx\nu'_\mathrm{k}$

\begin{equation}
Z \equiv \frac{\eta_\mathrm{eff}}{\eta_\mathrm{kin}} \approx \frac{ \Gamma^2(\frac43) 6^{2/3}I_D }{128\pi^4} \frac1{\nu_\mathrm{k}^{\prime 5/3}} = \frac{ \Gamma^2(\frac43) 6^{2/3}I_D }{128\pi^4} \left( \frac{\rho V_\mathrm{A}^2}{A\eta}\right)^{5/3}  \gg 1
\end{equation}

\begin{equation}
 \sigma_{xy} = \frac{ \Gamma(\frac43)^2 6^{2/3}I_D }{128\pi^4}  \frac{(\rho V_\mathrm{A}^2)^{5/3}}{(A\eta)^{2/3} } \ll P.
\end{equation}

In the next section we will calculate renorm factor of viscosity in approximation of evanescent viscosity $\nu'_\mathrm{k} \ll 1$ for who -> $Z \gg 1$.
Aim of the current work in with the nonlinear terms are introduced in MHD equations is to investigate in details simplest solvable case of self-sustain
wave turbulence, when effective viscosity dramatically increases. In this simplest 2D case we will consider solutions without rotating $\omega_\mathrm{c}=0$.
As we already mentioned the next section will have technically character.

\begin{eqnarray*}
&& Z=1+\frac{C_z}{\nu_\mathrm{tot}^{\prime 5/3}} \approx C_z\left( \frac{V_\mathrm{A}^2}{\nu_\mathrm{tot}A} \right)^{5/3} \gg 1, \quad \eta_\mathrm{eff}=Z(\eta=\rho_0\nu_\mathrm{k})=C_z\rho_0\nu_\mathrm{k}\left(\frac{V_\mathrm{A}^2}{\nu_\mathrm{k}A} \right)^{5/3} \gg 1 \\
&& \sigma_{xy}=\eta_\mathrm{eff}A=Z\eta A =C_z\rho_0 (\nu_\mathrm{k}A)\left(  \frac{V_\mathrm{A}^2}{\nu_\mathrm{k}A}\right) ^{5/3}=C_z\rho_0 \frac{V_\mathrm{A}^{10/3}}{\nu_\mathrm{k}^{2/3} A^{2/3}}\\
&& \tilde{Q}=A\sigma_{xy}=C_z\rho_0 \frac{V_\mathrm{A}^{10/3}A^{1/3}}{\nu_\mathrm{k}^{2/3}} = C_z\frac{B_0^{10/3}A^{1/3}}{\nu_\mathrm{k}^{2/3}\rho_0^{2/3}\mu_0^{5/3}}
\end{eqnarray*}

\textbf{Cold Plasma $\eta \approx 0$}

\begin{eqnarray}
 \eta_\mathrm{eff} = \frac{ \Gamma(\frac43)^2 6^{2/3}I_D }{128\pi^4} \rho\frac{V_\mathrm{A}^2}{A}\left( \frac { 6V_\mathrm{A}^2}{Ac^2\epsilon_0\rho_\Omega }\right)^{2/3} \\
 \sigma_{xy} = \eta_\mathrm{eff}A = \frac{ \Gamma(\frac43)^2 6^{2/3}I_D }{128\pi^4} \rho V_\mathrm{A}^{10/3} \left( \frac6{Ac^2\epsilon_0\rho_\Omega }\right)^{2/3} \ll P \\
 \tilde{Q}_\mathrm{tot}=  \frac{ \Gamma(\frac43)^2 6^{2/3}I_D }{128\pi^4} \rho V_\mathrm{A}^{10/3} A^{1/3} \left( \frac6{c^2\epsilon_0\rho_\Omega }\right)^{2/3}
\end{eqnarray}

Anomalous transport in accretion disks is created by self-sustained beams of SMW propagating in radial direction; figuratively lasing of alfvenons create compact astrophysical objects.

\section{Dimension analysis of MRI}

$$ \omega_\mathrm{c}=-\frac23,\quad \nu_\mathrm{tot} \rightarrow 0,\quad \partial_\tau w_\mathbf{Q}=0,\quad w_\mathbf{Q} \sim 1,\quad |Q| \sim 1$$

$$ \oint \mathbf{S}\cdot\mathrm{d}\mathbf{f} \sim 1,\quad \tilde{Q}=p_BA=\sigma A=\eta_\mathrm{eff}A^2,\quad V_\mathrm{A} \ll c_\mathrm{s},\quad p_B\ll p=nT$$
$$ p_B/p=1/\beta \ll 1 \qquad \mbox{high beta-plasmas} $$

\begin{equation}
 \sigma_{xy} \sim p_B,\quad \sigma_{xy}=C_\omega p_B=C_\omega \frac{p}{\beta}=\alpha_{s-s}p
\end{equation}

Conclusion: Dimension analysis shows that Shakura-Sunaev parameter is simple reciprocal of plasma beta.
$$ \alpha_{s-s}=C_w/\beta,\quad \mathcal{R}e=\frac{V_\mathrm{A}(\Lambda=V_\mathrm{A}/A)}{\nu},\quad \eta_\mathrm{eff}=p_B/A $$

$$ Z=\frac{\eta_\mathrm{eff}}{\eta}=\frac{B_0^2}{\mu_0\eta A}=\frac{B_0^2}{\mu_0 \rho \nu A}= \frac{V_\mathrm{A}}{\nu A} = \frac{C_z}{\nu_\mathrm{tot}^{\prime \kappa(\omega)}}$$

$$ \kappa(\omega_\mathrm{c}=0)=\frac53,\quad \kappa(\omega_\mathrm{c}=-\frac23)=1$$

\section{Self-Sustained Chain of Equation}

For the general 2D case the nonlinear terms in \Eqref{ch3bay_ivan} read:

\begin{eqnarray}
\mathbf{N}_v  = \Pi^{\perp\mathbf{Q}}\cdot\sum_{\mathbf{Q}'}
\left[
\left(\begin{array}{cc}
v_{\mathbf{Q}'}^xv_{\mathbf{Q}-\mathbf{Q}'}^x & v_{\mathbf{Q}'}^xv_{\mathbf{Q}-\mathbf{Q}'}^y  \\
v_{\mathbf{Q}'}^yv_{\mathbf{Q}-\mathbf{Q}'}^x & v_{\mathbf{Q}'}^yv_{\mathbf{Q}-\mathbf{Q}'}^y
\end{array} \right)
+
\left(\begin{array}{ccc}
b_{\mathbf{Q}'}^xb_{\mathbf{Q}-\mathbf{Q}'}^x & b_{\mathbf{Q}'}^xb_{\mathbf{Q}-\mathbf{Q}'}^y \\
b_{\mathbf{Q}'}^yb_{\mathbf{Q}-\mathbf{Q}'}^x & b_{\mathbf{Q}'}^yb_{\mathbf{Q}-\mathbf{Q}'}^y
\end{array} \right)
\right]
\cdot
\left(\begin{array}{c}
Q_x \\
Q_y
\end{array} \right)
\end{eqnarray}

\begin{eqnarray}
\mathbf{N}_b  = \Pi^{\perp\mathbf{Q}}\cdot\sum_{\mathbf{Q}'}
\left[
\left(\begin{array}{cc}
b_{\mathbf{Q}'}^xv_{\mathbf{Q}-\mathbf{Q}'}^x & b_{\mathbf{Q}'}^xv_{\mathbf{Q}-\mathbf{Q}'}^y \\
b_{\mathbf{Q}'}^yv_{\mathbf{Q}-\mathbf{Q}'}^x & b_{\mathbf{Q}'}^yv_{\mathbf{Q}-\mathbf{Q}'}^y
\end{array} \right)
-
\left(\begin{array}{cc}
v_{\mathbf{Q}'}^xb_{\mathbf{Q}-\mathbf{Q}'}^x & v_{\mathbf{Q}'}^xb_{\mathbf{Q}-\mathbf{Q}'}^y \\
v_{\mathbf{Q}'}^yb_{\mathbf{Q}-\mathbf{Q}'}^x & v_{\mathbf{Q}'}^yb_{\mathbf{Q}-\mathbf{Q}'}^y
\end{array} \right)
\right]
\cdot
\left(\begin{array}{c}
Q_x \\
Q_y
\end{array} \right),
\end{eqnarray}

where $\Pi^{\perp\mathbf{Q}}$ is the projection operator and have

\begin{eqnarray}
\Pi^{\perp\mathbf{Q}} = n_{yy} \left(\begin{array}{cc}
 1 & \tau \\
 \tau & \tau^2
\end{array} \right)
\end{eqnarray}

Taking into account the asymptotic

\begin{eqnarray}
 \lim\limits_{\tau \rightarrow \infty}{\mathbf{N}_b} &=&
\left(\begin{array}{c}
 0  \\
 \sum_{\mathbf{Q}'} (b_{\mathbf{Q}'}^yv_{\mathbf{Q}-\mathbf{Q}'}^y - v_{\mathbf{Q}'}^yb_{\mathbf{Q}-\mathbf{Q}'}^y)Q_y  \end{array} \right) \\
 \lim\limits_{\tau \rightarrow \infty}{\mathbf{N}_v} &=&
\left(\begin{array}{c}
 0  \\
 \sum_{\mathbf{Q}'} (v_{\mathbf{Q}'}^yv_{\mathbf{Q}-\mathbf{Q}'}^y + b_{\mathbf{Q}'}^yb_{\mathbf{Q}-\mathbf{Q}'}^y)Q_y  \end{array} \right)
\end{eqnarray}

To calculate the time derive of y component of nonlinear term $\mathbf{N}_b$ we use relation

\begin{eqnarray*}
&&\mathrm{d}_{\overline{\tau}}N_b^y = \mathrm{d}_{\overline{\tau}}\left[\sum_{\mathbf{Q}'} (b_{\mathbf{Q}'}^yv_{\mathbf{Q}-\mathbf{Q}'}^y - v_{\mathbf{Q}'}^yb_{\mathbf{Q}-\mathbf{Q}'}^y)Q_y \right]=[(b_{\mathbf{Q}'}^yb_{\mathbf{Q}-\mathbf{Q}'}^y + v_{\mathbf{Q}'}^yv_{\mathbf{Q}-\mathbf{Q}'}^y)(Q_y-2Q'_y)]Q_y ,  \\
&&\mathrm{d}_{\overline{\tau}} v_{\mathbf{Q}-\mathbf{Q}'}^y = (Q_y-Q'_y)b_{\mathbf{Q}-\mathbf{Q}'}^y,\quad \mathrm{d}_{\overline{\tau}} b_{\mathbf{Q}-\mathbf{Q}'}^y = -(Q_y-Q'_y)v_{\mathbf{Q}-\mathbf{Q}'}^y \nn
\end{eqnarray*}

The resulting external force acting on the out oscillator we have

\begin{eqnarray}
F_\mathrm{ext} &=& \mathrm{d}_{\overline{\tau}}N_b^y - Q_yN_v^y + N_b^yQ^2\nu'_\mathrm{k} + N_b^x  \\
 && = \sum_{\mathbf{Q}'}\left[(b_{\mathbf{Q}'}^yb_{\mathbf{Q}-\mathbf{Q}'}^y + v_{\mathbf{Q}'}^yv_{\mathbf{Q}-\mathbf{Q}'}^y)(Q_y^2-2Q'_yQ_y) - Q_y^2(v_{\mathbf{Q}'}^yv_{\mathbf{Q}-\mathbf{Q}'}^y + b_{\mathbf{Q}'}^yb_{\mathbf{Q}-\mathbf{Q}'}^y) + Q_y^3\overline{\tau}^2\nu'_\mathrm{k} (b_{\mathbf{Q}'}^yv_{\mathbf{Q}-\mathbf{Q}'}^y - v_{\mathbf{Q}'}^yb_{\mathbf{Q}-\mathbf{Q}'}^y)\right]  \nn \\
 && = \sum_{\mathbf{Q}'}\left[Q_y^3\overline{\tau}^2\nu'_\mathrm{k}(b_{\mathbf{Q}'}^yv_{\mathbf{Q}-\mathbf{Q}'}^y - v_{\mathbf{Q}'}^yb_{\mathbf{Q}-\mathbf{Q}'}^y) -2Q'_yQ_y(b_{\mathbf{Q}'}^yb_{\mathbf{Q}-\mathbf{Q}'}^y + v_{\mathbf{Q}'}^yv_{\mathbf{Q}-\mathbf{Q}'}^y)\right]. \nn
\end{eqnarray}

\begin{equation}
\lim\limits_{\nu'_\mathrm{k} \rightarrow 0}{\left(\frac{\int_0^\infty\nu'_\mathrm{k}\overline{\tau}^2e^{-\nu'_\mathrm{tot} Q_y^2\overline{\tau}^2/6}\mathrm{d}\overline{\tau}}{\int_0^\infty e^{-\nu'_\mathrm{tot} Q_y^2\overline{\tau}^2/6}\mathrm{d}\overline{\tau}} = \frac{6^{2/3}}{3\Gamma(\frac43)} \frac{\nu_\mathrm{k}^{_\prime 1/3}}{Q_y^{4/3}(1+\frac{1}{\mathrm{P}_\mathrm{m}})^{2/3} }\right)}  = 0
\end{equation}

Finaly the result in limit of hot plasma is

\begin{equation}
F_\mathrm{ext} \approx -2Q_y \sum_{\mathbf{Q}'}Q'_y\left(b_{\mathbf{Q}'}^yb_{\mathbf{Q}-\mathbf{Q}'}^y + v_{\mathbf{Q}'}^yv_{\mathbf{Q}-\mathbf{Q}'}^y\right)
\end{equation}

\begin{equation*}
b_y(\overline{\tau},Q_y)=
\int_{-\infty}^{\overline{\tau}} \frac1{Q_y}\sin[Q_y(\overline{\tau}-\overline{\tau}_0)]
\exp\left(-\frac12\int^{\overline{\tau}}_{\overline{\tau}_0} \mathrm{d}\tau \gamma(\tau) \right) \theta(\overline{\tau}-\overline{\tau}_0) 
\cdot -2Q_y \sum_{\mathbf{Q}'}Q'_y\left(b_{\mathbf{Q}'}^yb_{\mathbf{Q}-\mathbf{Q}'}^y + v_{\mathbf{Q}'}^yv_{\mathbf{Q}-\mathbf{Q}'}^y\right) \mathrm{d}\overline{\tau_0}
\end{equation*}


Now we will represent $b_y$ as a sum sine and cosine terms $ b_y(\overline{\tau},Q_y)= \sin(Q_y\overline{\tau})X(\overline{\tau}) - \cos(Q_y\overline{\tau})Y(\overline{\tau}) $

\begin{eqnarray*}
b_y(\overline{\tau},Q_y)=\int_{-\infty}^\infty\frac1{Q_y}\sin[Q_y\overline{\tau} -Q_y\overline{\tau_0}]\exp\left[-\frac{\nu^{\prime}_{tot}}{6}(\overline{\tau}^3 - \overline{\tau_0}^3)W_y^2\right]\theta(\overline{\tau}-\overline{\tau_0})\mathbf{N}_{bb}^{yy}(Qx,Qy)\mathrm{d}\overline{\tau_0}\nn \\
= \sin(Q_y\overline{\tau})\left\{\int^{\overline{\tau}}_{-\infty}\mathrm{d}\overline{\tau_0} \frac1{Q_y}\cos(Q_y\overline{\tau_0})\exp\left[ -\frac{\nu^{\prime}_{tot}}{6}(\overline{\tau}^3 - \overline{\tau_0}^3) \right] \mathbf{N}_{bb}^{yy}(\mathbf{Q})\equiv Y(\overline{\tau}) \right\}\nn \\
- \cos(Q_y\overline{\tau})\left\{\int^{\overline{\tau}}_{-\infty}\mathrm{d}\overline{\tau_0} \frac1{Q_y}\sin(Q_y\overline{\tau_0})\exp\left[ -\frac{\nu^{\prime}_{tot}}{6}(\overline{\tau}^3 - \overline{\tau_0}^3) \right] \mathbf{N}_{bb}^{yy}(\mathbf{Q})\equiv-X(\overline{\tau}) \right\}\nn \\
=\cos(Q_y\overline{\tau})X(\overline{\tau}) + \sin(Q_y\overline{\tau})Y(\overline{\tau}) \approx X_0 \cos(Q_y\overline{\tau}) + Y_0\sin(Q_y\overline{\tau})
\end{eqnarray*}

We can assume that the amplitude of the amplified wave as a function of $\overline{\tau}$ is approximately equal to amplitude in $\overline{\tau_0}=0$.

$$ WKB \rightarrow X(\overline{\tau}) \approx X(0),\quad Y(\overline{\tau}) \approx Y(0); \qquad \overline{\tau}-\overline{\tau}_0=0 $$

\begin{eqnarray}
X(0)=X_0=-\int^{0}_{-\infty}\mathrm{d}\overline{\tau_0}\frac1{Q_y}\sin(Q_y\overline{\tau_0})\exp\left[-\frac{\nu^{\prime}_{tot}}{6} (-\overline{\tau})^3Q_y^2 \right] \mathbf{N}_{bb}^{yy}(\mathbf{Q}) \\
Y_0=Y(\overline{\tau}=0)=\int^{0}_{-\infty}\mathrm{d}\overline{\tau_0}\frac1{Q_y}\cos(Q_y\overline{\tau_0})\exp\left[-\frac{\nu^{\prime}_{tot}}{6} (-\overline{\tau})^3Q_y^2 \right] \mathbf{N}_{bb}^{yy}(\mathbf{Q})
\end{eqnarray}

\begin{eqnarray}
 X(0) = -\int_{-\infty}^{0}\ \frac{\mathrm{d}Q_x}{Q_y} \sin(Q_x)\exp\left[-\frac{\nu^{\prime}_{tot}}{6}\overline{\tau}^3Q_y^2\right]
 \cdot \sum_{\mathbf{Q}'}Q'_y\left(b_{\mathbf{Q}'}^yb_{\mathbf{Q}-\mathbf{Q}'}^y + v_{\mathbf{Q}'}^yv_{\mathbf{Q}-\mathbf{Q}'}^y\right) \nn\\
 Y(0) = \int_{-\infty}^{0}  \frac{\mathrm{d}Q_x}{Q_y} \cos(Q_x) \exp\left[-\frac{\nu^{\prime}_{tot}}{6}\overline{\tau}^3 Q_y^2\right]
 \cdot \sum_{\mathbf{Q}'}Q'_y\left(b_{\mathbf{Q}'}^yb_{\mathbf{Q}-\mathbf{Q}'}^y + v_{\mathbf{Q}'}^yv_{\mathbf{Q}-\mathbf{Q}'}^y\right)
\end{eqnarray}

\begin{center}

\begin{tikzpicture}[
    media/.style={font={\footnotesize\sffamily}},
    wave/.style={decorate,decoration={snake,post length=1.4mm,amplitude=2mm, segment length=2mm},thick},
    awave/.style={decorate,decoration={snake,post length=1.4mm,amplitude=6mm, segment length=1.3mm},thick},
    interface/.style={ postaction={draw,decorate,decoration={border,angle=-45, amplitude=0.3cm,segment length=2mm}}},
    scale=1.5]

    \fill[gray!10] (0,3) rectangle (4,0);
    \fill[gray!10] (-4,-3) rectangle (0,0);

     \draw[dashed,gray](0,-3)--(0,3);
     \draw[dashed,gray](-4,0)--(4,0);

     \coordinate (Origin)   at (0,0);

    \coordinate (Bone) at (-0.8,1.2);
    \coordinate (Btwo) at (1.2,-0.8);

    \draw [ultra thick,-latex,red] (Origin) -- (Bone) node [below left] {$Q_1$};
    \draw [ultra thick,-latex,red] (Origin) -- (Btwo) node [above right] {$Q_2$};
    \draw [ultra thick,-latex,red] (Origin) -- (0.8,0.8) node [above right] {$Q_1+Q_2$};

    \draw [style=help lines,dashed,blue] (-1,3) -- (1,-3) node [below left] {$A$};
    \draw [ultra thick,-latex,blue] (0,2.7) -- (-0.9,2.7);
    \draw [ultra thick,-latex,blue] (0,2) -- (-0.7,2);
    \draw [ultra thick,-latex,blue] (0,1.3) -- (-0.45,1.3);

    \draw [ultra thick,-latex,blue] (0,-2.7) -- (0.9,-2.7);
    \draw [ultra thick,-latex,blue] (0,-2) -- (0.7,-2);
    \draw [ultra thick,-latex,blue] (0,-1.3) -- (0.45,-1.3);

    \draw[<->,line width=1pt] (4,0) node[above]{$Q_x$}-|(0,3) node[right]{$Q_y$};
    \draw[->,awave] (-1,2.2)--(-3,2.2)node[left]{$D^f$};
    \draw[->,wave]   (3,2.2)--(1,2.2)node[above]{$D^0$};
    \draw[->,wave] (-3,-2.2)--(-1,-2.2)node[above]{$D^0$};
    \draw[->,awave] (1,-2.2)--(3,-2.2)node[right]{$D^f$};

    \path[media] (-3,.6)  node {Amplified waves}
		  (3,-.6)  node {Amplified waves}
		  (3,.6)  node {Seeds}
                 (-3,-.6) node {Seeds};

    \filldraw[fill=white,line width=1pt](0,0)circle(.12cm);
    \draw[line width=.6pt] (0,0)
                          +(-135:.12cm) -- +(45:.12cm)
                          +(-45:.12cm) -- +(135:.12cm);

\end{tikzpicture}
\end{center}

In calculation of wave-wave interaction we will use the following scheme:

Amplified wave with wave-vector $\mathbf{Q}_1$ interact with another amplified wave with wave-vector $\mathbf{Q}_2$ and in the result we have wave with wave-vector $\mathbf{Q}=\mathbf{Q}_1+\mathbf{Q}_2$.

Domain of amplification
$$ Q_1^x < 0,\qquad Q_2^x > 0,\qquad Q_1^y > 0,\qquad Q_2^y < 0$$

$$ Q_1^x = \frac{Q_x}2 + P_x,\qquad Q_2^x = \frac{Q_x}2 - P_x,\qquad Q_1^y = \frac{Q_y}2 + P_y,\qquad Q_2^y = \frac{Q_y}2 - P_y $$

$$\mathbf{P} \rightarrow \mathbf{K} , (P_x,P_y)=(-K_x,K_y) $$

\begin{eqnarray}
Q_1^x = -\left( K_x - \frac{Q_x}2  \right) < 0,\qquad Q_2^x = +\left( K_x + \frac{Q_x}2  \right) > 0, \\
Q_1^y = +\left( K_y + \frac{Q_y}2  \right) > 0,\qquad Q_2^y = -\left( K_y - \frac{Q_y}2  \right) < 0 \\
K_x \subset \left(\frac{Q_x}2,\infty\right), \qquad K_y \subset \left(\frac{Q_y}2,\infty\right)
\end{eqnarray}

After change of variables we have

\begin{equation}
\label{I_initial}
 I=\int_{Q_x/2}^\infty \frac{\mathrm{d}K_x}{2\pi} \int_{Q_y/2}^\infty \frac{\mathrm{d}K_y}{2\pi} \, Q_1^y \left(b^y_{Q_1}b^y_{Q_2} + v^y_{Q_1}v^y_{Q_2}\right).
\end{equation}

Let us introduce some short notations

\begin{eqnarray}
 D^+ &\equiv& D(Q^y_1)= D\left(K_y+\frac{Q_y}{2}\right),\quad D^-\equiv D(Q^y_2) = D\left(\frac{Q_y}{2}-K_y\right), \nn \\
\phi^+ &\equiv& \phi(Q_1^y)=\phi\left(K_y+\frac{Q_y}{2}\right),\quad  \phi^-\equiv \phi(Q_2^y)=\left(\frac{Q_y}{2}-K_y\right).
\end{eqnarray}

Then in explicit form we have for $b^y_{Q_1}b^y_{Q_2}$
\begin{equation*}
   D^+\cos\left[Q^y_1\overline{\tau_1}+\phi^+\right]\exp[-\frac{\nu'_{tot}}{6}(Q_1^y)^2\overline{\tau_1}^3] \cdot
		       D^-\cos\left[Q^y_2\overline{\tau_2}+\phi^-\right]\exp[-\frac{\nu'_{tot}}{6}(Q_2^y)^2\overline{\tau_2}^3],
\end{equation*}
and respectively for $ v^y_{Q_1}v^y_{Q_2}$
\begin{equation*}
 D^+\sin\left[Q^y_1\overline{\tau_1}+\phi^+\right]\exp[-\frac{\nu'_{tot}}{6}(Q_1^y)^2\overline{\tau_1}^3] \cdot
 D^-\sin\left[Q^y_2\overline{\tau_2}+\phi^-\right]\exp[-\frac{\nu'_{tot}}{6}(Q_2^y)^2\overline{\tau_2}^3].
\end{equation*}

The argument of exponential function in both of relations above is strictly negative
\begin{eqnarray*}
 (Q_1^y)^2\overline{\tau_1}^3 &=& -\frac{(Q_1^x)^3}{Q_1^y} = \left( \frac{(K_x-Q_x/2)^3}{K_y+Q_y/2}\right) > 0,\\
 (Q_2^y)^2\overline{\tau_2}^3 &=& -\frac{(Q_2^x)^3}{Q_2^y} = \left( \frac{(K_x+Q_x/2)^3}{K_y-Q_y/2}\right) > 0.
\end{eqnarray*}

The resulting exponent function in expression $b^y_{Q_1}b^y_{Q_2} + v^y_{Q_1}v^y_{Q_2}$ is
\begin{equation*}
 \exp[-\frac{\nu'_{tot}}{6}(Q_1^y)^2\overline{\tau_1}^3] \cdot \exp[-\frac{\nu'_{tot}}{6}(Q_2^y)^2\overline{\tau_2}^3] =
 \exp\left\{ -\frac{\nu'_{tot}}{6}\left[\frac{(K_x-Q_x/2)^3}{K_y+Q_y/2}+\frac{(K_x+Q_x/2)^3}{K_y-Q_y/2}\right]\right\}
\end{equation*}

To calculate trigonometrical part of \Eqref{I_initial} we use formula
$$ \cos(\alpha)\cos(\beta) + \sin(\alpha)\sin(\beta) = \cos(\alpha - \beta), $$

and for out case obtain

\begin{eqnarray*}
&&\cos\left(Q^y_1\overline{\tau_1}+\phi^+\right)\cos\left(Q^y_2\overline{\tau_2}+\phi^-\right) + \sin\left(Q^y_1\overline{\tau_1}+\phi^+\right)\sin\left(Q^y_2\overline{\tau_2}+\phi^-\right) = \cos[ 2K_x + \phi^+ - \phi^-] \\
&&=\cos(2K_x)\left\{\cos\left[\phi(Q_1^y)\right] \cos[\phi(Q_2^y)] + \sin[\phi(Q_1^y)]\sin[\phi(Q_2^y)] \right\} \\
&\quad& -\sin(2K_x) \left\{\sin\left[\phi(Q_1^y)\right] \cos[\phi(Q_2^y)] - \cos[\phi(Q_1^y)]\sin[\phi(Q_2^y)] \right\}
\end{eqnarray*}

Finlay we can assemble all calculated terms for $  Q_1^y \left(b^y_{Q_1}b^y_{Q_2} + v^y_{Q_1}v^y_{Q_2}\right) $
\begin{eqnarray}
 Q_1^y D^+ D^- \cos[ 2K_x + \phi^+ - \phi^-] \exp\left\{ -\frac{\nu'_{tot}}{6}\left[\frac{(K_x-Q_x/2)^3}{K_y+Q_y/2}+\frac{(K_x+Q_x/2)^3}{K_y-Q_y/2}\right]\right\} \nn
\end{eqnarray}

Now our expression for I is
\begin{eqnarray*}
I= && \int_{Q_x/2}^\infty \exp\left\{ -\frac{\nu'_{tot}}{6}\left[\frac{(K_x-Q_x/2)^3}{K_y+Q_y/2}+\frac{(K_x+Q_x/2)^3}{K_y-Q_y/2}\right]\right\} \cos(2K_x) \frac{\mathrm{d}K_x}{2\pi} \cdot \\
   && \int_{Q_y/2}^\infty \frac{\mathrm{d}K_y}{2\pi} Q_1^y D^+ D^- \left\{\cos\phi^+  \cos\phi^- + \sin\phi^+\sin\phi^- \right\} \\
   &-&\int_{Q_x/2}^\infty \exp\left\{ -\frac{\nu'_{tot}}{6}\left[\frac{(K_x-Q_x/2)^3}{K_y+Q_y/2}+\frac{(K_x+Q_x/2)^3}{K_y-Q_y/2}\right]\right\} \sin(2K_x) \frac{\mathrm{d}K_x}{2\pi} \cdot \\
   && \int_{Q_y/2}^\infty \frac{\mathrm{d}K_y}{2\pi} Q_1^y D^+ D^- \left\{\sin\phi^+  \cos\phi^- - \cos\phi^+\sin\phi^- \right\}
\end{eqnarray*}

In order to extract the viscosity outside of the integral we introduce following variables

$$ K_x \equiv \left(\frac{6}{\nu'_\mathrm{tot}}\right)^{1/3} x, \qquad Q_x \equiv \left(\frac{6}{\nu'_\mathrm{tot}}\right)^{1/3} q_x,\quad \omega=2\left(\frac{6}{\nu'_\mathrm{tot}}\right)^{1/3}. $$

The integral over $\mathrm{d}K$ can be represent as a sum of fast oscillations integral form $0$ to $\infty$ and simple integral of trigonometrical function in finite interval.
In the second integral exponential function doesn't affect the result because the viscosity $\nu'_\mathrm{tot}$ tends to zero.

\begin{eqnarray}
&& \int_{Q_x/2}^\infty \exp\left\{ -\frac{\nu'_{tot}}{6}\left[\frac{(K_x-Q_x/2)^3}{K_y+Q_y/2}+\frac{(K_x+Q_x/2)^3}{K_y-Q_y/2}\right]\right\} \cos(2K_x) \frac{\mathrm{d}K_x}{2\pi} =\\
&& \frac1{2\pi}\left(\frac{6}{\nu'_\mathrm{tot}}\right)^{1/3}\int_{0}^\infty \exp\left\{-\left[\frac{(x-q_x/2)^3}{K_y+Q_y/2}+\frac{(x+q_x/2)^3}{K_y-Q_y/2}\right]\right\} \cos(2\omega x) \mathrm{d}x - \nn \\
&& \int_0^{Q_x/2} \exp\left\{ -\frac{\nu'_{tot}}{6}\left[\frac{(K_x-Q_x/2)^3}{K_y+Q_y/2}+\frac{(K_x+Q_x/2)^3}{K_y-Q_y/2}\right]\right\} \cos(2K_x) \frac{\mathrm{d}K_x}{2\pi}\nn
\end{eqnarray}

\begin{eqnarray}
&& \lim_{\omega \rightarrow \infty} \left\{ \frac1{2\pi}\left(\frac{6}{\nu'_\mathrm{tot}}\right)^{1/3}\int_{0}^\infty \exp\left\{-\left[\frac{(x-q_x/2)^3}{K_y+Q_y/2}+\frac{(x+q_x/2)^3}{K_y-Q_y/2}\right]\right\} \cos(\omega x) \mathrm{d}x  \right\} = \nn \\
&& \frac1{8\pi} \left(\frac{\nu'_\mathrm{tot}}{6}\right)^{1/3} \left. \frac{\mathrm{d}}{\mathrm{d}x}  \left(\exp\left\{-\left[\frac{(x-q_x/2)^3}{K_y+Q_y/2}+\frac{(x+q_x/2)^3}{K_y-Q_y/2}\right]\right\}\right) \, \right|_0 = 0
\end{eqnarray}

\begin{eqnarray}
&&\lim_{\nu'_{tot} \rightarrow 0} \left\{ \int_0^{Q_x/2} \exp\left\{ -\frac{\nu'_{tot}}{6}\left[\frac{(K_x-Q_x/2)^3}{K_y+Q_y/2}+\frac{(K_x+Q_x/2)^3}{K_y-Q_y/2}\right]\right\} \cos(2K_x) \frac{\mathrm{d}K_x}{2\pi} \right\} = \nn\\
&&\int_0^{Q_x/2} \cos(2K_x) \frac{\mathrm{d}K_x}{2\pi} = \frac1{4\pi}\sin(Q_x)
\end{eqnarray}

\begin{eqnarray}
&& \lim_{\omega \rightarrow \infty} \left\{ \frac1{2\pi}\left(\frac{6}{\nu'_\mathrm{tot}}\right)^{1/3}\int_{0}^\infty \exp\left\{-\left[\frac{(x-q_x/2)^3}{K_y+Q_y/2}+\frac{(x+q_x/2)^3}{K_y-Q_y/2}\right]\right\} \sin(\omega x) \mathrm{d}x  \right\} = \nn \\
&& \frac1{8\pi}  \left(\exp\left\{-\left[\frac{(-q_x/2)^3}{K_y+Q_y/2}+\frac{(+q_x/2)^3}{K_y-Q_y/2}\right]\right\}\right)=\frac1{8\pi} \exp\left\{-q_x^3 \frac{Q_y}{2(4K_y^2-Q_y^2)} \right\}
\end{eqnarray}

\begin{eqnarray}
&&\lim_{\nu'_{tot} \rightarrow 0} \left\{ \int_0^{Q_x/2} \exp\left\{ -\frac{\nu'_{tot}}{6}\left[\frac{(K_x-Q_x/2)^3}{K_y+Q_y/2}+\frac{(K_x+Q_x/2)^3}{K_y-Q_y/2}\right]\right\} \sin(2K_x) \frac{\mathrm{d}K_x}{2\pi} \right\} = \nn\\
&&\int_0^{Q_x/2} \sin(2K_x) \frac{\mathrm{d}K_x}{2\pi} = \frac1{4\pi} \left[\cos(Q_x)-1\right]
\end{eqnarray}

Finlay we can assemble previously calculated results for I
\begin{eqnarray}
I= && -\frac1{4\pi}\sin(Q_x) \int_{Q_y/2}^\infty \frac{\mathrm{d}K_y}{2\pi} Q_1^y D^+ D^- \left\{\cos\phi^+  \cos\phi^- + \sin\phi^+\sin\phi^- \right\}\nn +\\
   && \frac1{4\pi}\left[\cos(Q_x)-1\right]\int_{Q_y/2}^\infty \frac{\mathrm{d}K_y}{2\pi} Q_1^y D^+ D^- \left\{\sin\phi^+  \cos\phi^- - \cos\phi^+\sin\phi^- \right\}\nn
\end{eqnarray}

\begin{eqnarray}
 X &=& \frac1{4\pi Q_y}\left(\frac6{\nu'_\mathrm{tot}}\right)^{1/3} \int_0^\infty \frac12 \exp\left\{-\frac{q_x^3}{Q_y} \right\} \mathrm{d}q_x \int_{Q_y/2}^\infty \frac{\mathrm{d}K_y}{2\pi} Q_1^y D^+ D^- \left\{\cos\phi^+  \cos\phi^- + \sin\phi^+\sin\phi^- \right\}\nn \\
 Y &=& \frac1{4\pi Q_y}\left(\frac6{\nu'_\mathrm{tot}}\right)^{1/3} \int_0^\infty \frac12 \exp\left\{-\frac{q_x^3}{Q_y} \right\} \mathrm{d}q_x \int_{Q_y/2}^\infty \frac{\mathrm{d}K_y}{2\pi} Q_1^y D^+ D^- \left\{\sin\phi^+  \cos\phi^- - \cos\phi^+\sin\phi^- \right\}\nn
\end{eqnarray}

After integration over $\mathrm{d}q_x$ we have

\begin{eqnarray}
 X &=& \frac{\Gamma(\frac43)}{16\pi Q_y^{2/3}}\left(\frac6{\nu'_\mathrm{tot}}\right)^{1/3} \int_{Q_y/2}^\infty Q_1^y D^+ D^- \left\{\cos\phi^+  \cos\phi^- + \sin\phi^+\sin\phi^- \right\} \mathrm{d}K_y \nn \\
 Y &=& \frac{\Gamma(\frac43)}{16\pi Q_y^{2/3}}\left(\frac6{\nu'_\mathrm{tot}}\right)^{1/3} \int_{Q_y/2}^\infty Q_1^y D^+ D^- \left\{\sin\phi^+  \cos\phi^- - \cos\phi^+\sin\phi^- \right\} \mathrm{d}K_y \nn
\end{eqnarray}

$$ R \equiv \frac{\Gamma(\frac43)}{16\pi}\left(\frac6{\nu'_\mathrm{tot}}\right)^{1/3},\quad \tilde{X} \equiv XR,\quad \tilde{Y} \equiv YR,\quad \tilde{D} \equiv DR. $$
Here we us exclude viscosity in our equation using scale transformation with $R$

\begin{eqnarray}
 \tilde{X}_\mathrm{i}(Q_y) &=& \int_{Q_y/2}^\infty \left(\frac{K_y}{Q_y^{2/3}} + \frac{Q_y^{1/3}}2\right) \tilde{D}_+ \tilde{D}_- \left[\cos\phi_+  \cos\phi_- + \sin\phi_+\sin\phi_- \right] \mathrm{d}K_y  \\
 \tilde{Y}_\mathrm{i}(Q_y) &=& \int_{Q_y/2}^\infty \left(\frac{K_y}{Q_y^{2/3}} + \frac{Q_y^{1/3}}2\right) \tilde{D}_+ \tilde{D}_- \left[\sin\phi_+  \cos\phi_- - \cos\phi_+\sin\phi_- \right] \mathrm{d}K_y
\end{eqnarray}

\begin{equation}
\nu_\mathrm{wave}  \approx\frac{R}{4\pi^2}\int_0^\infty Q_y \tilde{D}_\mathrm{f}^2(Q_y)\,\mathrm{d}Q_y,\quad I_D=\int_0^\infty Q_y \tilde{D}_\mathrm{f}^2(Q_y)\,\mathrm{d}Q_y
\end{equation}

\begin{figure}
\hspace{-2cm}\includegraphics[height=10cm]{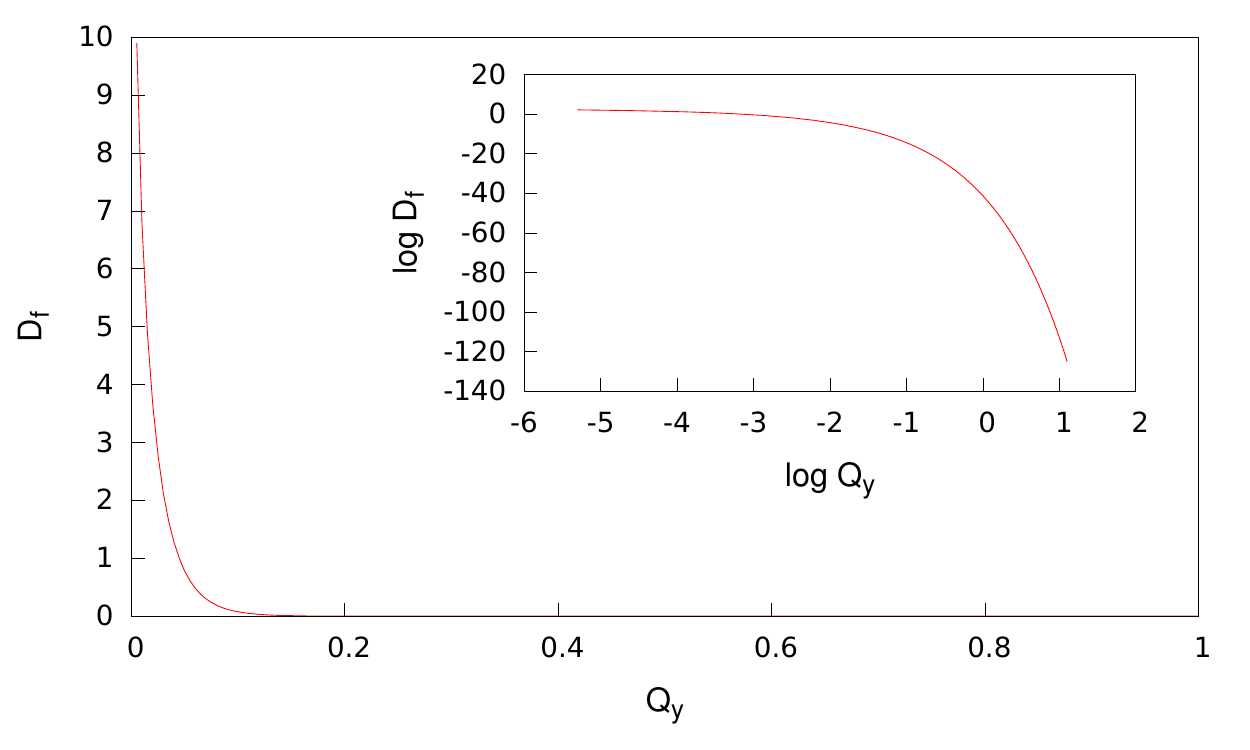}
\caption{$D_f(Q_y)$}
\end{figure}

\begin{figure}
\hspace{-2cm}\includegraphics[height=10cm]{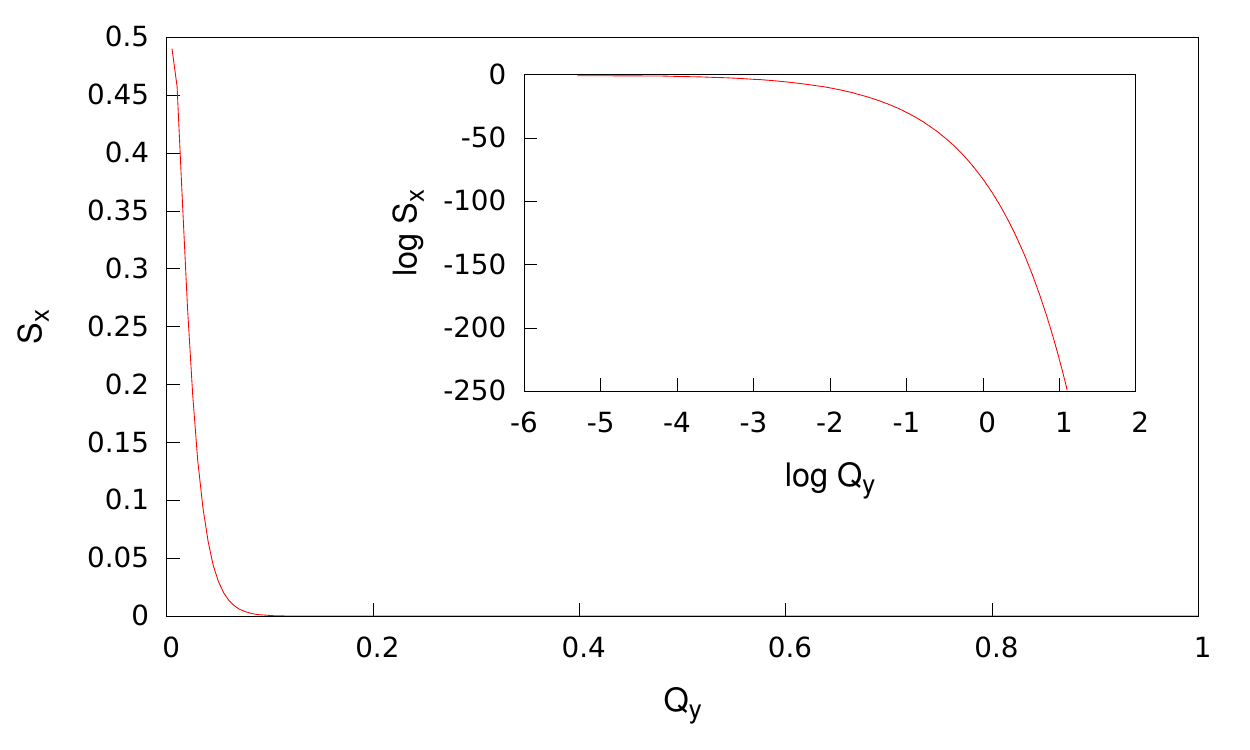}
\caption{$S_x(Q_y)$}
\end{figure}

We perform numerical analysis of the solution of the general MHD equations in the simplest case of 2D non-rotating plasma.
The solution obtained in approximation of small viscosity or large Reynolds numbers $\mathcal{R}$ show that self-sustained wave turbulence consist of two narrow beams with high wave density in $Q$-space.
For large enough Reynolds numbers this beams or jets are converted in rays of magnetohydrodynamic waves in $Q$-space.
We believe that this bright expressed phenomena qualitatively will be preserved in 3D  case when rotation of plasma is taking into account.
On this way solving of the MHD equations will reveal mechanism of anomalous transport in magnetized space plasmas.
It is matter of numerical analysis to check, that giant effective viscosity and friction forces in accretion disks is determined by lasing of alfvenons in radial direction.

\section{3D rotation case with Coriolis force}

\begin{eqnarray}
 &&\mathrm{d}^2_{\overline{\tau}}b_x + \left[\nu'_\mathrm{tot}Q^2-2n_xn_y(1+\omega_c)\right]\mathrm{d}_{\overline{\tau}}b_x +\left[Q^2_\alpha +2\tau \nu'_\mathrm{m}Q_y^2(1+\omega_c) +\nu'_\mathrm{k}\nu'_\mathrm{m}Q_y^4 \right]b_x\nonumber \\
 &&=\mathrm{d}_{\overline{\tau}}\mathbf{N}_b^x - Q_\alpha \mathbf{N}_v^x + 2\omega_c Q_\alpha v_y(n_xn_x +1) - \mathbf{N}_b^x[2n_xn_y(1+\omega_c) -\nu'_\mathrm{k}Q^2] \\
 &&\mathrm{d}^2_{\overline{\tau}}b_y + \left[\nu'_\mathrm{tot}Q^2+2n_xn_y\omega_c\right]\mathrm{d}_{\overline{\tau}}b_y +\left[Q^2_\alpha +2\tau \nu'_\mathrm{m}Q_y^2(1-\omega_c) +\nu'_\mathrm{k}\nu'_\mathrm{m}Q_y^4 \right]b_y\nonumber \\
 &&=\mathbf{N}_b^y - \mathbf{N}_v^yQ_\alpha + N_b^y(\nu'_\mathrm{k}Q^2+2\omega_c n_xn_y) + \mathbf{N}_b^x - \nu'_\mathrm{m}Q^2b_x -2Q_\alpha v_x[n_xn_y + \omega_c(n_yn_y-1)] \nonumber
\end{eqnarray}


Obviously the approximate solution \Eqref{HeunW} can't be use for further analytical analysis.
We can obtain asymptotic for the approximate solution in terms of spherical Bessel function of the second kind,
hypergeometric function, but using just sine asymptotic with taking into account phase and amplitude
is better way to solve the problem because we will have universal description and technique in terms of simple trigonometrical functions.

\begin{eqnarray}
 \lim\limits_{\tau \rightarrow \infty}{\mathbf{N}_b} &=&
\left(\begin{array}{c}
 0  \\
 \sum_{\mathbf{Q}'} (b_{\mathbf{Q}'}^yv_{\mathbf{Q}-\mathbf{Q}'}^y - v_{\mathbf{Q}'}^yb_{\mathbf{Q}-\mathbf{Q}'}^y)Q_y + (b_{\mathbf{Q}'}^yv_{\mathbf{Q}-\mathbf{Q}'}^z - v_{\mathbf{Q}'}^yb_{\mathbf{Q}-\mathbf{Q}'}^z)Q_z  \\
 \sum_{\mathbf{Q}'} (b_{\mathbf{Q}'}^zv_{\mathbf{Q}-\mathbf{Q}'}^y - v_{\mathbf{Q}'}^zb_{\mathbf{Q}-\mathbf{Q}'}^y)Q_y + (b_{\mathbf{Q}'}^zv_{\mathbf{Q}-\mathbf{Q}'}^z - v_{\mathbf{Q}'}^zb_{\mathbf{Q}-\mathbf{Q}'}^z)Q_z \end{array} \right) \\
  \lim\limits_{\tau \rightarrow \infty}{\mathbf{N}_v} &=&
\left(\begin{array}{c}
 0  \\
 \sum_{\mathbf{Q}'} (v_{\mathbf{Q}'}^yv_{\mathbf{Q}-\mathbf{Q}'}^y + b_{\mathbf{Q}'}^yb_{\mathbf{Q}-\mathbf{Q}'}^y)Q_y + (v_{\mathbf{Q}'}^yv_{\mathbf{Q}-\mathbf{Q}'}^z + b_{\mathbf{Q}'}^yb_{\mathbf{Q}-\mathbf{Q}'}^z)Q_z \\
 \sum_{\mathbf{Q}'} (v_{\mathbf{Q}'}^zv_{\mathbf{Q}-\mathbf{Q}'}^y + b_{\mathbf{Q}'}^zb_{\mathbf{Q}-\mathbf{Q}'}^y)Q_y + (v_{\mathbf{Q}'}^zv_{\mathbf{Q}-\mathbf{Q}'}^z + b_{\mathbf{Q}'}^zb_{\mathbf{Q}-\mathbf{Q}'}^z)Q_z \end{array} \right)
\end{eqnarray}

To calculate the time derive of y component of nonlinear term $\mathbf{N}_b$ we use relation

\begin{eqnarray*}
&&\mathrm{d}_{\overline{\tau}}N_b^y = \mathrm{d}_{\overline{\tau}}\left[\sum_{\mathbf{Q}'} (b_{\mathbf{Q}'}^yv_{\mathbf{Q}-\mathbf{Q}'}^y - v_{\mathbf{Q}'}^yb_{\mathbf{Q}-\mathbf{Q}'}^y)Q_y + (b_{\mathbf{Q}'}^yv_{\mathbf{Q}-\mathbf{Q}'}^z - v_{\mathbf{Q}'}^yb_{\mathbf{Q}-\mathbf{Q}'}^z)Q_z \right] \\
&& =\sum_{\mathbf{Q}'} [(b_{\mathbf{Q}'}^yb_{\mathbf{Q}-\mathbf{Q}'}^y + v_{\mathbf{Q}'}^yv_{\mathbf{Q}-\mathbf{Q}'}^y)Q_y + (b_{\mathbf{Q}'}^yb_{\mathbf{Q}-\mathbf{Q}'}^z + v_{\mathbf{Q}'}^yv_{\mathbf{Q}-\mathbf{Q}'}^z)Q_z](Q_\alpha-2Q'_\alpha) ,  \\
&&\mathrm{d}_{\overline{\tau}} v_{\mathbf{Q}-\mathbf{Q}'}^y = (Q_\alpha-Q'_\alpha)b_{\mathbf{Q}-\mathbf{Q}'}^y,\quad \mathrm{d}_{\overline{\tau}} b_{\mathbf{Q}-\mathbf{Q}'}^y = -(Q_\alpha-Q'_\alpha)v_{\mathbf{Q}-\mathbf{Q}'}^y \nn
\end{eqnarray*}

The resulting external force acting on the out oscillator we have

\begin{eqnarray}
F_\mathrm{ext} &=& \mathrm{d}_{\overline{\tau}}N_b^y - Q_\alpha N_v^y  \\
&&= -2\sum_{\mathbf{Q}'} Q'_{\alpha} \left[(b_{\mathbf{Q}'}^yb_{\mathbf{Q}-\mathbf{Q}'}^y + v_{\mathbf{Q}'}^yv_{\mathbf{Q}-\mathbf{Q}'}^y)Q_y + (b_{\mathbf{Q}'}^yb_{\mathbf{Q}-\mathbf{Q}'}^z + v_{\mathbf{Q}'}^yv_{\mathbf{Q}-\mathbf{Q}'}^z)Q_z\right]  \nn
\end{eqnarray}

Here we can use incompressibility condition to express $z$ component of magnetic filed and velocity from $x$ and $y$ components.
In order to simplify the consideration we can use phase portraits from previous chapter to motivate the approximation

\begin{equation}
F_\mathrm{ext} \approx -4Q_y \sum_{\mathbf{Q}'}Q'_\alpha\left(b_{\mathbf{Q}'}^yb_{\mathbf{Q}-\mathbf{Q}'}^y + v_{\mathbf{Q}'}^yv_{\mathbf{Q}-\mathbf{Q}'}^y\right)
\end{equation}

Thus we obtain important result: nonlinear terms for general 3D case with Coriolis force
have in approximation the same form as nonlinear term for 2D pure shear case.

\begin{equation*}
b_y(\overline{\tau},Q_y,Q_z)=
\int_{-\infty}^{\overline{\tau}} \frac{-4Q_y}{Q_\alpha}\sin[Q_\alpha(\overline{\tau}-\overline{\tau}_0)] \theta(\overline{\tau}-\overline{\tau}_0)
\mathrm{e}^{-\nu^{\prime}_{tot}(\overline{\tau}^3-\overline{\tau_0}^3)Q_y^2/6}
\cdot  \sum_{\mathbf{Q}'}Q'_\alpha\left(b_{\mathbf{Q}'}^yb_{\mathbf{Q}-\mathbf{Q}'}^y + v_{\mathbf{Q}'}^yv_{\mathbf{Q}-\mathbf{Q}'}^y\right) \mathrm{d}\overline{\tau_0}
\end{equation*}

\begin{eqnarray*}
 X(0) = \int_{0}^{\infty}\ \frac{\mathrm{d}Q_x}{Q_\alpha} \cos\left[Q_x\left(\sin(\theta)+\cos(\theta)\frac{Q_z}{Q_y}\right) \right]\exp\left[-\frac{\nu^{\prime}_{tot}}{6}\frac{Q_x^3}{Q_y}\right]
 \cdot \sum_{\mathbf{Q}'}Q'_\alpha\left(b_{\mathbf{Q}'}^yb_{\mathbf{Q}-\mathbf{Q}'}^y + v_{\mathbf{Q}'}^yv_{\mathbf{Q}-\mathbf{Q}'}^y\right) \nn\\
 Y(0) = -\int_{0}^{\infty}  \frac{\mathrm{d}Q_x}{Q_\alpha} \sin\left[Q_x\left(\sin(\theta)+\cos(\theta)\frac{Q_z}{Q_y}\right) \right] \exp\left[-\frac{\nu^{\prime}_{tot}}{6}\frac{Q_x^3}{Q_y}\right]
 \cdot \sum_{\mathbf{Q}'}Q'_\alpha\left(b_{\mathbf{Q}'}^yb_{\mathbf{Q}-\mathbf{Q}'}^y + v_{\mathbf{Q}'}^yv_{\mathbf{Q}-\mathbf{Q}'}^y\right)
\end{eqnarray*}

\begin{eqnarray}
I= && \int_{0}^\infty \frac{\mathrm{d}Q'_z}{2\pi} \int_{Q_y/2}^\infty  \frac{\mathrm{d}K_y}{2\pi} \int_{Q_x/2}^\infty \frac{\mathrm{d}K_x}{2\pi} Q'_\alpha D(Q'_\alpha) D(Q_\alpha-Q'_\alpha)  \\
   && \exp\left\{ -\frac{\nu'_{tot}}{6}\left[\frac{(K_x-Q_x/2)^3}{K_y+Q_y/2}+\frac{(K_x+Q_x/2)^3}{K_y-Q_y/2}\right]\right\} \cos\left(\hat{K_x} - \hat{Q_x} + \Delta\phi\right) \nonumber
\end{eqnarray}

\begin{eqnarray*}
&&\Delta\phi = \phi\left[ \sin(\theta)\left(K_y+\frac{Q_y}{2} \right) + \cos(\theta)(Q_z-Q'_z) \right] - \phi\left[ \sin(\theta)\left(\frac{Q_y}{2}-K_y \right) + \cos(\theta)Q'_z) \right] \\
&&\hat{K_x} = K_x\left(2\sin(\theta) +\cos(\theta)\frac{Q_z-Q'_z}{K_y+Q_y/2} - \cos(\theta)\frac{Q'_z}{K_y-Q_y/2} \right) \\
&&\hat{Q_x} = Q_x\frac12\cos(\theta)\left(\frac{Q_z-Q'_z}{K_y+Q_y/2} + \frac{Q'_z}{K_y-Q_y/2} \right)
\end{eqnarray*}

\begin{equation*}
b_y(\overline{\tau},Q_y,Q_z)=
\int_{-\infty}^{\overline{\tau}} \frac{-2Q_y}{Q_\alpha}\sin[Q_\alpha(\overline{\tau}-\overline{\tau}_0)] \theta(\overline{\tau}-\overline{\tau}_0)
\mathrm{e}^{-\nu^{\prime}_{tot}(\overline{\tau}^3-\overline{\tau_0}^3)Q_y^2/6} \cdot  I
\end{equation*}

\begin{eqnarray*}
I= && \int_{0}^\infty \frac{\mathrm{d}Q'_z}{2\pi} \int_{Q_y/2}^\infty  \frac{\mathrm{d}K_y}{2\pi} \int_{Q_x/2}^\infty \frac{\mathrm{d}K_x}{2\pi} \cos(\hat{K_x}) \exp\left\{ -\frac{\nu'_{tot}}{6}\left[\frac{(K_x-Q_x/2)^3}{K_y+Q_y/2}+\frac{(K_x+Q_x/2)^3}{K_y-Q_y/2}\right]\right\} \nonumber \\
   &&  Q'_\alpha D(Q'_\alpha) D(Q_\alpha-Q'_\alpha) \left[\cos(\hat{Q_x})\cos(\Delta\phi) + \sin(\hat{Q_x})\sin(\Delta\phi) \right] + \\
&& \int_{0}^\infty \frac{\mathrm{d}Q'_z}{2\pi} \int_{Q_y/2}^\infty  \frac{\mathrm{d}K_y}{2\pi} \int_{Q_x/2}^\infty \frac{\mathrm{d}K_x}{2\pi} \sin(\hat{K_x}) \exp\left\{ -\frac{\nu'_{tot}}{6}\left[\frac{(K_x-Q_x/2)^3}{K_y+Q_y/2}+\frac{(K_x+Q_x/2)^3}{K_y-Q_y/2}\right]\right\} \nonumber \\
&&  Q'_\alpha D(Q'_\alpha) D(Q_\alpha-Q'_\alpha) \left[\sin(\hat{Q_x})\cos(\Delta\phi) + \cos(\hat{Q_x})\sin(\Delta\phi) \right]
\end{eqnarray*}

\begin{eqnarray*}
&& I= \\
&& -\frac1{2\pi}\int_{0}^\infty \frac{\mathrm{d}Q'_z}{2\pi} \int_{Q_y/2}^\infty  \frac{\mathrm{d}K_y}{2\pi}
Q'_\alpha D(Q'_\alpha) D(Q_\alpha-Q'_\alpha) \left[\cos(\hat{Q_x})\cos(\Delta\phi) + \sin(\hat{Q_x})\sin(\Delta\phi) \right]
 \nonumber \\
   &&\frac{1}{2\sin(\theta) +\cos(\theta)\left(\frac{Q_z-Q'_z}{K_y+Q_y/2} -\frac{Q'_z}{K_y-Q_y/2}\right) } \\
   && \left\{ \sin\left[\frac{Q_x}2\left(2\sin(\theta) +\cos(\theta)\left(\frac{Q_z-Q'_z}{K_y+Q_y/2} - \frac{Q'_z}{K_y-Q_y/2}\right) \right) \right] + \exp\left[-\frac{q_x^3}{3} \cdot \frac{Q_y}{4K_y^2 - Q_y^2} \right]  \right\}\\
   && + \frac1{2\pi} \int_{0}^\infty \frac{\mathrm{d}Q'_z}{2\pi} \int_{Q_y/2}^\infty  \frac{\mathrm{d}K_y}{2\pi}
Q'_\alpha D(Q'_\alpha) D(Q_\alpha-Q'_\alpha) \left[\sin(\hat{Q_x})\cos(\Delta\phi) + \cos(\hat{Q_x})\sin(\Delta\phi) \right] \nonumber \\
   &&\frac{1}{2\sin(\theta) +\cos(\theta)\left(\frac{Q_z-Q'_z}{K_y+Q_y/2} -\frac{Q'_z}{K_y-Q_y/2}\right) } \\
   && \left\{ \cos\left[\frac{Q_x}2\left(2\sin(\theta) +\cos(\theta)\left(\frac{Q_z-Q'_z}{K_y+Q_y/2} - \frac{Q'_z}{K_y-Q_y/2}\right) \right) \right] -1 \right\}
\end{eqnarray*}

\begin{eqnarray*}
&& I = \\
    && \frac1{2\pi}\int_{0}^\infty \frac{\mathrm{d}Q'_z}{2\pi} \int_{Q_y/2}^\infty  \frac{\mathrm{d}K_y}{2\pi}
    Q'_\alpha D(Q'_\alpha) D(Q_\alpha-Q'_\alpha) \frac{1}{2\sin(\theta) +\cos(\theta)\left(\frac{Q_z-Q'_z}{K_y+Q_y/2} -\frac{Q'_z}{K_y-Q_y/2}\right) } \nonumber \\
    && \left\{ \sin\left[Q_x\left(\sin(\theta)  -\cos(\theta)\frac{Q'_z}{K_y-Q_y/2} \right) \right] - \sin(\hat{Q_x}) + \cos(\hat{Q_x})\exp\left[-\frac{q_x^3}{3} \cdot \frac{Q_y}{4K_y^2 - Q_y^2} \right]  \right\}\cos(\Delta\phi)\\
    && \left\{ \cos\left[Q_x\left(\sin(\theta) +\cos(\theta) \frac{Q_z-Q'_z}{K_y+Q_y/2} \right) \right] - \cos(\hat{Q_x}) +\sin(\hat{Q_x})\exp\left[-\frac{q_x^3}{3} \cdot \frac{Q_y}{4K_y^2 - Q_y^2} \right]  \right\} \sin(\Delta\phi)
\end{eqnarray*}


\begin{eqnarray*}
 X &=& -\frac1{2\pi}\int_{-\infty}^{0}\ \frac{\mathrm{d}Q_x}{Q_\alpha} \sin\left[Q_x\left(\sin(\theta) +\cos(\theta)\frac{Q_z}{Q_y} \right)\right]  \sin\left[Q_x\left(\sin(\theta)  -\cos(\theta)\frac{Q'_z}{K_y-Q_y/2} \right) \right] \\
 && \exp\left[-\frac{\nu^{\prime}_{tot}}{6} \cdot \frac{Q_x^3}{Q_y} \right]  \cdot \int_{0}^\infty \frac{\mathrm{d}Q'_z}{2\pi} \int_{Q_y/2}^\infty  \frac{\mathrm{d}K_y}{2\pi} \cos(\Delta\phi) \frac{Q'_\alpha D(Q'_\alpha) D(Q_\alpha-Q'_\alpha)}{2\sin(\theta) +\cos(\theta)\left(\frac{Q_z-Q'_z}{K_y+Q_y/2} -\frac{Q'_z}{K_y-Q_y/2}\right)}  \\
 Y &=& \frac1{2\pi}\int_{-\infty}^{0}\ \frac{\mathrm{d}Q_x}{Q_\alpha} \cos\left[Q_x\left(\sin(\theta) +\cos(\theta)\frac{Q_z}{Q_y} \right)\right] \cos\left[Q_x\left(\sin(\theta) +\cos(\theta) \frac{Q_z-Q'_z}{K_y+Q_y/2} \right) \right] \\
 && \exp\left[-\frac{\nu^{\prime}_{tot}}{6} \cdot \frac{Q_x^3}{Q_y} \right]  \cdot \int_{0}^\infty \frac{\mathrm{d}Q'_z}{2\pi} \int_{Q_y/2}^\infty  \frac{\mathrm{d}K_y}{2\pi} \sin(\Delta\phi) \frac{Q'_\alpha D(Q'_\alpha) D(Q_\alpha-Q'_\alpha)}{2\sin(\theta) +\cos(\theta)\left(\frac{Q_z-Q'_z}{K_y+Q_y/2} -\frac{Q'_z}{K_y-Q_y/2}\right)} \\
\end{eqnarray*}

\begin{eqnarray*}
 X &=& -\frac1{Q_\alpha}\frac{\Gamma(\frac43)}{(2\pi)^4} \left(\frac{6Q_y}{\nu_{tot}^\prime}\right)^{1/3} \int_{0}^\infty \mathrm{d}Q'_z \int_{Q_y/2}^\infty \mathrm{d}K_y  \frac{Q'_\alpha D(Q'_\alpha) D(Q_\alpha-Q'_\alpha) \cos(\Delta\phi) J_c}{2\sin(\theta) +\cos(\theta)\left(\frac{Q_z-Q'_z}{K_y+Q_y/2} -\frac{Q'_z}{K_y-Q_y/2}\right)}  \\
 Y &=& \frac1{Q_\alpha}\frac{\Gamma(\frac43)}{(2\pi)^4} \left(\frac{6Q_y}{\nu_{tot}^\prime}\right)^{1/3} \int_{0}^\infty \mathrm{d}Q'_z \int_{Q_y/2}^\infty \mathrm{d}K_y  \frac{Q'_\alpha D(Q'_\alpha) D(Q_\alpha-Q'_\alpha) \sin(\Delta\phi) J_s}{2\sin(\theta) +\cos(\theta)\left(\frac{Q_z-Q'_z}{K_y+Q_y/2} -\frac{Q'_z}{K_y-Q_y/2}\right)} \\
\end{eqnarray*}

 $$ R \equiv \frac{\Gamma(\frac43)}{(2\pi)^4}\left(\frac6{\nu'_\mathrm{tot}}\right)^{1/3},\quad \tilde{X} \equiv XR,\quad \tilde{Y} \equiv YR,\quad \tilde{D} \equiv DR. $$

\begin{eqnarray*}
\tilde{X} &=& - \frac{\sqrt[3]{Q_y}}{Q_\alpha} \int_{0}^\infty \mathrm{d}Q'_z \int_{Q_y/2}^\infty \mathrm{d}K_y  \frac{\left[\sin(\theta)(K_y + Q_y/2) + \cos(\theta)Q'_z\right] J_c}{2\sin(\theta) +\cos(\theta)\left(\frac{Q_z-Q'_z}{K_y+Q_y/2} -\frac{Q'_z}{K_y-Q_y/2}\right)} D(Q'_\alpha) D(Q_\alpha-Q'_\alpha) \cos(\Delta\phi) \\
\tilde{Y} &=& \frac{\sqrt[3]{Q_y}}{Q_\alpha} \int_{0}^\infty \mathrm{d}Q'_z \int_{Q_y/2}^\infty \mathrm{d}K_y  \frac{\left[\sin(\theta)(K_y + Q_y/2) + \cos(\theta)Q'_z\right]  J_s}{2\sin(\theta) +\cos(\theta)\left(\frac{Q_z-Q'_z}{K_y+Q_y/2} -\frac{Q'_z}{K_y-Q_y/2}\right)} D(Q'_\alpha) D(Q_\alpha-Q'_\alpha) \sin(\Delta\phi) \\
\end{eqnarray*}

\begin{equation}
\nu_\mathrm{wave}^\prime  \approx\frac{R}{(2\pi)^3} I_{3D} ,\quad I_{3D} = \int_0^\infty Q_y \tilde{D}_\mathrm{f}^2(Q_y,Q_z)\,\mathrm{d}Q_y\mathrm{d}Q_z
\end{equation}

\begin{eqnarray}
\tilde{D}_\mathrm{i} &=& \sqrt{\tilde{X}_\mathrm{i}^2+\tilde{Y}_\mathrm{i}^2},\quad \phi_\mathrm{i}=\arctan{\frac{\tilde{Y}_\mathrm{i}}{\tilde{X}_\mathrm{i}}} \nonumber \\
D_\mathrm{f}  &=&  D_\mathrm{i}G(Q_y,Q_z),\quad \phi_\mathrm{f} = F(\phi_\mathrm{i})\equiv \arctan \mathrm{\frac{s_{ig} s_{u} + s_{iu} s_{g}}{s_{ig} c_{u} + s_{iu} c_{g}}}
\end{eqnarray}

\appendix
\addappheadtotoc

\chapter{Source Code}

In the computer algebra system Maple the confluent Heun function reference is
$\mathrm{HeunC}(\alpha,\beta,\gamma,\delta,\eta,z)$.  This special
function obeys the equation
\be
z(z-1)y^{\prime\prime}
+ [\mathsf{A}z^2+\mathsf{B}z+\mathsf{C}]\,y'
+[\mathsf{D}z+\mathsf{E}]\,y=0,
\ee
where
\begin{eqnarray}
&&\mathsf{A}=\alpha,\qquad
\mathsf{B}=2+\beta+\gamma-\alpha, \qquad
\mathsf{C}=-1-\beta,\qquad
\\
&& \mathsf{D}=\frac{1}{2}[(2+\gamma+\beta)\alpha+2\delta],\qquad
\mathsf{E}=\frac{1}{2}[-\alpha(1+\beta)+(1+\gamma)\beta + \gamma +2\eta)].
\end{eqnarray}
For the series expansion
\be
y=\sum_{n=0}^\infty c_n =\sum_{n=0}^\infty a_n z^n,\qquad c_n\equiv a_nz^n
\ee
we arrive at the recursion
\be
\mathtt{F}a_{n-1}+\mathtt{G}a_n+\mathtt{H}a_{n+1}=0,
\ee
where
\begin{eqnarray}
&&\mathtt{F}=\delta +\frac{1}{2}(2n+\beta\gamma),\\
&&\mathtt{G}=n^2+(1+\gamma+\beta-\alpha)n
+\frac{1}{2}[\gamma+2\eta-\alpha +\beta(1-\alpha+\gamma)]\nn\\
&&\mathtt{H}=-(n+1)(n+1+\beta).
\end{eqnarray}
Supposing $a_{-1}=0$ and $a_0=1$ we use the recursion
\be
a_{n+1}=-(\mathtt{F}a_{n-1}+\mathtt{G}a_n)/\mathsf{H}
\ee
which, for example, gives
\be
a_1=\frac{1}{2}\left[\beta(1+\gamma-\alpha)+\gamma+2\eta-\alpha\right]/(1+\beta).
\ee
As the effective Schr\"odinger equation has a solution for
arbitrary $\xi\in(-\infty,\;+\infty)$, the formal series for Heun
function have convergent Pad\'e approximants.

Those Pad\'e approximants can be calculated by the well-known
$\varepsilon$-algorithm. First we calculate the series of the
partial sums in zeroth $(0)$ approximation
\be
S_i^{(0)}=S_{i-1}^{(0)}+c_i
\ee
and for the first $N$ terms we get
\begin{eqnarray}
&&S_0^{(0)}=c_0, \qquad S_1^{(0)}= c_0+c_1,\qquad\dots,\qquad
S_N^{(0)}=c_0+c_1+\dots +c_N.
\end{eqnarray}
Our problem is to calculate the limit of the sequence
\be
S=\lim_{n\rightarrow\infty} S_n.
\ee
For this calculation we generate the auxiliary sequence
\be
H_i^{(0)}=1/c_{i+1}
\ee
recalling that $1/0=0$, i.e., using pseudoinverse numbers if we have
to divide by zero:
\be
H_0^{(0)}=1/c_{1},\qquad
H_1^{(0)}=1/c_{2},
\qquad \dots, \qquad
H_{N-1}^{(0)}=1/c_{N}.
\ee

The epsilon-algorithm is the calculation of the recursion for
the series
\begin{eqnarray}
&&S_i^{(k)}=S_{i+1}^{(k-1)}+1/\left(H_{i+1}^{(k-1)}-H_i^{(k-1)}\right),\\
&&H_i^{(k)}=H_{i+1}^{(k-1)}+1/\left(S_{i+1}^{(k-1)}-S_i^{(k-1)}\right)
\end{eqnarray}
for all indices for which these relations make sense.

The maximal in modulus auxiliary element
$H_\mathrm{max}=\left|H_{I+1}^{(K)}\right|$ gives the best Pad\'e
approximant $S_{I}^{(K)}$ for the searched limes and the accuracy is
of the order of $1/H_\mathrm{max}$.

For each accuracy of the final result $\epsilon$ we can calculate
$S$ and $H$ sequences with an accuracy $\delta$ to assure that
$1/H_\mathrm{max}< \epsilon$.  In such a way we obtain a method for
calculating the confluent Heun function and the solution to the MHD
equation in power series of time. If from physical arguments we know that a
solution exists, the divergent series can be
summed. Having a method for calculating the confluent Heun
function, we can calculate both the even and odd solutions. The accuracy in calculating Heun functions is
controlled by the Wronskian
\be
W(\psi_\mathrm{g},\psi_\mathrm{u})
=
\left|\begin{array}{cc}
\psi_\mathrm{g}(\xi)&\psi_\mathrm{u}(\xi)\\
\mathrm{d}_\xi\psi_\mathrm{g}(\xi)&\mathrm{d}_\xi\psi_\mathrm{u}(\xi)
\end{array}\right|
=1.
\ee
The constants from the general solution are also given by the
Wronskians
\begin{eqnarray}
&& C_\mathrm{g}=W(\psi,\psi_\mathrm{u})
=\psi(\xi_0)\mathrm{d}_\xi\psi_\mathrm{u}(\xi_0)
-\psi_\mathrm{u}(\xi_0)\mathrm{d}_\xi\psi(\xi_0)
,\\
&& C_\mathrm{u}=W(\psi_\mathrm{g},\psi)
=\psi_\mathrm{g}(\xi_0)\mathrm{d}_\xi\psi(\xi_0)
-\psi(\xi_0)\mathrm{d}_\xi\psi_\mathrm{g}(\xi_0)
.
\label{ch2CoefficientsByWronskians}
\end{eqnarray}
Those formulae generally apply for a Cauchi problem where the
initial conditions are imposed on the function being
sought, i.e., on $\psi(\xi_0)$ and its derivative
$\mathrm{d}_\xi\psi_\mathrm{g}(\xi_0).$


\begin{verbatim}
////////////////////////////////////////
// 
//  After the Fortran90 program by E. Penev published in
//  T. Mishonov and E. Penev
//  "Thermodynamics of Gaussian fluctuations and paraconductivity in 
//  layered superconductors"  International Journal of 
//  Modern Physics B, Vol. 14, No. 32 (2000) 3831-3879.
//  
//  Former C version of this program is given in the preprint:
//  T.M. Mishonov, S.I. Klenov, E.S. Penev,
//  "Temperature dependence of the specific heat and the penetration
//  depth of anisotropic-gap BCS superconductors for a factorizable
//  pairing potential"  http://arxiv.org/abs/cond-mat/0212491v5
//  [cond-mat.supr-con] [v5] Wed, 28 Apr 2004
//
//  Owners: Todor M. Mishonov & Zlatan D. Dimitrov 
// 
//  Description: 
//  Finds the limit of a series in the case where only
//  the first N+1 terms b[i] are known.
//
//  Method:
//  The routine operates by applying the epsilon-algorithm
//  to the sequence of partial sums of a series supplied on input.
//  For desciption of the algorithm, please see 
//  G. Baker, Jr., and P. Graves-Morris, Pad´ Approximants,
//  G.-C. Rota editor, 
//  Encyclopedia of Mathematics and its Applications, 
//  Vol. 13, (Addison-Wesley, London,1981), Table 3, p. 78; 
//  C. Brezinski, Pad´e-type Approximation and General  Orthogonal
//  Polynomials (Birkhauser, 1980);
//  P. Wynn, Math. Tables Aids Comput. 10, 91 (1956);
//  D. Shanks, J. Math. Phys. 34, 1 (1955).
//
//////////////////////////////////////////////////////////////////////////////

                                                                                                                                                                                                                                                                                                                                                                                                                                                                                                                        #include <stdio.h>
#include <math.h>
#include <string.h>


//////////////////////////////////////////////////////////////////////////////
//////////    EPSILON ALGORITH    ////////////////////////////////////////////
//////////////////////////////////////////////////////////////////////////////
long double summa( int n, long double b[n], int *iP, int *kP, 
 long double *p_err,long double *pLimes,int *p_OK,int *p_nk,int *p_ierr)
// a[i], i=0 ... n
// Routine for calculation of optimal Pade approximant of series
{
int mnz=2;// Maximal Number of Zeroes in a[i]
int nz=0; // current Number of Zero a[i] at the moment

int i=0,k, imax, kPade=0, iPade=0, ierr=7,nPade, iOK=0, ndz=0;
// Pade approximant aLimes[iPade/kPade] +/- err;
// err == empirical evaluation of the error << 1
long double bmax=0, s[n],t[n][n],ew,aLimes=0,err,eps=1,norma=0,temp,scale=0;
float epsf=1,fLimes;
char report[100];

int mdl=1;// (modul) number of unified terms
int nk=(n+1)/mdl-1;
long double a[nk];


for(k=0;k<=nk;k++){
   a[k]=0;
   for(i=0;i<mdl;i++) 
     {a[k]+=b[mdl*k+i];}
                  }
n=nk;

// double precision machine epsilon should be a global variable
for(eps=1;eps+1>1;eps/=2) 
epsf=sqrt(eps); // atifitial single precision

// PROTECTION

for(i=0;i<=n;i++) norma+=fabs(a[i]);
if(norma==0) {ierr=2; strcpy(report,"norma==0"); goto label;} 
// this line is "insgurance" against all terms to be == 0

if( fabs(a[0]) > fabs(a[n]) ) for(i=0;i<=n;i++) 
	  {for(temp=0,k=i;k>=0;temp+=a[k],s[i]=temp,k--);}
                 else for(s[0]=a[0],i=1;i<=n;i++) s[i]=s[i-1]+a[i];
			      // summation starts from smallest terms
// implicitely we suppose s[-1]=0; the s[0]=s[-1]+a[0];
// then a[i]=s[i]-s[i-1] for i=0,1,2,3, ... 
// in Fortran program Limes the array s[i] is the imput variable

for(i=1;i<=n;i++)
{
if( a[i] != 0) { nz=0; if( 1/fabs(a[i])>bmax ) { 
bmax=1/fabs(a[i]); err=fabs(a[i]); iPade=i;aLimes=s[i];ierr=1;
strcpy(report,"Usual Taylor summation");
						}
		}
if( a[i] == 0 ) {nz++; if( nz == mnz ) 
    {aLimes=s[i-mnz]; iPade=i-nz, err=0; ierr=3; 
    strcpy(report,"a[i]==0,..., a[i-mnz] ==0; polynom?; no need of epsilon summation");
    goto label;} 
           }
} 
// the series could be fastly covergent -- no need of epsilon
// algorith the label means exit the end of the routine

for(i=0;i<=mnz;i++) scale+=fabs(a[i]);
for(i=0;i<=n;i++){ if(fabs(a[i])>scale/eps) {n=i; break;} } 
// good working line

//printf("n=%i, ierr=%i  \n",n,ierr);
if(n<3) {ierr=5; strcpy(report,"n<3"); aLimes=s[n]; iPade=n; goto label;} 
// this line is an "insgurance" for small n
// all partial summs sould be "comparable" within mashibe epsilon; 
// a new technolosical detail

// a lot of unnecessary details above!

// INITIALIZATION 
for(i=0;i<=n;t[0][i]=s[i],i++);// epsilon table
for(i=0;i<=n-1;i++)
{
if(a[i+1] != 0) {b[i]=1/a[i+1]; 
                 t[1][i]=b[i]; // epsilon table
                } 
else {b[i]=0; 
      t[1][i]=b[i]; // epsilon table
      ndz++; 
      ierr= 4; 
      strcpy(report,"if( a[i+1] == 0) then 1/0=0");}
      // should be harmless division by zero
}

// BEGINING OF EPSILON ALGORITH
// epsilon algorith beginning with 2nd row: (E-W)*(S-N) ==1;
imax=n-2;
for(k=2;k<=n+1;k++) // odd k means pade row s[i]; even k help row b[i]
	{       
	
	if(k%2==0) { for(i=0;i<=imax;i++) 
			{s[i]=s[i+1]; 
                         ew=b[i+1]-b[i];
			 if(ew !=0) s[i]+=1/ew; 
			 t[k][i]=s[i]; // epsilon table
			}
		   }
         
	if(k%2==1) { for(i=0;i<=imax;i++) 
			{b[i]=b[i+1]; 
		         ew=s[i+1]-s[i];
			 if(ew !=0) b[i]+=1/ew; 		
                         t[k][i]=b[i]; // epsilon table
                                               if ( bmax < fabs(b[i]) ) {
						bmax=fabs(b[i]); 
	      // maximal help row element criterion; the MAIN detail! 
						aLimes=s[i];
						kPade=(k-1)/2;
						iPade=i+kPade;
						err=1/bmax;
						ierr=0;	
		strcpy(report,"ierr=0;The regular work of epsilon algorith");
			 						  }

                        }
		   }

		imax--; 
	if(imax<0) break;
	}
label: nPade=iPade+kPade+1;// label for emrgency exit
if(err/norma<epsf) {fLimes= (float) aLimes; iOK=1;}
printf("norma>>>>%Le\n",norma);
puts(report);
printf("iPade=%i,kPade=%i,err=%Le,ndz=%i,nz=%i,i=%i,nPade=%i,n=%i\n",
iPade,kPade,err,ndz,nz,i,nPade,n);
printf(">>>>>iOK=%i<<<<<\n",iOK);
*iP=iPade;
*kP=kPade;
*p_err=err;
*pLimes=aLimes;
*p_OK=iOK;
*p_nk=nk;
*p_ierr=ierr;
return aLimes;
}


/////////////////////////////////////////////////////////////////////////////
// FUNCTION

long double function (long double x)
{
int n=51,iPade, *p=&iPade ; //n=7 min pri sdvoiavane(mdl=2) ili n=3 
int kPade,iOK,nk,ierr;
long double result, err,aLimes,sign;
long double a[99];
int i=0;
  
/////
long double v[100],suma=0;
/long double alpha=0,beta=-0.5,gamma=0,delta=-0.025,eta=0.275,az=0,buki=0,vedi=0;
////

//HeunC(alpha,beta,gamma,delta,eta;Z)

/*v[0]=1;
v[1]=(gamma - alpha + beta*(gamma-alpha+1) + 2*eta)/(2*(beta+1));
for(i=2;i<n+1;i++)
{
	m=i-1;
			az=delta + (alpha*0.5)*(m+m + beta + gamma);
buki=m*(m+1+beta+gamma-alpha) + 0.5*(gamma+eta+eta-alpha+beta*(1-alpha+gamma));
			vedi=-(m+1)*(m+1+beta);
			v[i]=-(v[i-1]*buki + v[i-2]*az)/vedi;
}

*/
sign=1;
v[0]=0;
for(i=1;i<=n;i++)
{
v[i]=sign/i;
sign=-sign;
//printf("v[%i]=%Lg\n",i,v[i]);
}

long double xn;
xn=1;
for(i=0;i<=n;i++)
{
a[i]=v[i]*xn;xn*=x;
//printf("a[%i]=%Lg\n",i,a[i]);
suma+=a[i];
}
//printf("\n suma= %.12Lf\n ",suma);
// ln(1+x)=x-x^2/2+x^3/3-x^4/4+x^5/5-...


/*
// here cut %%%%%%%%%%%%%%%%%%%%%%%%%%%%%%%
int mdl=2;// number of unified terms
int nk=n/mdl;
for(k=0;k<=nk;k++){
   b[k]=0;
   for(i=0;i<mdl;i++) 
     {b[k]+=a[mdl*k+i];}
                  }
// here cut %%%%%%%%%%%%%%%%%%%%%%%%%%%%%%%%
*/

result=summa(n,a,p,&kPade,&err,&aLimes,&iOK,&nk,&ierr);

return result;

}



int main(void)
{
long double x,z;

x=exp(1)-1;
//for(x=0.1;x<15;x+=0.1)
//{
z=function(x);//z=ln(e)=1;
printf("%Lg %.12Lf\n",x,z);
printf("ln(1+x)=%.12f\n",log(M_E));
//}

return 0;
}


\end{verbatim}

\chapter{Integral Equations}

\begin{verbatim}
 
#include <stdio.h>
#include <math.h>

double Q0=M_PI/2;

double deltag( double q)
{
return ( -M_PI/2. + atan(fabs(q)/Q0) );
}

double deltau( double q)
{
return ( -M_PI/2. );
}

int main(void)
{
int N=500,N1=N+1,i,j,k,ja,ka,iter,Niter=65;
double Qmax=5.,DQ=Qmax/N, Q,P,X,Y,XQ,YQ,x,y,sig,siu,aN,rN=1.,rNnew,rN2=1.,zN2,zN,d_reg,err1,err2,err3=0.,rID,Qav;
double D[N1],D_i[N1],phi[N1],phi_i[N1],c[N1],s[N1],q[N1],dg[N1],du[N1],sg[N1],cg[N1],su[N1],cu[N1],sug[N1],G[N1],phinew[N1],Dnew[N1],Sx[N1];

 for(i=1;i<=N;i++)
	{
	q[i]=DQ*i;
	dg[i]=deltag(q[i]);
        sg[i]=sin(dg[i]);
        cg[i]=cos(dg[i]);
        du[i]=deltau(q[i]);
        su[i]=sin(du[i]);
        cu[i]=cos(du[i]);
        sug[i]=sin(du[i]-dg[i]);
	}

 for(i=1;i<=N;i++)
	{
	D[i]=1.;
	phi[i]=0.;
        c[i]=cos(phi[i]);
        s[i]=sin(phi[i]);
	}
        
for(iter=0;iter<=Niter;iter++)// beginning of iteration procedure
{
	for(i=1;i<=N;i++)
           {// tova e cycle po Q
	    Q=q[i];  
            XQ=0.; YQ=0.;
		  for(j=-N;j<=N;j++)
		    {// tova e cycle po P in (-infty,+infty)
		      ja=abs(j);
		      P=DQ*j;
		      k=i-j; ka=abs(k);//QP=Q-P=DQ*(i-j)=DQ*k 
		      if(ka>N) continue;
		      if(ka==0) continue; // i,ja,ka in [1,N]]
		      if(ja==0) continue; // i,j,k in [-N, -1] U [1, N]
		      XQ+=P*D[ja]*D[ka]*(c[ja]*c[ka]+s[ja]*s[ka]);//*DQ
		      YQ+=P*D[ja]*D[ka]*(s[ja]*c[ka]-c[ja]*s[ka]);//*DQ
	              		    }// next j
	   d_reg=pow(fabs(Q),2./3);          		        
           X=(XQ*DQ)/ d_reg;//*DQ
           Y=(YQ*DQ)/ d_reg;//*DQ
           D_i[i]=sqrt(X*X+Y*Y);// Dnew_initial
           phi_i[i]=atan2(Y,X);
           sig=sin(phi_i[i]-dg[i]);
           siu=sin(phi_i[i]-du[i]);
           
	   x=sig*cos(du[i])+sin(phi_i[i]-du[i])*cos(dg[i]);
           y=sin(phi_i[i]-dg[i])*sin(du[i])+siu*sin(dg[i]);
           phi[i]=atan2(y,x);       
	   aN=(sig*su[i]+siu*sg[i])*(sig*su[i]+siu*sg[i])
             +(sig*cu[i]+siu*cg[i])*(sig*cu[i]+siu*cg[i]); // a Numerator
           G[i]=sqrt(aN)/fabs(sug[i]);
	   Dnew[i]=G[i]*D_i[i];
	}// next i  
	
        rN2=0.;// real norma D	
        for(i=1;i<=N;i++) {rN2+=Dnew[i]*Dnew[i]*DQ;}
        rNnew=sqrt(rN2); 
        if(iter==Niter-1){err2=(rNnew-1/zN)*2/(rNnew+1/zN);}
        err1=(rNnew-rN)*2./(rNnew+rN);
        rN=rNnew;

        for(i=1;i<=N;i++) {Dnew[i]/=rN;}// Normalization. Now |Dnew|=1
        
        for(i=1;i<=N;i++) 
	   {
	     
	    if(iter == Niter-1) { err3+=fabs(D[i]-Dnew[i])/fabs(D[i]); }//err3+=fabs(D[i]-Dnew[i]);}
	    D[i]=Dnew[i];
            c[i]=cos( phi[i]);
            s[i]=sin( phi[i]);
           }// getting old

        if(iter==Niter-1) {zN=rN; for(i=1;i<=N;i++) D[i]/=zN; D_i[i]=D[i]/G[i]; }// |D|=1/zN

} // next iter


		 //for(i=1;i<=N;i++)
			//printf("%g %g\n",q[i],D[i]*D[i]*q[i]);
			//printf("%g %g\n",q[i],D[i]*D[i]*q[i] );
			//printf("%g %g\n",q[i],exp(0.69286746320344-40.915167099567*q[i]));
			

       for(rID=0.,Qav=0.,i=1;i<=N;i++)
       {
	Sx[i]=D[i]*D[i]*q[i];
        rID+=Sx[i]*DQ;
	Qav+=q[i]*Sx[i]*DQ;
	//printf("%g %g\n",q[i],D[i]*q[i]);
       } 
       Qav/=rID;
       printf("Niter=%i  rID=%g  Qav=1/%g, Qav/DQ=%g,  err3=%g \n",Niter,rID,1/Qav, Qav/DQ, err3);


return 0;
}



\end{verbatim}

\chapter{Fast oscillation function}

\begin{eqnarray}
 I=\int_0^\infty \cos(kx)f(x)\mathrm{d}x=\frac1k\int_0^\infty f(x)\mathrm{d}\sin(kx)=\left.\frac1k f(x)\sin(kx)\right|_0^\infty - \frac1k\int_0^\infty\sin(kx)\mathrm{d}f(x)  \nn \\ 
  =-\frac1k\int_0^\infty f'(x)\sin(kx)\mathrm{d}x = \frac1{k^2}\int_0^\infty f'(x)\mathrm{d}\cos(kx)=\left.\frac1{k^2}f'(x)\cos(kx)\right|_0^\infty - \frac1{k^2}\int_0^\infty\cos(kx)f''(x)\mathrm{d}x \nn\\
  =-\frac{f'(0)}{k^2} - \frac1{k^2}\int_0^\infty\cos(kx)f''(x)\mathrm{d}x=\int_0^\infty\cos(kx)f(x)\mathrm{d}x
\end{eqnarray}

\begin{eqnarray}
 I=-\frac{f'(0)}{k^2} -\frac1{k^2}\left[-\frac{f'''(0)}{k^2} -\frac1{k^2}\int_0^\infty\cos(kx)f^{(4)}(x)\mathrm{d}x \right] \nn\\
  = \frac{f'(0)}{k^2} +\frac1{k^2}\left[-\frac{f^{(3)}(0)}{k^2} +\frac1{k^2}\left\{ \frac{f_0^{(5)}}{k^2} + \frac1{k^2}\int_0^\infty\cos(kx)f^{(6)}(x)\mathrm{d}x \right\}\right] \nn\\
  = \frac{f_0^{(1)}}{k^2} + \frac{f_0^{(3)}}{(k^2)^2} - \frac{f_0^{(5)}}{(k^2)^3} + \frac{f_0^{(7)}}{(k^2)^4} - \frac{f_0^{(9)}}{(k^2)^5} + \dots \nn\\
  = -\left.\frac1{k^2}\left[1 - \frac1{(k^2)^1}\mathrm{d}^2_x + \frac1{(k^2)^2}(\mathrm{d}^2_x)^2 - \frac1{(k^2)^3}(\mathrm{d}^2_x)^3 + \dots  \right] \mathrm{d}_xf(x)\right|_0
\end{eqnarray}

\begin{equation}
 \frac1{1+q}=1-q+q^2-q^3+q^4-q^5
\end{equation}

\begin{equation}
 I=-\left.\frac1{k^2}\left[\frac1{1+\frac{\mathrm{d}_x^2}{k^2}} \right] \mathrm{d}_xf(x)\right|_0 = -\left.\left[\frac{\mathrm{d}_x}{k^2+\mathrm{d}_x^2} \right]f(x)\right|_0
\end{equation}

Example:

\begin{eqnarray}
 f(x)=\mathrm{e}^{-\alpha x},\quad \mathrm{d}_xf(x)=-\alpha f(x),\quad \mathrm{d}_x^n f(x)=(-\alpha)^n f(n) \\
\frac1{k^2+\mathrm{d}_x^2}\mathrm{e}^{-\alpha x}=\frac1{k^2+(-\alpha)^2}\mathrm{e}^{-\alpha x}
\end{eqnarray}

\begin{eqnarray}
 J=\int_0^\infty \sin(kx)f(x)\mathrm{d}x=-\frac1k\int_0^\infty f(x)\mathrm{d}\cos(kx)=-\left.\frac1k f(x)\cos(kx)\right|_0^\infty + \frac1k\int_0^\infty\cos(kx)\mathrm{d}f(x)  \nn \\ 
  =\frac{f_0}{k} + \frac1{k^2}\int_0^\infty f'(x)\mathrm{d}\sin(kx)=-\left.\frac1{k^2}f'(x)\sin(kx)\right|_0^\infty + \frac1{k^2}\int_0^\infty\sin(kx)f''(x)\mathrm{d}x \nn\\
  =\frac{f'(0)}{k^2} + \frac1{k^2}\int_0^\infty\sin(kx)f''(x)\mathrm{d}x=-\int_0^\infty\sin(kx)f(x)\mathrm{d}x
\end{eqnarray}

\begin{eqnarray}
 J=-\frac{f'(0)}{k^2} - \frac1{k^2}\left[-\frac{f'''(0)}{k^2} -\frac1{k^2}\int_0^\infty\cos(kx)f^{(4)}(x)\mathrm{d}x \right] \nn\\
  = \frac{f'(0)}{k^2} + \frac1{k^2}\left[-\frac{f^{(3)}(0)}{k^2} +\frac1{k^2}\left\{ \frac{f_0^{(5)}}{k^2} + \frac1{k^2}\int_0^\infty\cos(kx)f^{(6)}(x)\mathrm{d}x \right\}\right] \nn\\
  = \frac{f_0^{(1)}}{k^2} + \frac{f_0^{(3)}}{(k^2)^2} - \frac{f_0^{(5)}}{(k^2)^3} + \frac{f_0^{(7)}}{(k^2)^4} - \frac{f_0^{(9)}}{(k^2)^5} + \dots \nn\\
  = -\left.\frac1{k^2}\left[1 - \frac1{(k^2)^1}\mathrm{d}^2_x + \frac1{(k^2)^2}(\mathrm{d}^2_x)^2 - \frac1{(k^2)^3}(\mathrm{d}^2_x)^3 + \dots  \right] \mathrm{d}_xf(x)\right|_0
\end{eqnarray}

\setsecnumdepth{none}
\maxsecnumdepth{none}
\chapter{Acknowledgement}

This work was partially supported by Scientific fund of St. Clement of Ohrid University at Sofia
under grant number 147-2012.
\maxsecnumdepth{subsubsection}
\setsecnumdepth{subsubsection}
\backmatter

\end{document}